\documentclass[a4paper,fleqn]{cas-sc}

\usepackage[numbers]{natbib}

\begin{document}
\let\WriteBookmarks\relax
\def\floatpagepagefraction{1}
\def\textpagefraction{.001}
\shorttitle{Relativistic quantum theory and algorithms}
\shortauthors{S. Taioli et~al.}

\title [mode = title]{Relativistic quantum theory and algorithms: a toolbox for modeling many-fermion systems in different scenarios}                      
\tnotemark[1]

\tnotetext[1]{This document is the results of the research project PANDORA funded by the National Institute of Nuclear Physics. S.T is also funded by the Bruno Kessler Foundation.}


\author[1,2,3]{Simone Taioli}[orcid=0000-0003-4010-8000]
\cormark[1]
\ead{taioli@ectstar.eu}


\address[1]{European Centre for Theoretical Studies in Nuclear Physics and Related Areas (ECT*-FBK), Trento, Italy}
\address[2]{Trento Institute for Fundamental Physics and Applications (TIFPA-INFN), Trento, Italy}
\address[3]{Peter the Great St. Petersburg Polytechnic University, Russia}

\author[4,5]{Stefano Simonucci}[orcid=0000-0001-9287-8651]

\cormark[2]


\address[4]{School of Science and Technology, University of Camerino, Italy}
\address[5]{INFN, Sezione di Perugia, Italy}

\ead{stefano.simonucci@unicam.it}


\cortext[cor1]{Principal Corresponding author}
\cortext[cor2]{Corresponding author}


\begin{abstract}
In this chapter we focus first on the theoretical methods and relevant computational approaches to calculate the electronic structure of atoms, molecules, and clusters containing heavy elements for which relativistic effects become significant. In particular, we discuss the mean-field approximation of the Dirac equation for many-electron systems, and its self-consistent numerical solution by using either radial mesh or Gaussian basis sets. The former technique is appropriate for spherical symmetric problems, such as atoms, while the latter approach is better suited to study non-spherical non-periodic polycentric systems, such as molecules and clusters.
We also outline the pseudopotential approximation in relativistic context to deal with the electron-ion interaction in extended systems, where the unfavourable computational scaling with system size makes it necessary. As test cases we apply our theoretical and numerical schemes to the calculation of the electronic structure i) of the gold atom, and ii) of the superatom W@Au$_{12}$, where the inclusion of spin-orbit effects is crucial to the accurate understanding of the electronic properties.
Furthermore, we describe the extension of our relativistic approach to deal with nuclear reactions driven by the weak force, such as the electron capture and $\beta$-decay, also at finite temperature in astrophysical scenarios, using the Fermi--Dirac statistics. The latter processes are indeed major drivers of the nucleosynthesis of the elements in stars and, thus, their understanding is crucial to model the chemical evolution of the Universe.
Finally, we show the application of our relativistic quantum mechanical framework to the assessment of the elastic differential scattering cross section of electrons impinging on molecular targets, notably liquid water. The latter process, together with several inelastic scattering collisions by which secondary electrons deposit their energy, represents a fundamental event of the chain of the physico-chemical mechanisms initiated by the passage of fast ion beams through a bio-medium. This technique is used in hadrontherapy for cancer cure.
\end{abstract}



\begin{keywords}
Relativistic quantum mechanics \sep Computational methods \sep Gaussian functions \sep Real-space grids \sep Gold atom and W@Au$_{12}$ \sep $\beta$-decay \sep Stars \sep Cancer cure \sep Electron-molecule elastic scattering
\end{keywords}

\maketitle

\section{Introduction}

The relativistic extension of quantum mechanics to deal with many-particle systems, such as nuclei, atoms, molecules and clusters, represents the most fundamental theory of all molecular sciences \cite{rehher2009}. Indeed, it combines the laws of special relativity, to which natural systems obey, with the realm of quantum mechanics. \\
\indent Nonetheless, the description of the motion of valence electrons, which drive the chemistry of materials, typically relies on a phenomenological approach whereby the Schr{\"o}dinger equation is solved with the possible addition of a spin-orbit coupling term to the effective quantum-mechanical Hamiltonian for a system of electrons. The latter term corrects the quantum-mechanical theory for relativistic effects e.g. due to the presence of heavy elements or large magnetic moments leading to significant spin-orbit interactions in the system. However, this phenomenological approach, whose attractiveness lays essentially in its affordable computational cost with respect to the fully relativistic quantum electrodynamics methods, fails to deliver accurate results comparable to experiments, most notably in heavy element compounds where electrons attain sufficient speeds that the four-dimensional nature of the basic constituents of matter cannot be neglected. In fact, the theory of light interacting with many-electrons systems must adopt a rigorous gauge invariant and Lorentz covariant form to describe both electromagnetic and fermionic degrees of freedom and their interaction simultaneously \cite{rehher2009}. \\
\indent This chapter aims to describe the theoretical and computational methods to simulate the electronic structure of many-electron systems for which relativistic effects are significant. Specifically, we present advanced algorithms to find the numerical solution of the relativistic quantum-mechanical Dirac equation using a mean field approximation. This theoretical and computational framework, which is called the Dirac--Hartree--Fock (DHF) method, is implemented  either using real-space grids or by introducing functional spaces of square-integrable ($L^2$) functions. \\
\indent The former numerical method consists in the discretization in space by using a grid of points to sample the (electron or nucleon) wavefunctions to a desired accuracy. While grids can be used for problems with any number of dimensions, they are typically preferred for dealing with spherical symmetric systems as they reduce to the simple one-dimensional case upon factorizing out the wavefunction angular dependence.
The big advantages of real-space grid calculations are their simplicity, versatility, and the possibility to systematically improve the calculation accuracy by increasing the mesh density as the error is proportional to a power law of the step size (defined as the distance between neighboring points of the grid). Furthermore, boundary conditions, particularly for solving the DHF equation for scattering states, are easily imposed. Finally, the use of pseudopotentials to limit the explicit treatment of the electron-electron and electron-ion interaction to outer electrons only, which we will describe in the context of the Hermite Gaussian Basis Functions \cite{taioli2009surprises,taioli2010electron,taioli2015computational} (HGBF), can be also efficiently implemented in grid-space methods.\\
\indent The grid-based approach can be considered a limiting case (for infinitesimally small Gaussian widths) of the second numerical method that we describe here, which relies on multi-centered HGBF -- chosen as prototypes of $L^2$ functions -- spanning the linear vector space $\cal G$ of bound and scattering states \cite{taioli2009surprises,taioli2010electron,taioli2015computational,taioli2009mixed,taioli2021resonant}. Despite the solution cannot be systematically improved, at variance with space grid methods (or plane-waves), due to Basis Set Superimposition (or over-linear) errors \cite{TZELI201042}, an advantage of using HGBF is that it is possible to derive explicit analytical expressions of the integrals necessary to carry out atomic and molecular calculations.\\
\indent Applications will be focused on non periodic many-body systems interacting via the Coulomb force, thus mainly atoms, molecules, and clusters. However, we will show how our theoretical and computational methods are general enough to deal with any (pseudo-) scalar (e.g. the pion field) and any (pseudo-) vector potential (e.g. the Yukawa interaction between nucleons). In this regard, we will use our approach also in problems typical of the nuclear physics, such as the theory of beta decay of unstable nuclei induced by the electroweak force (for which we refer the reader to Ref. \cite{morresi2018nuclear}). This approach has been implemented in a simulation package called DIRECT (DIRac Equation from Camerino and Trento). \\
\indent In particular, this fully relativistic quantum-mechanical framework will be tested first to carry out the all-electron calculation of the electronic structure of the gold ion (Au$^+$). We compare the numerical results of some electronic structure properties, such as the ionization potential (\`a la Koopmans), by using either radial or HGBF basis sets. The chemistry of gold is indeed shaped by strong relativistic effects, whereby the radius orbital of the core electrons shrinks by more than 20\%. \\
\indent Furthermore, as a second test case we will discuss the W@Au$_{12}$ superatomic system, where the inclusion of spin-orbit effects is necessary to describe the spectroscopic features. The 13-atom W@Au$_{12}$ cluster has been experimentally observed as an highly stable icosahedral cage of 12 gold atoms containing a central tungsten, and characterised by photoelectron spectroscopy (PES) as a closed-shell electron configuration with 18 valence electrons and an HOMO--LUMO gap around 3 eV \cite{https://doi.org/10.1002/anie.200290048,C5NR07246H}. This makes it chemically inert.
Gold-based nanoclusters of different size and shape have in general attracted increasing interest, primarily due to their unique and tunable electronic and optical properties. They can find application in the development of photovoltaic cells or batteries, of biological and chemical sensors, or could replace toxic or rare elements in biomedical devices, being biocompatible and chemically inert. More recently, the use in radiation therapy for cancer cure of high-Z materials, such as gold, platinum, gadolinium, and iron nanoparticles (NPs) with a size that ranges between 1 to 100 nm, has been investigated and shown to increase secondary electrons and free radical production \cite{Jain,Porcel_2010,doi:10.1259/bjr.20140134,doi:10.1063/1.3589914,Kim_2012,Kuncic_2018,Schuemann_2020,https://doi.org/10.1002/wnan.1656}, eventually enhancing the Relative Biological Effectiveness (RBE) of hadrontherapy. By selectively accumulating near the tumour region gold NPs, the RBE can be potentially enhanced by 15–20\% due to an increase in both direct and indirect damage. However, while some of these materials can be toxic for humans, gold NPs might offer a viable biocompatible alternative to enhance localized dose and biodamage under photon \cite{BRIGGS20131098} and proton irradiation \cite{https://doi.org/10.1002/ppsc.201300276}, once an adequate concentration of NPs in the tumour is achieved. In this context, we will describe also the recent implementation of a fully relativistic theory of the pseudopotential approach, which is a necessary approximation to achieve an efficient scaling in the electronic structure calculations of heavy-element compounds. \\
\indent \indent A second application of our fully relativistic framework, which goes  beyond the treatment of bound states, concerns the interpretation of beta-decay spectra of heavy nuclei in both Earth conditions \cite{morresi2018nuclear} and astrophysical \cite{Simonucci_2013} scenarios.\\
\indent 
The use of radioisotopes that decay via beta-emission is widespread in science and medicine. On the one hand, $\beta$-decay processes are the dominant source of background in liquid Xenon dark matter detectors \cite{hayen2020detailed} or in cerium-based LaBr$_{3}$-based crystals that are used to fabricate scintillators for applications in gamma-ray spectroscopy and, more generally, to characterise ionizing radiation \cite{morresi2018nuclear}. $\beta$-emitting radiopharmaceuticals are widely used also in clinical oncology to diagnose cancer and to deliver radiation therapy in cancer treatment, or in the field of PET imaging \cite{Welsh}.\\
\indent On the other hand, stars have been known since decades to be the cauldrons 
where heavy chemical elements are forged and matter becomes complex starting from building blocks such as H, He and
other very light elements ($A<8$) produced by Big Bang nucleosynthesis (BBN) in the early stage of the Universe \cite{RevModPhys.29.547,alma998733153405216,Sean}. This 
understanding relies on the discovery that some old stars host isotopes, whose lifetime requires in situ formation. 
Theoretical models and observations thus support the conclusion that reactions occurring within stars are responsible for heavy elements formation. Nevertheless, the solution to the global problem of studying the chemical evolution of the Universe, from the evolutionary path after core He burning  to the mechanisms driving core-collapse supernovae and gamma ray bursts, is still missing and we are only now obtaining first rough matches between models and observations \cite{doi:10.1146/annurev.aa.23.090185.001535}.
Indeed, while robust confirmation of the validity of the solar model can be derived by the analysis of solar neutrinos, helio-seismology and improved experimental nuclear physics data \cite{RevModPhys.83.195},
a thorough understanding of the processes of nucleosynthesis occurring at advanced stages of star evolution is uncertain \cite{Palmerini_2013}, even for masses low enough to experience only H- and He-burning phases (Low and Intermediate Mass Stars, with $M < 6-8$ solar masses). This could be also the case of the cosmological Li abundance, which is still unexplained within the framework of the Big Bang Standard Model, despite
almost all its predictions on isotopic abundances are confirmed \cite{Simonucci_2013,Nakamura_2010}. \\
\indent 
In fact, the problem of determining an ample selection of stellar elemental abundances, the isotopic composition of meteorites and of the interstellar medium, is cumbersome and its solution inherently needs an interdisciplinary look. Indeed, this description must integrate a variety of approaches, ranging from the description of nuclear reaction cross sections to the fine tuning of evolutionary models. In particular, a thorough understanding of the nucleosynthesis in stars must include the accurate determination of $\beta$-decay and electron capture rates (other than slow and rapid neutron capture processes) in astrophysical scenarios. Here we discuss the application of our fully-relativistic quantum method for calculating weak interaction processes in nuclear astrophysics context (typically occurring in post-main sequence long-lived stages of stars).
These methodological advances to determine reaction rates will lead to a more quantitative-based knowledge of the advanced evolutionary stages in massive stars and core-collapse supernovae.
 \\
\indent 
A last example to demonstrate the diverse applicability of our relativistic quantum mechanical scheme will be the determination of the elastic mean free path (EMFP) of electrons moving within a medium. This approach, based on scattering theory, will be used in particular to assess the electron EMFP in liquid water. The latter study is relevant for modelling the biodamage induced by ion irradiation in materials of biological relevance \cite{Nikjoo2016RepProgPhys,Kraft2000}, such as the human body tissue typically mimicked by liquid water \cite{doi:10.1021/acs.jpclett.0c03250}. Indeed, accurate ab initio calculations of elastic events, together with the assessment of several other chemical-physical mechanisms of electron energy loss (ionization, excitation, dissociative electron attachment \cite{taioli2006waterwaves,taioli2006wave}) initiated by ion irradiation, make it possible to obtain reliable probabilities to assess clustered DNA damage \cite{doi:10.1021/acs.jpclett.0c03250}.\\   
\indent 
This chapter is organized as follows: section 2 outlines the theoretical foundations of our relativistic quantum chemical method; section III deals with the computational algorithms devised to find the numerical solution of the DHF equation using either a space-grid or HBGF basis sets; sections IV, V, and VI are devoted to the application of our method to the calculations of the electronic properties of Au$^+$ and W@Au$_{12}$, to the simulation of $\beta-$decay of $^{31}$Si, $^{89}$Sr and $^{134}$Cs, and to the electron EMFP calculation in liquid water, respectively.

\section{Wavefunction calculation within the mean field approximation}

\subsection{The Dirac-Hartree-Fock method}

\indent The many-particle Dirac equation reads \cite{rehher2009} (everywhere we use atomic units otherwise stated):

\small
\begin{equation}\label{diroc}
\left[\sum_{i}\left(c{\bf{\alpha}}_{i}\cdot\mathbf{p}_{i}+\beta_{i}mc^{2}+V_{i}\right)+\sum_{i<j}\left( 1-\mathbf{\alpha}_{i}\cdot \mathbf{\alpha}_{j}\right)g_{ij}\right]\psi\left(\mathbf{r}_{1},\cdots\mathbf{r}_{N}\right)=E\psi\left(\mathbf{r}_{1},\cdots\mathbf{r}_{N}\right)
\end{equation}
\normalsize
where $\alpha_i$ and $\beta$ are the following Hermitian involutory 4 $\times$ 4 matrices
\begin{eqnarray}
\mathbf{\alpha}=\left(\begin{array}{cc}
0 & \mathbf{\sigma}\\
\mathbf{\sigma} & 0
\end{array}\right),\qquad\beta=\left(\begin{array}{cc}
I & 0\\
0 & -I
\end{array}\right)
\end{eqnarray}
that mutually anticommute, $\sigma_i$ are the Pauli matrices, and $I$ is the 2$\times$2 identity matrix. In equation (\ref{diroc}) $V_{i}$ is an external potential, which may represent a general interaction between fermions.

When both scalar ($g_S$) and vector ($g_V$) potentials are in place, the Dirac equation (\ref{diroc}) can be written \cite{morresi2018nuclear}: 

\begin{equation}\label{direq}
\left\{ \sum_i \left(c\mathbf{\alpha}_i \cdot \mathbf{p}_{i}+\beta_i mc^2 + V_i \right) + \sum_{i<j}\left[\beta_i \beta_j g_{S,ij} + \left(1-\mathbf{\alpha}_i \cdot \mathbf{\alpha}_j \right)g_{V,ij}\right]\right\} \psi\left(\mathbf{r}_1, \cdots \mathbf{r}_N \right)=E\psi\left(\mathbf{r}_1,\cdots \mathbf{r}_N \right)
\end{equation}

It is convenient to write the Dirac Hamiltonian in second quantization, as fermionic anticommutation rules are therein automatically included in the formalism. The Dirac equation (\ref{direq}) reads:

\begin{eqnarray}
\label{genH}
&H= \sum_{s_{1}s_{2}}\int d \mathbf{r} \hat{\psi}_{s_{1}}^{+}(\mathbf{r})\left[-ic\mathbf{\alpha}_{s_1 s_2} \cdot \mathbf{\nabla}+\beta_{s_1 s_2} mc^{2} +\delta_{s_1 s_2} V(\mathbf{r})\right]\hat{\psi}_{s_{2}}(\mathbf{r}) + \nonumber \\
&\frac{1}{2}\sum_{s_1 s_2 s_1^{\prime}s_2^{\prime}}\int d\mathbf{r} d\mathbf{r}^{\prime} \ \hat{\psi}_{s_1}^+(\mathbf{r})\hat{\psi}_{s_1^{\prime}}^+(\mathbf{r}^{\prime})\times 
\left[\beta_{s_1 s_2 }\beta_{s_{1}^{\prime}s_{2}^{\prime}}g_{S}\left(\mathbf{r},\mathbf{r}^{\prime}\right)   +\left(\delta_{s_1s_2}\delta_{s_1^{\prime}s_2^{\prime}}-\mathbf{\alpha}_{s_1 s_2} \cdot \mathbf{\alpha}_{s_1^{\prime}s_2^{\prime}}^{\prime} \right) g_V \left(\mathbf{r},\mathbf{r}^{\prime}\right) \right] \hat{\psi}_{s_{2}^{\prime}}(\mathbf{r}^{\prime})\hat{\psi}_{s_{2}}(\mathbf{r})\nonumber \\
\end{eqnarray}
\normalsize
where $s_{1},s_{2},s_{1}^{\prime},s_{2}^{\prime}$ label the bispinor upper and lower components.\\
\indent The Hartree-Fock (HF) approximation to the exact Dirac Hamiltonian (\ref{genH}) is introduced by assuming that:
\begin{equation}
\left\langle \hat{\psi}_{s_{1}}^{+}(\mathbf{r})\hat{\psi}_{s_{1}^{\prime}}^{+}(\mathbf{r}^{\prime})\hat{\psi}_{s_{2}^{\prime}}(\mathbf{r}^{\prime})\hat{\psi}_{s_{2}}(\mathbf{r})\right\rangle =\left\langle \hat{\psi}_{s_{1}}^{+}(\mathbf{r})\hat{\psi}_{s_{2}}(\mathbf{r})\right\rangle \left\langle \hat{\psi}_{s_{1}^{\prime}}^{+}(\mathbf{r}^{\prime})\hat{\psi}_{s_{2}^{\prime}}(\mathbf{r}^{\prime})\right\rangle -  \left\langle \hat{\psi}_{s_{1}}^{+}(\mathbf{r})\hat{\psi}_{s_{2}^{\prime}}(\mathbf{r}^{\prime})\right\rangle \left\langle \hat{\psi}_{s_{1}^{\prime}}^{+}(\mathbf{r}^{\prime})\hat{\psi}_{s_{2}}(\mathbf{r})\right\rangle
\end{equation}

\noindent By defining the ($4\times4$) density matrix
\begin{equation}
\rho_{s_{2}^{\prime}s_{1}}\left(\mathbf{r}^{\prime},\mathbf{r}\right)=\left\langle \hat{\psi}_{s_{1}}^{+}(\mathbf{r})\hat{\psi}_{s_{2}^{\prime}}(\mathbf{r}^{\prime})\right\rangle 
\end{equation}

\noindent the DHF Hamiltonian (\ref{genH}) in second quantization can be written:

\begin{eqnarray}
\label{DHFimplicit}
& H_{DHF}=  \sum_{s_{1}s_{2}}\int d\mathbf{r}\hat{\psi}_{s_{1}}^{+}(\mathbf{r}) \left\{ -\mathbf{\alpha}_{s_{1}s_{2}}\cdot\left[ic\mathbf{\nabla}+\mathbf{A}_{H}\left(\mathbf{r}\right)\right]+\beta_{s_{1}s_{2}}\left[mc^{2}+V_{HS}\left(\mathbf{r}\right)\right]+\delta_{s_{1}s_{2}}\left[V(\mathbf{r})+V_{H}(\mathbf{r})\right]\right\} \hat{\psi}_{s_{2}}(\mathbf{r}) \nonumber \\
&-\sum_{s_{1}^{\prime}s_{2}}\int d\mathbf{r}d\mathbf{r}^{\prime}\hat{\psi}_{s_{1}^{\prime}}^{+}(\mathbf{r}^{\prime})\left[V_{FS,s_{1}^{\prime}s_{2}}\left(\mathbf{r}^{\prime},\mathbf{r}\right)-A_{F,s_{1}^{\prime}s_{2}}\left(\mathbf{r}^{\prime},\mathbf{r}\right)+V_{F,s_{1}^{\prime}s_{2}}\left(\mathbf{r}^{\prime},\mathbf{r}\right)\right]\hat{\psi}_{s_{2}}(\mathbf{r})   
\end{eqnarray}

\noindent where

\begin{eqnarray}\label{avh}
\mathbf{A}_{H}\left(\mathbf{r}\right) & =\int d\mathbf{r}^{\prime}\sum_{s_{1}^{\prime}s_{2}^{\prime}}\left[\rho_{s_{2}^{\prime}s_{1}^{\prime}}\left(\mathbf{r}^{\prime},\mathbf{r}^{\prime}\right)\mathbf{\alpha}_{s_{1}^{\prime}s_{2}^{\prime}}\right]g_{V}\left(\mathbf{r},\mathbf{r}^{\prime}\right )\\
V_{H}\left(\mathbf{r}\right) & =\int d\mathbf{r}^{\prime}\sum_{s_{1}^{\prime}s_{2}^{\prime}}\left[\rho_{s_{2}^{\prime}s_{1}^{\prime}}\left(\mathbf{r}^{\prime},\mathbf{r}^{\prime}\right)\delta_{s_{1}^{\prime}s_{2}^{\prime}}\right]g_{V}\left(\mathbf{r},\mathbf{r}^{\prime}\right) \label{H} \\
V_{HS}\left(\mathbf{r}\right) & =\int d\mathbf{r}^{\prime}\sum_{s_{1}^{\prime}s_{2}^{\prime}}\left[\rho_{s_{2}^{\prime}s_{1}^{\prime}}\left(\mathbf{r}^{\prime},\mathbf{r}^{\prime}\right)\beta_{s_{1}^{\prime}s_{2}^{\prime}}\right]g_{S}\left(\mathbf{r},\mathbf{r}^{\prime}\right)\label{ki} 
\end{eqnarray}
and
\begin{eqnarray}
\label{af}
A_{F,s_{1}^{\prime}s_{2}}\left(\mathbf{r}^{\prime},\mathbf{r}\right) & =\sum_{s_{1}s_{2}^{\prime}}\mathbf{\alpha}_{s_{1}s_{2}}\cdot\mathbf{\alpha}_{s_{1}^{\prime}s_{2}^{\prime}}\rho_{s_{2}^{\prime}s_{1}}\left(\mathbf{r},\mathbf{r}^{\prime}\right)g_{V}\left(\mathbf{r},\mathbf{r}^{\prime}\right)\\ \label{vf}
V_{F,s_{1}^{\prime}s_{2}}\left(\mathbf{r}^{\prime},\mathbf{r}\right) & =\sum_{s_{1}s_{2}^{\prime}}\delta_{s_{1}s_{2}}\delta_{s_{1}^{\prime}s_{2}^{\prime}}\rho_{s_{2}^{\prime}s_{1}}\left(\mathbf{r},\mathbf{r}^{\prime}\right)g_{V}\left(\mathbf{r},\mathbf{r}^{\prime}\right)\\ \label{vs}
V_{FS,s_{1}^{\prime}s_{2}}\left(\mathbf{r}^{\prime},\mathbf{r}\right) & =\sum_{s_{1}s_{2}^{\prime}}\beta_{s_{1}s_{2}}\beta_{s_{1}^{\prime}s_{2}^{\prime}}\rho_{s_{2}^{\prime}s_{1}}\left(\mathbf{r},\mathbf{r}^{\prime}\right)g_{S}\left(\mathbf{r},\mathbf{r}^{\prime}\right)
\end{eqnarray}

The sums over $s_1$ and $s_2^{\prime}$ in equations (\ref{af}-\ref{vs}) must be understood as matrix product. By writing the density matrix as a $2\times2$ bloc matrix ($L~(S)$ labels the large (small) component of the Dirac bispinor)

\begin{equation}
\rho\left(\mathbf{r}^{\prime},\mathbf{r}\right)=\left(\begin{array}{cc}
\rho_{LL}\left(\mathbf{r}^{\prime},\mathbf{r}\right) & \rho_{LS}\left(\mathbf{r}^{\prime},\mathbf{r}\right)\\
\rho_{SL}\left(\mathbf{r}^{\prime},\mathbf{r}\right) & \rho_{SS}\left(\mathbf{r}^{\prime},\mathbf{r}\right)
\end{array}\right)
\end{equation}
Equations (\ref{avh}-\ref{vs}) can be written explicitly:
\begin{eqnarray}
\mathbf{A}_{H}(\mathbf{r}) & =\int d\mathbf{r}^{\prime}\mathrm{Tr}\left[\left(\begin{array}{cc}
\rho_{LL}\left(\mathbf{r}^{\prime},\mathbf{r}^{\prime}\right) & \rho_{LS}\left(\mathbf{r}^{\prime},\mathbf{r}^{\prime}\right) \nonumber \\
\rho_{SL}\left(\mathbf{r}^{\prime},\mathbf{r}^{\prime}\right) & \rho_{SS}\left(\mathbf{r}^{\prime},\mathbf{r}^{\prime}\right)
\end{array}\right)\left(\begin{array}{cc}
0 & \mathbf{\sigma}\\
\mathbf{\sigma} & 0
\end{array}\right)\right]g_{V}\left(\mathbf{r},\mathbf{r}^{\prime}\right)\\
 & =\int d\mathbf{r}^{\prime}\mathrm{Tr}\left\{ \left[\rho_{LS}\left(\mathbf{r}^{\prime},\mathbf{r}^{\prime}\right)+\rho_{SL}\left(\mathbf{r}^{\prime},\mathbf{r}^{\prime}\right)\right]\mathbf{\sigma}\right\} g_{V}\left(\mathbf{r},\mathbf{r}^{\prime}\right)
\end{eqnarray}
 
\begin{eqnarray}\label{HHH}
V_{H}(\mathbf{r}) & =\int d\mathbf{r}^{\prime}\mathrm{Tr}\left[\left(\begin{array}{cc}
\rho_{LL}\left(\mathbf{r}^{\prime},\mathbf{r}^{\prime}\right) & \rho_{LS}\left(\mathbf{r}^{\prime},\mathbf{r}^{\prime}\right)  \nonumber \\
\rho_{SL}\left(\mathbf{r}^{\prime},\mathbf{r}^{\prime}\right) & \rho_{SS}\left(\mathbf{r}^{\prime},\mathbf{r}^{\prime}\right)
\end{array}\right)\right]g_{V}\left(\mathbf{r},\mathbf{r}^{\prime}\right)\\
 & =\int d\mathbf{r}^{\prime}\mathrm{Tr}\left[\rho_{LL}\left(\mathbf{r}^{\prime},\mathbf{r}^{\prime}\right)+\rho_{SS}\left(\mathbf{r}^{\prime},\mathbf{r}^{\prime}\right)\right]g_{V}\left(\mathbf{r},\mathbf{r}^{\prime}\right)
\end{eqnarray}

\begin{eqnarray}\label{bo1}
V_{HS}\left(\mathbf{r}\right) & =\int d\mathbf{r}^{\prime}\mathrm{Tr}\left[\left(\begin{array}{cc}
\rho_{LL}\left(\mathbf{r}^{\prime},\mathbf{r}^{\prime}\right) & \rho_{LS}\left(\mathbf{r}^{\prime},\mathbf{r}^{\prime}\right) \nonumber\\
\rho_{SL}\left(\mathbf{r}^{\prime},\mathbf{r}^{\prime}\right) & \rho_{SS}\left(\mathbf{r}^{\prime},\mathbf{r}^{\prime}\right)
\end{array}\right)\left(\begin{array}{cc}
1 & 0\\
0 & -1
\end{array}\right)\right]g_{S}\left(\mathbf{r},\mathbf{r}^{\prime}\right)\\
 & =\int d\mathbf{r}^{\prime}\mathrm{Tr}\left[\rho_{LL}\left(\mathbf{r}^{\prime},\mathbf{r}^{\prime}\right)-\rho_{SS}\left(\mathbf{r}^{\prime},\mathbf{r}^{\prime}\right)\right]g_{S}\left(\mathbf{r},\mathbf{r}^{\prime}\right)
\end{eqnarray}

\begin{eqnarray}
A_{F}\left(\mathbf{r}^{\prime},\mathbf{r}\right) & =\left(\begin{array}{cc}
0 & \mathbf{\sigma} \\
\mathbf{\sigma} & 0
\end{array}\right)\cdot\left(\begin{array}{cc}
\rho_{LL}\left(\mathbf{r}^{\prime},\mathbf{r}\right) & \rho_{LS}\left(\mathbf{r}^{\prime},\mathbf{r}\right)\\
\rho_{SL}\left(\mathbf{r}^{\prime},\mathbf{r}\right) & \rho_{SS}\left(\mathbf{r}^{\prime},\mathbf{r}\right)
\end{array}\right)\cdot\left(\begin{array}{cc}
0 & \mathbf{\sigma}\\
\mathbf{\sigma} & 0
\end{array}\right)g_{V}\left(\mathbf{r}^{\prime},\mathbf{r}\right) \nonumber \\
 & =\left(\begin{array}{cc}
\mathbf{\sigma}\cdot\rho_{SS}\left(\mathbf{r}^{\prime},\mathbf{r}\right)\cdot\mathbf{\sigma} & \mathbf{\sigma}\cdot\rho_{LS}\left(\mathbf{r}^{\prime},\mathbf{r}\right)\cdot\mathbf{\sigma} \\
\mathbf{\sigma}\cdot\rho_{SL}\left(\mathbf{r}^{\prime},\mathbf{r}\right)\cdot\mathbf{\sigma} & \mathbf{\sigma}\cdot\rho_{LL}\left(\mathbf{r}^{\prime},\mathbf{r}\right)\cdot\mathbf{\sigma}
\end{array}\right)g_{V}\left(\mathbf{r}^{\prime},\mathbf{r}\right)
\end{eqnarray}

\begin{equation}\label{uno}
V_{F}\left(\mathbf{r}^{\prime},\mathbf{r}\right)=\left(\begin{array}{cc}
\rho_{LL}\left(\mathbf{r}^{\prime},\mathbf{r}\right) & \rho_{SL}\left(\mathbf{r}^{\prime},\mathbf{r}\right)\\
\rho_{LS}\left(\mathbf{r}^{\prime},\mathbf{r}\right) & \rho_{SS}\left(\mathbf{r}^{\prime},\mathbf{r}\right)
\end{array}\right)g_{V}\left(\mathbf{r}^{\prime},\mathbf{r}\right)
\end{equation}

\begin{eqnarray}
V_{FS}\left(\mathbf{r}^{\prime},\mathbf{r}\right) & =\left(\begin{array}{cc}
1 & 0\\
0 & -1
\end{array}\right)\cdot\left(\begin{array}{cc}
\rho_{LL}\left(\mathbf{r}^{\prime},\mathbf{r}\right) & \rho_{LS}\left(\mathbf{r}^{\prime},\mathbf{r}\right)\\
\rho_{SL}\left(\mathbf{r}^{\prime},\mathbf{r}\right) & \rho_{SS}\left(\mathbf{r}^{\prime},\mathbf{r}\right)
\end{array}\right)\cdot\left(\begin{array}{cc}
1 & 0\\
0 & -1
\end{array}\right)g_{S}\left(\mathbf{r}^{\prime},\mathbf{r}\right) \nonumber \\
 & =\left(\begin{array}{cc}
\rho_{LL}\left(\mathbf{r}^{\prime},\mathbf{r}\right) & -\rho_{SL}\left(\mathbf{r}^{\prime},\mathbf{r}\right)\\
-\rho_{LS}\left(\mathbf{r}^{\prime},\mathbf{r}\right) & \rho_{SS}\left(\mathbf{r}^{\prime},\mathbf{r}\right)
\end{array}\right)g_{S}\left(\mathbf{r}^{\prime},\mathbf{r}\right)
\end{eqnarray}

With these definitions in place, the most general Dirac equation (\ref{DHFimplicit}) including both scalar and vector potentials can be written in matrix form as follows:

\begin{equation}\label{DHFexplicit}
\left(\begin{array}{ll}
mc^{2}+W_{V}+W_{S}+\mathbf{A}_{P}\cdot\mathbf{\sigma}-E & -c\mathbf{\sigma}\cdot i\mathbf{\nabla}-\mathbf{\sigma}\cdot\mathbf{A}+W_{PS}\\
-c\mathbf{\sigma}\cdot i\mathbf{\nabla}-\mathbf{\sigma}\cdot\mathbf{A}+W_{PS} & -mc^{2}+W_{V}+\mathbf{A}_{P}\cdot\mathbf{\sigma}-W_{S}-E
\end{array}\right)\left(\begin{array}{l}
\psi_{L}\\
\psi_{S}
\end{array}\right)=0
\end{equation}
where $W_{S}, W_{PS}$ are generic scalar and pseudo-scalar potentials, while $W_{V}, \mathbf{A}_{P}$ are vector and pseudo-vector potentials. We notice that the scalar and vector potentials sum up ($W_{S}+W_{V}$) in the head of the Hamiltonian matrix; they act on the upper part of the spinor (the particle). At variance, the scalar and vector potentials are subtracted ($W_{V}-W_{S}$) in the tail of the Hamiltonian matrix; they act on the lower part of the spinor (the anti-particle): this element thus accounts for the relativistic corrections, that is the spin-orbit interaction.

\subsection{Rotational invariant systems}

The DHF equation can be written specifically to deal with spherical symmetric systems for which both the electronic density $\rho(\mathbf{r})=\rho(r)$ and the HF potential $V(\mathbf{r})=V(r)$ are rotational invariant and depend only on the radial coordinate modulus.\\ \noindent Furthermore, one has: 
\begin{equation}\label{plo}
g_{V,S}\left(\mathbf{r},\mathbf{r}^{\prime}\right)=g_{V,S}\left(\left|\mathbf{r}-\mathbf{r}^{\prime}\right|\right)= \sum_{l=0}^{\infty}\frac{4\pi}{2l+1}g_{V,S,l}\left(r_{<},r_{>}\right)\sum_{m=-l}^{l}Y_{lm}^{*}\left(\vartheta,\varphi\right)Y_{lm}\left(\vartheta^{\prime},\varphi^{\prime}\right)
\end{equation}
where $r_{<}= \mathrm{min}(r, r')$, $r_>= \mathrm{max}(r, r')$, and $Y_{lm}$ are spherical harmonics.\\
The solution to the DHF equation in spherical symmetry can be written as:
\begin{eqnarray}\label{rotinv}
H\psi_{\alpha,\kappa,m}(\mathbf{r})&=&E_{\alpha,\kappa}\psi_{\alpha,\kappa,m}(\mathbf{r})\\ \psi_{\alpha,\kappa,m}(\mathbf{r})&=& \left(\begin{array}{c}
\psi_{\alpha,\kappa,L}\left(r\right)\chi_{\kappa,m}\left(\vartheta,\varphi\right)\\
\psi_{\alpha,\kappa,S}\left(r\right)\chi_{-\kappa,m}\left(\vartheta,\varphi\right)
\end{array}\right)\label{rotinv1}
\end{eqnarray}
where the angular eigenfunctions fulfill the following completeness relation
\begin{equation}
\sum_{m}\frac{4\pi}{2l+1}\int\chi_{\kappa,m}^{+}\left(\vartheta^{\prime},\varphi^{\prime}\right)\chi_{\kappa,m}\left(\vartheta^{\prime},\varphi^{\prime}\right)Y_{l,0}^{*}\left(\vartheta^{\prime},\varphi^{\prime}\right)d\Omega^{\prime}=2\left|\kappa\right|\sqrt{4\pi}\delta_{l0}
\end{equation}
The electron density also simplifies to:
\begin{equation}
\rho\left(\mathbf{r}^{\prime},\mathbf{r}\right)=\sum_{\alpha,\kappa,m}n\left(E_{\alpha,\kappa}\right)\left(\begin{array}{c}
\psi_{\alpha,\kappa,L}\left(r^{\prime}\right)\chi_{\kappa,m}\left(\vartheta^{\prime},\varphi^{\prime}\right)\\
\psi_{\alpha,\kappa,S}\left(r^{\prime}\right)\chi_{-\kappa,m}\left(\vartheta^{\prime},\varphi^{\prime}\right)
\end{array}\right)\left(\begin{array}{cc}
\psi_{\alpha,\kappa,L}^{*}\left(r\right)\chi_{\kappa,m}^{+}\left(\vartheta,\varphi\right) & \psi_{\alpha,\kappa,S}^{*}\left(r\right)\chi_{-\kappa,m}^{+}\left(\vartheta,\varphi\right)\end{array}\right)
\end{equation}
where
\begin{equation}
  n\left(E_{\alpha,\kappa}\right)=\frac{1}{e^{\frac{E_{\alpha,\kappa}-\mu}{K_\mathrm{B}T}}+1}  
\end{equation}
is the Fermi-Dirac distribution function for the grand canonical ensemble relative to a chemical potential $\mu$, $T$ is the temperature, and $K_\mathrm{B}$ the Boltzmann constant.\\
\indent The Hartree potential (see equation \ref{HHH}) reads:
\begin{eqnarray}\label{hartreerot}
& V_{H}\left(\mathbf{r}\right)  = \int d\mathbf{r}^{\prime}\mathrm{Tr}\left[\rho_{LL}\left(\mathbf{r}^{\prime},\mathbf{r}^{\prime}\right)+\rho_{SS}\left(\mathbf{r}^{\prime},\mathbf{r}^{\prime}\right)\right]g_{V}\left(\mathbf{r},\mathbf{r}^{\prime}\right)=  \nonumber \\
&\sum_{\alpha,\kappa,m}n\left(E_{\alpha,\kappa}\right)\int r^{\prime2}dr^{\prime}\left|\psi_{\alpha,\kappa,L}\left(r^{\prime}\right)\right|^{2}\sum_{l=0}^{\infty}\frac{4\pi}{2l+1}g_{V,l}\left(r_{<},r_{>}\right)\times \nonumber \\
 & 
\int\chi_{\kappa,m}^{+}\left(\vartheta^{\prime},\varphi^{\prime}\right)\chi_{\kappa,m}\left(\vartheta^{\prime},\varphi^{\prime}\right)Y_{l,0}^{*}\left(\vartheta^{\prime},\varphi^{\prime}\right)d\Omega^{\prime}Y_{l0}\left(\vartheta,\varphi\right)+ \nonumber \\
 & \sum_{\alpha,\kappa,m}n\left(E_{\alpha,\kappa}\right)\int r^{\prime2}dr^{\prime}\left|\psi_{\alpha,\kappa,S}\left(r^{\prime}\right)\right|^{2}\sum_{l=0}^{\infty}\frac{4\pi}{2l+1}g_{V,l}\left(r_{<},r_{>}\right)\times \nonumber \\
 & 
 \int\chi_{-\kappa,m}^{+}\left(\vartheta^{\prime},\varphi^{\prime}\right)\chi_{-\kappa,m}\left(\vartheta^{\prime},\varphi^{\prime}\right)Y_{l,0}^{*}\left(\vartheta^{\prime},\varphi^{\prime}\right)d\Omega^{\prime}Y_{l0}\left(\vartheta,\varphi\right)  \nonumber\\
&  =\int r^{\prime2}dr^{\prime}g_{V,l}\left(r_{<},r_{>}\right)\sum_{\alpha,\kappa}n\left(E_{\alpha,\kappa}\right)\left[\left|\psi_{\alpha,\kappa,L}\left(r^{\prime}\right)\right|^{2}+\left|\psi_{\alpha,\kappa,S}\left(r^{\prime}\right)\right|^{2}\right] 
\end{eqnarray}

\noindent while $V_{HS}\left(\mathbf{r}\right)$  (see equation \ref{bo1}) turns out
\begin{eqnarray}
V_{HS}\left(\mathbf{r}\right) & =&\int d\mathbf{r}^{\prime}\mathrm{Tr}\left[\rho_{LL}\left(\mathbf{r}^{\prime},\mathbf{r}^{\prime}\right)-\rho_{SS}\left(\mathbf{r}^{\prime},\mathbf{r}^{\prime}\right)\right]g_{S}\left(\mathbf{r},\mathbf{r}^{\prime}\right)\nonumber \\
 & =&\int r^{\prime2}dr^{\prime}g_{S,l}\left(r_{<},r_{>}\right)\sum_{\alpha,\kappa}n\left(E_{\alpha,\kappa}\right)\left[\left|\psi_{\alpha,\kappa,L}\left(r^{\prime}\right)\right|^{2}-\left|\psi_{\alpha,\kappa,S}\left(r^{\prime}\right)\right|^{2}\right]
\end{eqnarray}
Moreover, the Fock term in the rotational invariant case becomes (see equation \ref{uno}):
\begin{equation}\label{pot}
V_{F}\left(\mathbf{r}^{\prime},\mathbf{r}\right)=\left(\begin{array}{cc}
\rho_{LL}\left(\mathbf{r}^{\prime},\mathbf{r}\right) & \rho_{LS}\left(\mathbf{r}^{\prime},\mathbf{r}\right)\\
\rho_{SL}\left(\mathbf{r}^{\prime},\mathbf{r}\right) & \rho_{SS}\left(\mathbf{r}^{\prime},\mathbf{r}\right)
\end{array}\right)g_{V}\left(\mathbf{r},\mathbf{r}^{\prime}\right)
\end{equation}

\noindent In Equation (\ref{pot}) the diagonal term can be written explicitly:

\begin{eqnarray}\label{put}
& \int d\mathbf{r}\rho_{LL}\left(\mathbf{r}^{\prime},\mathbf{r}\right)g_{V}\left(\mathbf{r},\mathbf{r}^{\prime}\right)\psi\left(r\right)\chi_{\kappa,m}\left(\vartheta,\varphi\right)=\nonumber \\
&\sum_{\alpha,\kappa^{\prime},m^{\prime}}n\left(E_{\alpha,\kappa^{\prime}}\right)\psi_{\alpha,\kappa^{\prime},L}\left(r^{\prime}\right)\chi_{\kappa^{\prime},m^{\prime}}\left(\vartheta^{\prime},\varphi^{\prime}\right)\int r^{2}drd\Omega\psi_{\alpha,\kappa^{\prime},L}^{*}\left(r\right)\psi(r)\chi_{\kappa^{\prime},m^{\prime}}^{+}\left(\vartheta,\varphi\right)\chi_{\kappa,m}\left(\vartheta,\varphi\right)\times \nonumber \\
 & \sum_{l=0}^{\infty}\frac{4\pi}{2l+1}f_{V,S,l}\left(r_{<}\right)h_{V,S,l}\left(r_{>}\right)Y_{l,m-m^{\prime}}^{*}\left(\vartheta,\varphi\right)Y_{l,m-m^{\prime}}\left(\vartheta^{\prime},\varphi^{\prime}\right)
\end{eqnarray}
By defining
\begin{eqnarray}
&\int\chi_{\kappa_{1},m_{1}}^{+}(\Omega)\chi_{\kappa_{2},m_{2}}(\Omega)Y_{lm}^{*}(\Omega)d\Omega  =\int\chi_{\kappa_{2},m_{2}}^{+}(\Omega)\chi_{\kappa_{1},m_{1}}(\Omega)Y_{lm}(\Omega)d\Omega=\mathcal{F}_{m_{1},m_{2},m}^{\kappa_{1},\kappa_{2},l}= \nonumber \\
 & \sqrt{\frac{(2l+1)(2j_{1}+1)(2j_{2}+1)}{4\pi}}\delta_{m,m_{2}-m_{1}}(-1)^{m_{2}+\frac{1}{2}}\left(\begin{array}{ccc}
j_{2} & l & j_{1}  \\
-m_{2} & m & m_{1}
\end{array}\right)\left(\begin{array}{ccc}
j_{2} & l & j_{1}\\
\frac{1}{2} & 0 & -\frac{1}{2}
\end{array}\right)
\end{eqnarray}

\noindent one can project equation (\ref{put}) on $\chi_{\kappa^{\prime},m^{\prime}}$, obtaining:

\begin{eqnarray}
&\int d\Omega^{\prime}\chi_{\kappa^{\prime},m^{\prime}}^{+}\left(\vartheta^{\prime},\varphi^{\prime}\right)\int d\mathbf{r}\rho_{LL}\left(\mathbf{r}^{\prime},\mathbf{r}\right)g_{V}\left(\mathbf{r},\mathbf{r}^{\prime}\right)\psi\left(r\right)\chi_{\kappa,m}\left(\vartheta,\varphi\right) =  \nonumber \\ &\sum_{\alpha,\kappa^{\prime\prime}}n\left(E_{\alpha,\kappa^{\prime\prime}}\right)\psi_{\alpha,\kappa^{\prime\prime},L}\left(r^{\prime}\right)\int r^{2}dr\psi_{\alpha,\kappa^{\prime\prime},L}^{*}\left(r\right)\psi(r)\sum_{l=0}^{\infty}\frac{4\pi}{2l+1}f_{V,S,l}\left(r_{<}\right)h_{V,S,l}\left(r_{>}\right)\sum_{m^{\prime\prime}}^{\infty}\mathcal{F}_{m^{\prime\prime},m,m-m^{\prime\prime}}^{\kappa^{\prime\prime},\kappa,l}\mathcal{F}_{m^{\prime\prime},m^{\prime},m-m^{\prime\prime}}^{\kappa^{\prime\prime},\kappa^{\prime},l} \nonumber \\
 & =\sum_{\alpha,\kappa^{\prime\prime}}n\left(E_{\alpha,\kappa^{\prime\prime}}\right)\psi_{\alpha,\kappa^{\prime\prime},L}\left(r^{\prime}\right)\int r^{2}dr\psi_{\alpha,\kappa^{\prime\prime},L}^{*}\left(r\right)\psi(r)\sum_{l=0}^{\infty}f_{V,S,l}\left(r_{<}\right)h_{V,S,l}\left(r_{>}\right)\frac{\left|\kappa^{\prime\prime}\right|}{2l+1}\delta_{\kappa^{\prime},\kappa}
\end{eqnarray}
Similar expression can be obtained for $\rho_{SS}$.
The off-diagonal term in equation (\ref{pot}) can be explicitly written:
\begin{eqnarray}
&\int d\mathbf{r}\rho_{LS}\left(\mathbf{r}^{\prime},\mathbf{r}\right)g_{V}\left(\mathbf{r},\mathbf{r}^{\prime}\right)\psi\left(r\right)\chi_{\kappa,m}\left(\vartheta,\varphi\right)  =
\nonumber \\ & \sum_{\alpha,\kappa^{\prime},m^{\prime}}n\left(E_{\alpha,\kappa^{\prime}}\right)\psi_{\alpha,\kappa^{\prime},L}\left(r^{\prime}\right)\chi_{\kappa^{\prime},m^{\prime}}\left(\vartheta^{\prime},\varphi^{\prime}\right)\int r^{2}drd\Omega\psi_{\alpha,-\kappa^{\prime},S}^{*}\left(r\right)\psi(r)\chi_{-\kappa^{\prime},m^{\prime}}^{+}\left(\vartheta,\varphi\right)\chi_{\kappa,m}\left(\vartheta,\varphi\right)\nonumber \\
 & \times\sum_{l=0}^{\infty}\frac{4\pi}{2l+1}f_{V,S,l}\left(r_{<}\right)h_{V,S,l}\left(r_{>}\right)Y_{l,m-m^{\prime}}^{*}\left(\vartheta,\varphi\right)Y_{l,m-m^{\prime}}\left(\vartheta^{\prime},\varphi^{\prime}\right)
\end{eqnarray}
which can be projected on $\chi_{\kappa^{\prime},m^{\prime}}$, obtaining:

\begin{eqnarray}
&\int d\Omega^{\prime}\chi_{\kappa^{\prime},m^{\prime}}^{+}\left(\vartheta^{\prime},\varphi^{\prime}\right)\int d\mathbf{r}\rho_{LS}\left(\mathbf{r}^{\prime},\mathbf{r}\right)g_{V}\left(\mathbf{r},\mathbf{r}^{\prime}\right)\psi\left(r\right)\chi_{\kappa,m}\left(\vartheta,\varphi\right)  = \nonumber \\ & \sum_{\alpha,\kappa^{\prime\prime}}n\left(E_{\alpha,\kappa^{\prime\prime}}\right)\psi_{\alpha,\kappa^{\prime\prime},L}\left(r^{\prime}\right)\int r^{2}dr\psi_{\alpha,-\kappa^{\prime\prime},S}^{*}\left(r\right)\psi(r)\sum_{l=0}^{\infty}\frac{4\pi}{2l+1}f_{V,S,l}\left(r_{<}\right)h_{V,S,l}\left(r_{>}\right)\sum_{m^{\prime\prime}}^{\infty}\mathcal{F}_{m^{\prime\prime},m,m-m^{\prime\prime}}^{-\kappa^{\prime\prime},\kappa,l}\mathcal{F}_{m^{\prime\prime},m^{\prime},m-m^{\prime\prime}}^{\kappa^{\prime\prime},\kappa^{\prime},l}\nonumber \\
 & =\sum_{\alpha,\kappa^{\prime\prime}}n\left(E_{\alpha,\kappa^{\prime\prime}}\right)\psi_{\alpha,\kappa^{\prime\prime},L}\left(r^{\prime}\right)\int r^{2}dr\psi_{\alpha,-\kappa^{\prime\prime},S}^{*}\left(r\right)\psi(r)\sum_{l=0}^{\infty}f_{V,S,l}\left(r_{<}\right)h_{V,S,l}\left(r_{>}\right)\frac{\left|\kappa^{\prime\prime}\right|}{2l+1}\delta_{\kappa^{\prime},-\kappa}
\end{eqnarray}

 Similar expression can be derived for $\rho_{SL}$. Finally, these analytical formulae can be used to assess the Hartree and Fock potential energy terms (see equations \ref{hartreerot},\ref{pot}).

\subsubsection{Coulomb and Yukawa potentials}

An important rotational--invariant scattering potential for applications in atomic and nuclear physics is the Yukawa potential \cite{193548}:
\begin{equation}\label{Yukawa}
    V(r)=\frac{V_0 e^{\mu r}}{\mu r}
\end{equation}
where the potential strength $V_0$ is independent of $r$, and $1/\mu$ represents the effective range of the interaction. Owing to the fact that $V$ goes to zero rapidly for $r>>1/\mu$, the Yukawa potential represents the epitome of screened Coulomb interaction in both nuclei and atoms. Moreover, it reduces to the bare Coulomb potential for $\mu \rightarrow 0$, provided that the ratio $V_0/\mu=QQ'e^2$ is constant ($Q,Q'$ are the nuclear charges of the interacting particles, e.g. the projectile and the target).
For electrons, the central point is the expansion of the bare Coulomb operator $\frac{1}{\left|\mathbf{r}-\mathbf{r}^{\prime}\right|}$. It can be performed by the following Laplace expansion in spherical polar coordinates ${\displaystyle (r,\theta ,\varphi)}$:

\begin{equation}\label{coulexp}
\frac{1}{\left|\mathbf{r}-\mathbf{r}^{\prime}\right|}=\sum_{l=0}^{\infty}\frac{4\pi}{2l+1}\frac{r_{<}^{l}}{r_{>}^{l+1}}\sum_{m=-l}^{l}Y_{lm}^{*}\left(\vartheta,\varphi\right)Y_{lm}\left(\vartheta^{\prime},\varphi^{\prime}\right)
\end{equation}

A similar expression can be obtained for the Yukawa potential (\ref{Yukawa}):

\begin{equation}\label{Yukexp}
\frac{\mathrm{e}^{-\alpha\left|\mathbf{r}-\mathbf{r}^{\prime}\right|}}{\left|\mathbf{r}-\mathbf{r}^{\prime}\right|}=\sum_{l=0}^{\infty}\frac{4\pi}{2l+1}f_{\alpha,l}\left(r_{<}\right)h_{\alpha,l}\left(r_{>}\right)\sum_{m=-l}^{l}Y_{lm}^{*}\left(\vartheta,\varphi\right)Y_{lm}\left(\vartheta^{\prime},\varphi^{\prime}\right)
\end{equation}
where
\begin{equation}
f_{\alpha,l}\left(r\right)=\frac{\left(2l+1\right)!!}{\alpha^{2l}}\left(\frac{1}{r}\frac{d}{dr}\right)^{l}\frac{\sinh\left(\alpha r\right)}{\alpha r}
\end{equation}
and
\begin{equation}
h_{\alpha,l}\left(r\right)=\frac{1}{\left(2l-1\right)!!}\left(\frac{1}{r}\frac{d}{dr}\right)^{l}\frac{\mathrm{e}^{-\alpha r}}{r}
\end{equation}
are spherical Bessel functions with the following limiting properties:
\begin{equation}
\lim_{\alpha\rightarrow0}f_{\alpha,l}\left(r\right)=r^{l},\qquad\lim_{\alpha\rightarrow0}h_{\alpha,l}\left(r\right)=\frac{1}{r^{l+1}}
\end{equation}
Furthermore, $f_{\alpha,l}\left(r\right)$ e $h_{\alpha,l}\left(r\right)$ satisfy the following recursive equations:
\begin{eqnarray}
f_{\alpha,l-1}\left(r\right)-\frac{\alpha^{2}}{\left(2l+1\right)\left(2l+3\right)}f_{\alpha,l+1}\left(r\right) & =\frac{1}{r}f_{\alpha,l}\left(r\right)\\
-\frac{\alpha^{2}}{\left(2l-1\right)\left(2l+1\right)}h_{\alpha,l-1}\left(r\right)+h_{\alpha,l+1}\left(r\right) & =\frac{1}{r}h_{\alpha,l}\left(r\right)
\end{eqnarray}
To solve the Dirac equation for a Yukawa potential, one needs to calculate integrals such as: 
\begin{equation}
\int\limits _{0}^{\infty}dr\int\limits _{0}^{\infty}dr^{\prime}f_{\alpha}\left(r_{<}\right)h_{\alpha}\left(r_{>}\right)\xi_{1}\left(r\right)\xi_{2}\left(r^{\prime}\right)
\end{equation}
where $\xi(r)$ is the product of two wavefunctions. These integrals can be decomposed as follows:

\begin{eqnarray}
I & =\int\limits _{0}^{\infty}dr\int\limits _{0}^{r}dr^{\prime}f_{\alpha}\left(r^{\prime}\right)h_{\alpha}\left(r\right)\xi_{1}\left(r\right)\xi_{2}\left(r^{\prime}\right)+\int\limits _{0}^{\infty}dr\int\limits _{r}^{\infty}dr^{\prime}f_{\alpha}\left(r\right)h_{\alpha}\left(r^{\prime}\right)\xi_{1}\left(r\right)\xi_{2}\left(r^{\prime}\right) \nonumber \\
 & =\int\limits _{0}^{\infty}dr\left[\mathcal{F}_{f_{\alpha}\xi_{2}}\left(r\right)-\mathcal{F}_{f_{\alpha}\xi_{2}}\left(0\right)\right]h_{\alpha}\left(r\right)\xi_{1}\left(r\right)-\int\limits _{0}^{\infty}dr\left[\mathcal{F}_{h_{\alpha}\xi_{2}}\left(r\right)-\mathcal{F}_{h_{\alpha}\xi_{2}}\left(\infty\right)\right]h_{\alpha}\left(r\right)\xi_{1}\left(r\right)
\end{eqnarray}
where $\mathcal{F}_{\xi}\left(r\right)$ is the antiderivative of $\xi(r)$. In the next section we will show a numerical method to compute this integral in a radial grid.

\subsection{Computational methods for spherically symmetric systems}

The general form of the Dirac equation ((\ref{DHFimplicit})) for a spherical symmetric problem, such as particles interacting via mean-field or Yukawa potentials, can be written as follows:

\begin{equation}\label{difeq}
\frac{d\psi}{dr}=F_{E}(\psi,r)
\end{equation}

\noindent where $\psi$ is a two-component vector. $F_E$, being a linear function of both $\psi$ and of the energy $E$ (see Equation (\ref{DHFexplicit})), can be written as $F_{E}(\psi,r)=A(r)\psi$ where $A(r)$ is a $2 \times 2$ linear operator. 
The difficulty to find the numerical solution of Equation (\ref{difeq}) for many-fermion systems is due to the fact that $F_E$ contains a non-local term, related to the Fock exchange (see equation \ref{pot}).
To circumvent this issue, the non-local Fock term is replaced by an expression derived using the local density approximation (LDA) for the electron gas \cite{Slater1951, Salvat1987}

\begin{equation}\label{rho13}
    V_{ex}=\frac{3}{2} \alpha_X \Big[ \frac{3}{\pi} \rho (r) \Big]^{1/3}
\end{equation}

\noindent where $\rho(r)$ is the electron density and $\alpha_X= 3/2$. 

According to (\ref{rotinv1})
the solutions of the Dirac equation  (\ref{DHFexplicit}) in a spherically symmetric potential can be factorised into the product of monodimensional radial and angular functions \cite{Greiner}:

\begin{equation}\label{psi2}
\psi (\mathbf{r})=\psi_{\kappa,\mu} (\mathbf{r})=
\left(\begin{array}{ll} \frac{u_{\kappa}(r)}{r} \  \chi_{\kappa,\mu} (\Omega) \\ i \frac{v_{\kappa}(r)}{r} \  \chi_{-\kappa,\mu}(\Omega) \end{array}\right)
\end{equation}

\noindent where $\chi_{\kappa,\mu}$ are the tensor product of orbital and spin spherical harmonics characterised by the Dirac quantum numbers $\kappa,\mu$, respectively.

By using the factorization (\ref{psi2}) and considering for the sake of simplicity a system of many-fermions interacting via a mean-field potential, the Dirac equation (\ref{DHFexplicit}) reads:

\begin{equation}
\frac{\partial}{\partial r}\left(
\begin{array}{c}
u_{\kappa}(r)\\
v_{\kappa}(r)
\end{array}\right)=\left(\begin{array}{cc}
-\frac{\kappa}{r} & \frac{E}{c}-\frac{V(r)}{c}+c\\
\frac{V(r)}{c}+c-\frac{E}{c} & \frac{\kappa}{r}
\end{array}\right)\left(\begin{array}{c}
u_{\kappa}(r)\\
v_{\kappa}(r)
\end{array}\right)\label{DHF}
\end{equation}

\noindent where the total potential is given by ($Z$ is the atomic number):  

\begin{equation}\label{potdhf}
V(r)=-\frac{Z}{r}+\int \frac{\rho(r')}{r_>}d^3 r'-V_{ex}(r)
\end{equation}

\noindent We notice that the Dirac equation (\ref{DHF}) is valid for any central symmetric potential. For example, by using the Wood-Saxon potential \cite{Woodsaxon}, which typically models the nucleon-nucleon interaction, the calculation of the nuclear wavefunctions can in principle be carried out \cite{morresi2018nuclear}. \\
\indent Equations (\ref{DHF}) with the local interaction (\ref{potdhf}) are solved self-consistently. We stress that, as for the non-relativistic HF, the eigenvalues of the radial equation (\ref{DHF}) represent the electron binding energies. \\
\indent To integrate numerically the equation (\ref{DHF}), we introduce now a modified Runge-Kutta method, which is known as the method of collocations. 
This method proceeds by dividing the interval $[0,R]$ over which the solution is sought for ($R$ can even tend to $\infty$) in a number of grid points $[0=r_{0},\cdots r_{j},\cdots r_{N}=R]$. The latter can be determined as the roots of the $n-$th Legendre polynomial.
Each sub-interval $[r_{j-1},r_{j}]$ is further discretised in a number of $s_{j}$ points $\{x_{j,l}\}$, where the derivative of the wavefunction appearing in the general Dirac equation (\ref{difeq}) can 
be approximated by

\begin{equation}\label{derwf}
\frac{d\psi}{dr}\simeq\sum_{l=1}^{s_{j}}q_{j,l}\phi_{j,l}^{\prime}(r)
\end{equation}

\noindent and $\psi$ is obtained by integrating the equation (\ref{derwf}) as follows:

\begin{eqnarray}
\psi(r)&=&\psi(r_{j-1})+\int\limits_{r_{j-1}}^{r}\sum_{l=1}^{s_{j}}q_{j,l}\phi_{j,l}^{\prime}(x)dx \nonumber \\ &=& \sum_{i=1}^{j-1}\sum_{l=1}^{s_{i}}q_{i,l}\int\limits_{r_{i-1}}^{r_{i}}\phi_{i,l}^{\prime}(x)dx+\sum_{l=1}^{s_{j}}q_{j,l}\int\limits _{r_{j-1}}^{r}\phi_{j,l}^{\prime}(x)dx
\end{eqnarray}

\noindent imposing $\phi_{j,l}^{\prime}(x_{j,m})=\delta_{lm}$.
By defining:

\begin{equation}
\phi_{j,l}(r)=\int\limits _{r_{j-1}}^{r}\phi_{j,l}^{\prime}(x)dx,\qquad a_{j,lm}=\int\limits _{r_{j-1}}^{x_{j,l}}\phi_{j,m}^{\prime}(x)dx,\qquad b_{j,l}=\int\limits _{r_{j-1}}^{r_{j}}\phi_{j,l}^{\prime}(x)dx
\end{equation}

\noindent one obtains: 

\begin{equation}
\frac{d\psi}{dr}(x_{j,l})=q_{j,l}=F_{E}\left(\sum_{i=1}^{j-1}\sum_{l=1}^{s_{i}}b_{i,m}q_{i,m}+\sum_{l=1}^{s_{j}}a_{j,lm}q_{j,m},x_{j,l}\right)
\end{equation}

\noindent The Dirac equation (\ref{DHF}) can now be recast as an eigenvalue problem

\begin{equation}\label{eigenv}
c\gamma\left(\begin{array}{c}
q_{1}\\
q_{2}\\
q_{3}\\
\cdots\\
q_{N}
\end{array}\right)=\left(E-\beta mc^{2}-V\right)\left(\begin{array}{ccccc}
a_{1} & 0 & 0 & \cdots & 0\\
b_{1} & a_{2} & 0 & \cdots & 0\\
b_{1} & b_{2} & a_{3} & \cdots & 0\\
\cdots & \cdots & \cdots & \cdots & \cdots\\
b_{1} & b_{2} & b_{3} & \cdots & a_{N}
\end{array}\right)\left(\begin{array}{c}
q_{1}\\
q_{2}\\
q_{3}\\
\cdots\\
q_{N}
\end{array}\right)
\end{equation}

\noindent where $\gamma$ and $\beta$ are $2\times2$ matrices. In compact form, the equation (\ref{DHF})
can be written:
\begin{equation}\label{eigenpro}
\left[c\gamma+\left(V+\beta mc^{2}\right)A-EA\right]q=0
\end{equation}

\noindent where  
\begin{equation}
A=\left(\begin{array}{ccccc}
a_{1} & 0 & 0 & \cdots & 0\\
b_{1} & a_{2} & 0 & \cdots & 0\\
b_{1} & b_{2} & a_{3} & \cdots & 0\\
\cdots & \cdots & \cdots & \cdots & \cdots\\
b_{1} & b_{2} & b_{3} & \cdots & a_{N}
\end{array}\right)
\end{equation}

\noindent Finally, by transforming $A$ into a lower bidiagonal matrix, the equation (\ref{eigenpro}) reads:

\begin{equation}\label{tridiag}
\left[c\gamma Q+\left(V+\beta mc^{2}\right)AQ-EAQ\right]q^{\prime}=0
\end{equation}

\noindent where $q^{\prime}=Q^{-1}q$, $\psi=Aq=AQq^{\prime}$, and

\begin{equation}
Q=\left(\begin{array}{rrcrrr}
1 & 0 & \cdots & 0 & 0 & 0\\
-b_{2}^{-1}b_{1} & 1 & \cdots & 0 & 0 & 0\\
0 & -b_{3}^{-1}b_{2} & \cdots & 1 & 0 & 0\\
\cdots & \cdots & \cdots & \cdots & \cdots & \cdots\\
0 & 0 & \cdots & -b_{N-1}^{-1}b_{N-2} & 1 & 0\\
0 & 0 & \cdots & 0 & 0 & 1
\end{array}\right)
\end{equation}
\begin{equation}
AQ=\left(\begin{array}{rrcrrr}
a_{1} & 0 & 0 &\cdots  & 0 & 0\\
b_{1}-a_{2}b_{2}^{-1}b_{1} & a_{2} & 0 &\cdots & 0 & 0 \\
0 & b_{2}-a_{3}b_{3}^{-1}b_{2} & a_{3} & \cdots & 0 & 0\\
\cdots & \cdots & \cdots & \cdots & \cdots & \cdots\\
0 & 0 & \cdots & b_{N-2}-a_{N-1}b_{N-1}^{-1}b_{N-2} & a_{N-1} & 0\\
0 & 0 & \cdots & 0 & b_{N-1} & a_{N}
\end{array}\right)
\end{equation}

The tridiagonal form (\ref{tridiag}) is computationally efficient, particularly when we are dealing with quasi-local potentials, as  in that case the matrices $Q$, $AQ$, $V$, and $VAQ$ are almost diagonal. The eigenvalues $E$ can be indeed reckoned via the $LDU$ decomposition. The spatial extension of the region $[0,R_{max}]$ where the differential equation (\ref{difeq}) is solved depends  
weather we are dealing with bound or continuum state problems. Bound state wavefunctions typically die off after a few atomic radii ($R_{max}=25 $ a.u.), while continuum states extend to hundreds atomic units ($R_{max}\ge 100$ a.u.). In general, if one is to solve a continuum scattering state problem, e.g. the simulation of electron emission from heavy atoms,
the continuum wavefunction asymptotic behaviour ($R_{max} \ge r \rightarrow \infty$) must be matched via Coulomb functions \cite{Greiner}, which represent the solution for a charged particle in an ionic field: 

\begin{equation}\label{asym}
u \to \frac{\sqrt{\epsilon +c^2}}{\sqrt{\pi p}}\mathrm{cos}(pr-\eta\ln(pr) +\delta) \\
 v \to -\frac{\sqrt{\epsilon -c^2}}{\sqrt{\pi p}}\mathrm{sin}(pr -\eta\ln(pr)+\delta)
\end{equation} 

\noindent where $\eta=Zm/p$ ($Z$=atomic number, $m$=electron mass=1), and  $\epsilon=\sqrt{p^2c^2+c^4}$ is the free electron kinetic energy. This leads to the normalization condition of the continuum wavefunctions, $\frac{\pi p}{\sqrt{\epsilon+c^2}} \Big(u^2+\sqrt{\frac{\epsilon+c^2}{\epsilon-c^2}}v^2 \Big)=1$. To avoid a critical dependence of the wavefunction normalization on small changes of the box size due to the $u$ and $v$ oscillating behaviour at the borders of the integration region, we average over an entire oscillation period of the wavefunction. \\
\indent We point out that this numerical technique, based on the assessment of the wavefunctions on a radial grid, scales almost linearly with the number of points in the mesh.  
This approach is used below to calculate the electronic structure of the Au$^+$ ion.

\section{Methods for polycentric, non-spherically symmetric, non-periodic systems}

A second approach, which can be easily extended to polycentric molecular systems, relies on the projected potential method \cite{taioli2010electron,taioli2015computational}. Within this framework, the non-kinetic terms of equation (\ref{DHFexplicit}) are projected into a finite functional space $\cal{G}$ spanned by a number of $L^2$-functions ($\lambda$), which define a projector 
\begin{equation}\label{proj}
\pi=\sum_{\lambda}|\lambda\rangle \langle \lambda|
\end{equation}
where $g_j$ are $L^2$ basis functions.
Inside that subspace the equation of motion can be written: 

\begin{equation}\label{hamnosimm}
\left(H_{0}-E+\pi V\pi\right)\psi=0
\end{equation}
or, equivalently,

\begin{equation}
\pi\psi=\pi\frac{1}{E-H_{0}}\pi V\pi\psi\label{eq_sec_pot_proiettati}
\end{equation}
where

\begin{equation}\label{kinterm}
H_{0}=\left(\begin{array}{cc}
mc^{2} & -c\boldsymbol{\sigma}\cdot i\boldsymbol{\nabla}\\
-c\boldsymbol{\sigma}\cdot i\boldsymbol{\nabla} & -mc^{2}
\end{array}\right) 
\end{equation}

\noindent and 

\begin{equation}\label{projectpot}
V=\pi\left(\begin{array}{cc}
W_{V}+W_{S}+\mathbf{A}_{P}\cdot\boldsymbol{\sigma} & -\boldsymbol{\sigma}\cdot\mathbf{A}+W_{PS}\\
-\boldsymbol{\sigma}\cdot\mathbf{A}+W_{PS} & W_{V}+\mathbf{A}_{P}\cdot\boldsymbol{\sigma}-W_{S}
\end{array}\right)\pi
\end{equation}

The kinetic term (\ref{kinterm}) is not projected into the Hilbert subspace spanned by the projector in order to recover the continuum for dealing with scattering states. 
At variance, the potential terms in equation (\ref{projectpot})
are projected into the functional space using the projector operator (\ref{proj}). 
For many-electron systems the total Coulomb potential (\ref{potdhf}) upon projection reads:
\begin{equation}\label{projectpot1}
V_{\alpha}=\sum_{\lambda\mu\nu\tau}|\lambda\rangle S^{-1}_{\lambda\mu}\langle \mu|V|\nu\rangle S^{-1}_{\nu\tau}\langle \tau|; \quad
S_{\lambda\mu} = \langle\lambda|\mu\rangle
\end{equation}
The projected potential (\ref{projectpot1}) is well-suited for bound-state calculations, where wavefunctions are appreciably confined into a finite region. To model scattering states we require that the projected potential $V_{\alpha}$ correctly reproduces the effect of the long range part of the true potential at least in a part of
the asymptotic region, where $V_{\alpha}(r)\simeq V(r)$ (for $r$ large). This fact guarantees that the scattering wavefunction far from the scattering center has  
the correct form  \cite{taioli2010electron,taioli2015computational}.\\
\indent 
The solution of equation (\ref{hamnosimm}), or equivalently of equation (\ref{eq_sec_pot_proiettati}), of course delivers the eigenvalues of the projected Hamiltonian $H_{0}+\pi V\pi$ rather than those of the complete Hamiltonian $H_{0}+V$. 
However, we notice that the eigenvalues of the projected and complete Hamiltonian coincide provided that the vectors $\psi$ and $V\psi$ belong to the projected functional subspace. 
Equation (\ref{eq_sec_pot_proiettati}) can then be solved if the matrix elements

\begin{equation}\label{resolvent}
\pi\frac{1}{E^{2}-m^{2}c^{4}-c^{2}p^{2}}\pi, \quad \pi\frac{1}{E^{2}-m^{2}c^{4}-c^{2}p^{2}}\vec{p}\pi
\end{equation}

\noindent of the projected Dirac Hamiltonian are reckoned within the chosen functional space. To find the elements of these Green's operator we refer the interested reader to Refs. \cite{taioli2010electron,taioli2015computational}. 

\subsection{Gaussian functions}

The functions that we use to project the equation (\ref{hamnosimm}) and to expand the wavefunctions are HGBF. 
Gaussian functions (GF) are particularly suited for representing both the bound and continuum orbitals \cite{taioli2010electron} in multicentric molecular systems, as analytic expressions of monoelectronic
and bielectronic integrals, necessary for the calculations of the spectral quantities,
can be explicitly derived. 
These integrals are the average values of different operators
evaluated using a mixed basis set of HGBF of every order and centrature. 
The structure of the HGBF is
the following:
\begin{equation}
g(\mathbf{r})=g(u,v,w;a,\mathbf{R};\mathbf{r})=N{\frac{\partial^{u+v+w}}
{\partial X^u \partial Y^v\partial Z^w}}(\frac{2\alpha}{\pi})^{3/4}
exp[-\alpha (\mathbf{r}-\mathbf{R})^2]
\end{equation}
where $\mathbf{R}\equiv(X,Y,Z)$ gives the position where $g$ is centered, $\alpha$ is a coefficient determining the GF width, the order of derivation ($u,v,w$) determines the symmetry type, and $N$ is a
normalization factor
$$N=[\alpha^l(2u-1)!!(2v-1)!!(2w-1)!!]^{-1/2}{,}\qquad l=u+v+w$$

GFs build up a $L^2$ non-orthogonal basis set  characterized by the remarkable property that the product of GFs (represented by coefficients $\alpha_1,\alpha_2$ and centers $R_1,R_2$, respectively) is a GF itself:

\begin{equation}
g\left(\alpha_{1},\mathbf{R}_{1};\mathbf{r}\right)g\left(\alpha_{2},\mathbf{R}_{2};\mathbf{r}\right)=g\left(\alpha_{1}+\alpha_{2},\frac{\alpha_{1}\mathbf{R}_{1}+\alpha_{2}\mathbf{R}_{2}}{\alpha_{1}+\alpha_{2}};\mathbf{r}\right)e^{-\frac{\alpha_{1}\alpha_{2}}{\alpha_{1}+\alpha_{2}}\left(\mathbf{R}_{1}-\mathbf{R}_{2}\right)}
\end{equation}

\noindent with normalization condition given by 

\[
\left\langle g_{1}|g_{2}\right\rangle =\left(\frac{2\alpha_{1}}{\pi}\right)^{\frac{3}{4}}\left(\frac{2\alpha_{2}}{\pi}\right)^{\frac{3}{4}}\left(\frac{\pi}{\alpha_{1}+\alpha_{2}}\right)^{\frac{3}{2}}e^{-\frac{\alpha_{1}\alpha_{2}}{\alpha_{1}+\alpha_{2}}\left(\mathbf{R}_{1}-\mathbf{R}_{2}\right)}=\left(\frac{2\sqrt{\alpha_{1}\alpha_{2}}}{\alpha_{1}+\alpha_{2}}\right)^{\frac{3}{2}}e^{-\frac{\alpha_{1}\alpha_{2}}{\alpha_{1}+\alpha_{2}}\left(\mathbf{R}_{1}-\mathbf{R}_{2}\right)}
\]

\subsection{Mono-electronic integrals}

GFs allow the analytical calculation of the projected Coulomb potential (Hartree term):

\begin{equation}
\int d\mathbf{r}\frac{1}{r}g\left(\alpha,\mathbf{R};\mathbf{r}\right)=\left(\frac{\pi}{\alpha}\right)^{\frac{3}{2}}\frac{\mathrm{erf}\left(\sqrt{\alpha}R\right)}{R}
\end{equation}

\noindent This task can be achieved by exploting the fact that the Yukawa potential reduces to the electron-electron Coulomb potential in the short wave-length limit ($\mu \rightarrow 0$, see equation \ref{Yukawa}).
Mono-electronic integrals for the Yukawa potential can be written as follows:

\begin{eqnarray}\label{Yuko}
\small
\left\langle m\frac{e^{-\frac{r}{\xi}}}{r}\right\rangle= \int d\mathbf{r}\frac{e^{-\frac{r}{\xi}}}{r}g\left(\alpha,\mathbf{R};\mathbf{r}\right)  =2\pi\int\limits _{0}^{\infty}r^{2}dr\left[e^{-\alpha\left(r^{2}-2rR+R^{2}\right)}-e^{-\alpha\left(r^{2}+2rR+R^{2}\right)}\right]\frac{e^{-\frac{r}{\xi}}}{r}\nonumber \\
=  \frac{1}{2}  \left(\frac{\pi}{\alpha}\right)^{\frac{3}{2}}\frac{\left[-\mathrm{erfc}\left(\frac{1}{2\sqrt{\alpha}\xi}+\sqrt{\alpha}R\right)e^{\left(\frac{1}{4\alpha\xi^{2}}+\frac{R}{\xi}\right)}+\mathrm{erfc}\left(\frac{1}{2\sqrt{\alpha}\xi}-\sqrt{\alpha}R\right)e^{\left(\frac{1}{4\alpha\xi^{2}}-\frac{R}{\xi}\right)}\right]}{R} \nonumber \\
=  \frac{1}{2} \left(\frac{\pi}{\alpha}\right)^{\frac{3}{2}}\frac{\left[-\mathrm{erfcx}\left(\frac{1}{2\sqrt{\alpha}\xi}+\sqrt{\alpha}R\right)+\mathrm{erfcx}\left(\frac{1}{2\sqrt{\alpha}\xi}-\sqrt{\alpha}R\right)\right]e^{-\alpha R^{2}}}{R}
\end{eqnarray}

\noindent where $\mathrm{erfcx}(x)=\mathrm{erfc}(x)e^{x^{2}}$ and 

\[
\mathrm{erfc}(x)=\frac{2}{\sqrt{\pi}}\int\limits _{x}^{\infty}e^{-u^{2}}du,\qquad\mathrm{erfcx}(x)=\frac{2}{\sqrt{\pi}}\int\limits _{x}^{\infty}e^{x^{2}-u^{2}}du
\]
By Taylor expanding Equation (\ref{Yuko}) and observing that 

\begin{eqnarray}
\frac{d}{dx}\mathrm{erfcx}(a+x)e^{-x^{2}}=-\frac{2}{\sqrt{\pi}}e^{-x^{2}}+2a \ \mathrm{erfcx}(a+x)e^{-x^{2}} \nonumber \\ \frac{d}{dx}\mathrm{erfcx}(a-x)e^{-x^{2}}=\frac{2}{\sqrt{\pi}}e^{-x^{2}}-2a \ \mathrm{erfcx}(a-x)e^{-x^{2}}
\end{eqnarray}

\noindent we obtain finally:

\begin{eqnarray}
\left\langle m\frac{e^{-\frac{r}{\xi}}}{r}\right\rangle=\frac{2\pi^{\frac{3}{2}}}{\alpha} \Big[ \frac{1}{\sqrt{\pi}}-a \ \mathrm{erfcx}(a)+\frac{1}{3}\left(-2a^{3}\mathrm{erfcx}(a)+\frac{2a^{2}-1}{\sqrt{\pi}}\right)x^{2}+ \nonumber \\
+\frac{2}{15}\left(-4a^{5}\mathrm{erfcx}(a)+\frac{4a^{4}-2a^{2}+3}{\sqrt{\pi}}\right)+\cdots \Big] 
\end{eqnarray}

\noindent where $a=\frac{1}{2\sqrt{\alpha}\xi}$ and $x=\sqrt{\alpha}R$.\\
\indent Furthermore, one can also find analytically the integral of a Coulomb and Gaussian potentials between Gaussian functions as:

\begin{equation}
\left\langle g_{1}|\frac{1}{\left|\vec{r}-\vec{C}\right|}|g_{2}\right\rangle =\left(\frac{2\sqrt{\alpha_{1}\alpha_{2}}}{\pi}\right)^{\frac{3}{2}}e^{-\frac{\alpha_{1}\alpha_{2}}{\alpha_{1}+\alpha_{2}}\left(\vec{R}_{1}-\vec{R}_{2}\right)^{2}}\left(\frac{\pi}{\alpha_{1}+\alpha_{2}}\right)^{\frac{3}{2}}\frac{\mathrm{erf}\left(\sqrt{\left(\alpha_{1}+\alpha_{2}\right)\left|\frac{\alpha_{1}\vec{R}_{1}+\alpha_{2}\vec{R}_{2}}{\alpha_{1}+\alpha_{2}}-\vec{C}\right|^{2}}\right)}{\left|\frac{\alpha_{1}\vec{R}_{1}+\alpha_{2}\vec{R}_{2}}{\alpha_{1}+\alpha_{2}}-\vec{C}\right|}
\end{equation}
\noindent and:

\begin{eqnarray}
&\left\langle g_{1}|e^{-\gamma\left(\vec{r}-\vec{C}\right)^{2}}|g_{2}\right\rangle  =\left(\frac{2\sqrt{\alpha_{1}\alpha_{2}}}{\pi}\right)^{\frac{3}{2}}e^{-\frac{\alpha_{1}\alpha_{2}}{\alpha_{1}+\alpha_{2}}\left(\vec{R}_{1}-\vec{R}_{2}\right)^{2}}\int d^{3}re^{-\left(\alpha_{1}+\alpha_{2}\right)\left(\vec{r}-\frac{\alpha_{1}\vec{R}_{1}+\alpha_{2}\vec{R}_{2}}{\alpha_{1}+\alpha_{2}}\right)^{2}}e^{-\gamma\left(\vec{r}-\vec{C}\right)^{2}}\nonumber \\
 & =\left(\frac{2\sqrt{\alpha_{1}\alpha_{2}}}{\pi}\right)^{\frac{3}{2}}e^{-\frac{\alpha_{1}\alpha_{2}}{\alpha_{1}+\alpha_{2}}\left(\vec{R}_{1}-\vec{R}_{2}\right)^{2}}e^{-\frac{\left(\alpha_{1}+\alpha_{2}\right)\gamma}{\alpha_{1}+\alpha_{2}+\gamma}\left(\frac{\alpha_{1}\vec{R}_{1}+\alpha_{2}\vec{R}_{2}}{\alpha_{1}+\alpha_{2}}-\vec{C}\right)^{2}}\int d^{3}re^{-\left(\alpha_{1}+\alpha_{2}+\gamma\right)\left(\vec{r}-\frac{\alpha_{1}\vec{R}_{1}+\alpha_{2}\vec{R}_{2}+\gamma\vec{C}}{\alpha_{1}+\alpha_{2}+\gamma}\right)^{2}}\nonumber \\
 & =\left(\frac{2\sqrt{\alpha_{1}\alpha_{2}}}{\pi}\right)^{\frac{3}{2}}e^{-\frac{\alpha_{1}\alpha_{2}}{\alpha_{1}+\alpha_{2}}\left(\vec{R}_{1}-\vec{R}_{2}\right)^{2}}e^{-\frac{\left(\alpha_{1}+\alpha_{2}\right)\gamma}{\alpha_{1}+\alpha_{2}+\gamma}\left(\frac{\alpha_{1}\vec{R}_{1}+\alpha_{2}\vec{R}_{2}}{\alpha_{1}+\alpha_{2}}-\vec{C}\right)^{2}}\left(\frac{\pi}{\alpha_{1}+\alpha_{2}+\gamma}\right)^{\frac{3}{2}}
\end{eqnarray}

\subsection{Green's function matrix elements.}

The Green's function projected on GFs in equation (\ref{eq_sec_pot_proiettati}) reads:  
\begin{equation}\label{green}
\left\langle g_{1}|\frac{1}{p_{0}^{2}+\nabla^{2}}|g_{2}\right\rangle 
\end{equation}
where $p_{0}\in\mathbb{C}$. The integral in Equation (\ref{green}) can be more easily calculated in momentum space, as the kinetic operator ($\nabla^{2}$) is diagonal:

\begin{eqnarray}
&\left\langle g_{1}|\frac{1}{p_{0}^{2}+\nabla^{2}}|g_{2}\right\rangle = \nonumber \\ &\left(\frac{2\pi}{\sqrt{\alpha_{1}\alpha_{2}}}\right)^{\frac{3}{2}}\int d\mathbf{p}\frac{e^{-\frac{\alpha_{1}+\alpha_{2}}{4\alpha_{1}\alpha_{2}}p^{2}+i\mathbf{p}\cdot\left(\mathbf{R}_{1}-\mathbf{R}_{2}\right)}}{p_{0}^{2}-p^{2}}  
=\left(\frac{2\pi}{\sqrt{\alpha_{1}\alpha_{2}}}\right)^{\frac{3}{2}}4\pi\int\limits _{0}^{\infty}\frac{p^{2}dp}{p\left|\mathbf{R}_{1}-\mathbf{R}_{2}\right|}\frac{e^{-\frac{\alpha_{1}+\alpha_{2}}{4\alpha_{1}\alpha_{2}}p^{2}}\sin\left(p\left|\mathbf{R}_{1}-\mathbf{R}_{2}\right|\right)}{p_{0}^{2}-p^{2}} = \nonumber \\
&\left(\frac{2\pi}{\sqrt{\alpha_{1}\alpha_{2}}}\right)^{\frac{3}{2}}2\pi\int\limits _{-\infty}^{\infty}\frac{pdp}{\left|\mathbf{R}_{1}-\mathbf{R}_{2}\right|}\frac{e^{-\frac{\alpha_{1}+\alpha_{2}}{4\alpha_{1}\alpha_{2}}p^{2}+ip\left|\mathbf{R}_{1}-\mathbf{R}_{2}\right|}}{p_{0}^{2}-p^{2}}
\end{eqnarray}

These integrals can be reduced to integrals of the $\mathrm{erf}$ function in the complex space by considering that $w(z)=e^{-z^{2}}\mathrm{erfc}(-iz)$
and that

\begin{equation}
\int\limits _{-\infty}^{\infty}\frac{e^{-t^{2}}}{z-t}dt =
\left\{ \begin{array}{@{\kern2.5pt}lL}
    \hfill -i\pi w(z) & if Im(z) $>$ 0 \\
    \hfill -i\pi w(z)-2e^{-z^{2}} & if Im(z) $<$ 0.
\end{array}\right.
\end{equation}

\subsection{The two-step recurrence relation of $G_{m}(x)$.}

We observe that the integrals involving GFs of higher order can be obtained
through the differentiation of $s$-type GFs,
because of the Leibnitz's rule that allows one
to take the differentiation operator out of the integral sign.
In particular, the Yukawa potential matrix element can be calculated for Gaussians of any order ($s,p,d$ symmetry) by a recurrence relation.\\ \indent To demonstrate that, we define the following quantities:

\begin{equation}
G_{m}(x)=\left(-\frac{1}{2x}\frac{d}{dx}\right)^{m}\left(\frac{\alpha}{\pi}\right)^{\frac{3}{2}}\left\langle m\frac{e^{-\frac{r}{\xi}}}{r}\right\rangle =\left(-\frac{1}{2x}\frac{d}{dx}\right)^{m}\frac{\sqrt{\alpha}}{2}\frac{\left[-\mathrm{erfcx}\left(a+x\right)+\mathrm{erfcx}\left(a-x\right)\right]e^{-x^{2}}}{x} 
\end{equation}

\begin{equation}
I_{0}(a,x)=\left[-\mathrm{erfcx}\left(a+x\right)+\mathrm{erfcx}\left(a-x\right)\right]e^{-x^{2}}
\end{equation}
\begin{equation}
I_{1}(a,x)=\left[\mathrm{erfcx}\left(a+x\right)+\mathrm{erfcx}\left(a-x\right)\right]e^{-x^{2}}
\end{equation}
The first element of the recurrence relation reads: 
\begin{equation}
\small
G_{0}(x)=\frac{\sqrt{\alpha}}{2}\frac{I_{0}(a,x)}{x}
\end{equation}
while the second one can be obtained by observing that $\frac{dI_{0}}{dx}=\frac{4}{\sqrt{\pi}}e^{-x^{2}}-2aI_{1}$
and $\frac{dI_{1}}{dx}=-2aI_{0}$:

\begin{eqnarray}
G_{1}(x) & =\left(\frac{1}{2x}\frac{d}{dx}\right)G_{0}(x)=\frac{\sqrt{\alpha}}{4x^{2}}\left\{ -\frac{4}{\sqrt{\pi}}e^{-x^{2}}+2aI_{1}(a,x)+\frac{I_{0}(a,x)}{x}\right\} \nonumber \\
 & =\frac{1}{x^{2}}\left[\frac{1}{2}G_{0}(x)-\sqrt{\frac{\alpha}{\pi}}e^{-x^{2}}+\frac{\sqrt{\alpha}}{2}aI_{1}(a,x)\right]
\end{eqnarray}

\noindent Finally, the recurrence relation can be generally written as:

\begin{equation}
G_{m+2}(x)=\frac{1}{x^{2}}\left[\left(m+\frac{3}{2}\right)G_{m+1}(x)-\sqrt{\frac{\alpha}{\pi}}e^{-x^{2}}+a^{2}G_{m}(x)\right]
\end{equation}

\subsection{Bi-electronic integrals}

Bi-electronic integrals, which are the matrix elements of the screened Coulomb potential between GFs centered in different positions, can be written:

\begin{eqnarray}
&\left\langle s_{1}s_{3}|\frac{e^{-\frac{\left|\mathbf{r}-\mathbf{r}'\right|}{\xi}}}{\left|\mathbf{r}-\mathbf{r}'\right|}|s_{2}s_{4}\right\rangle  =\nonumber \\
& \left(\frac{2\alpha_{1}}{\pi}\right)^{\frac{3}{4}}\left(\frac{2\alpha_{2}}{\pi}\right)^{\frac{3}{4}} \left(\frac{2\alpha_{3}}{\pi}\right)^{\frac{3}{4}}\left(\frac{2\alpha_{4}}{\pi}\right)^{\frac{3}{4}} \int d\mathbf{r}\int d\mathbf{r}'e^{-\alpha_{1}\left(\mathbf{r}-\mathbf{R}_{1}\right)^{2}}e^{-\alpha_{2}\left(\mathbf{r}-\mathbf{R}_{2}\right)^{2}}e^{-\alpha_{3}\left(\mathbf{r}'-\mathbf{R}_{3}\right)^{2}}e^{-\alpha_{4}\left(\mathbf{r}'-\mathbf{R}_{4}\right)^{2}}= \nonumber \\
& \left[\frac{2\left(\alpha_{1}\alpha_{2}\alpha_{3}\alpha_{4}\right)^{\frac{1}{4}}}{\pi}\right]^{3}e^{-\frac{\alpha_{1}\alpha_{2}}{\alpha_{1}+\alpha_{2}}\left(\mathbf{R}_{1}-\mathbf{R}_{2}\right)}e^{-\frac{\alpha_{3}\alpha_{4}}{\alpha_{3}+\alpha_{4}}\left(\mathbf{R}_{3}-\mathbf{R}_{4}\right)}  \int d\mathbf{r}\int d\mathbf{r}'e^{-\left(\alpha_{1}+\alpha_{2}\right)\left(\mathbf{r}-\mathbf{P}_{12}\right)^{2}}e^{-\left(\alpha_{3}+\alpha_{4}\right)\left(\mathbf{r}'-\mathbf{P}_{34}\right)^{2}}\frac{e^{-\frac{\left|\mathbf{r}-\mathbf{r}'\right|}{\xi}}}{\left|\mathbf{r}-\mathbf{r}'\right|}
\end{eqnarray}
where $\mathbf{P}_{12}=\frac{\alpha_{1}\mathbf{R}_{1}+\alpha_{2}\mathbf{R}_{2}}{\alpha_{1}+\alpha_{2}}$ and $\mathbf{P}_{34} =\frac{\alpha_{3}\mathbf{R}_{3}+\alpha_{4}\mathbf{R}_{4}}{\alpha_{3}+\alpha_{4}}$.\\
\indent By defining $\mathbf{s}=\frac{\mathbf{r}+\mathbf{r}'}{2}$ and $\mathbf{q}=\mathbf{r}-\mathbf{r}'$ one obtains:

\begin{eqnarray}
& \int d\mathbf{r}\int  d\mathbf{r}'e^{-\left(\alpha_1+\alpha_2\right)\left(\mathbf{r}-\mathbf{P}_{12}\right)^2} e^{-\left(\alpha_3+\alpha_4\right)\left(\mathbf{r}'-\mathbf{P}_{34}\right)^{2}} \frac{e^{-\frac{\left|\mathbf{r}-\mathbf{r}'\right|}{\xi}}}{\left|\mathbf{r}-\mathbf{r}'\right|} =\int d\mathbf{s}\int d\mathbf{q}e^{-\left(\alpha_{1}+\alpha_{2}\right)\left(\mathbf{s}+\frac{\mathbf{q}}{2}-\mathbf{P}_{12}\right)^{2}}e^{-\left(\alpha_{3}+\alpha_{4}\right)\left(\mathbf{s}-\frac{\mathbf{q}}{2}-\mathbf{P}_{34}\right)^{2}}\frac{e^{-\frac{q}{\xi}}}{q} = \nonumber \\ & \left(\frac{\pi}{\alpha_{1}+\alpha_{2}+\alpha_{3}+\alpha_{4}}\right)^{\frac{3}{2}}\int d\mathbf{q}e^{-\frac{\left(\alpha_{1}+\alpha_{2}\right)\left(\alpha_{3}+\alpha_{4}\right)}{\alpha_{1}+\alpha_{2}+\alpha_{3}+\alpha_{4}}\left[\mathbf{q}-\left(\mathbf{P}_{12}-\mathbf{P}_{34}\right)\right]^{2}}\frac{e^{-\frac{q}{\xi}}}{q}
\end{eqnarray}
which can be reduced to an integral of the type reported in equation (\ref{Yuko}). 
Further details on the calculation of mono- and bi-electronic integrals using GFs of different orders (modified Gaussian functions) can be found in \cite{gollebiewski}.\\
\indent We point out that the self-consistent solution of the Dirac equation using the HGFB numerical approach scales from cubically, owing to the matrix diagonalization, to quartically, due to the calculation of the bi-electronic integrals, as a function of the number of Gaussians in the basis set. This approach is used below to calculate the electronic structure of the W@Au$_{12}$ nanocluster.

\section{Radial functions and Gaussian basis set at work: application to the electronic structure calculation of heavy-element atoms and clusters}

To show the potential of our relativistic approach, we calculate first the electronic structure of the Au$^+$ ion, by comparing the results obtained by both the radial and Gaussian basis sets, and second of the W@Au$_{12}$ superatomic system, where the inclusion of relativistic effects is necessary to describe accurately the electronic structure \cite{C8CP04651D}. 

\subsection{Au and W atoms from relativistic HF using radial grids and Gaussian function basis sets}

The calculations were carried out for Au$^{+}$, as it is a closed-shell system (neutral gold has indeed 79 electrons), using i) a contracted $ANO-R$ basis set  \cite{zobel2020a,pritchard2019a,feller1996a,Schuchardt} with (26$s$, 20$p$, 15$d$, 10$f$, 4$g$) GFs;
ii) a modified $ANO-R$ basis set obtained by uncontraction of the previous primitive $ANO-R$ functions to (26$s$, 26$p$, 20$d$, 15$f$, 10$g$) GFs;
iii) tempered GFs, whose exponent follow a geometric progression; iv) $aug-cc-pVTZ-PP$ and $cc-pVTZ-PP$ GF basis sets \cite{peterson2005a,figgen2005a}, which are two-component relativistic pseudo-potentials (PP) optimized to calculate ground and excited state spectroscopic properties. 
We notice that the procedure to uncontract the $ANO-R$ basis set is carried out to obtain GFs optimized also for the small component of the Dirac spinor and not only for the large one (for which this basis set is generated). Indeed, we have found out empirically that uncontracted basis sets achieve better accuracy in fully relativistic simulations. The calculations performed with both $ANO-R$ and tempered GF basis sets are all-electron, while those labelled by $-PP$ are characterised by semi-local PPs that mimic the ionic core-valence electrons interaction to achieve high computational efficiency in DHF calculations. The number of electrons for both Au and W included in the PP core is 60, and the remaining electrons are treated explicitly. The semi-local pseudo-potential reads \cite{figgen2005a}:
\begin{equation}\label{PPs}
V_{PP}=\frac{Q}{r}+\sum_{l,j} V_{lj}(r){\cal{P}}_{lj}    =\frac{Q}{r}+\sum_{ij}\sum_k B_{lj,k}e^{-\beta_{lj,k}r^2}
\end{equation}
where $\frac{Q}{r}$ represents the long-range part of the PP carrying the charge $Q$, while $V_{lj}(r)$ is the short-range radial (spherical-symmetric) potential characterised by the parameters $B_{lj,k}, \beta_{lj,k}$ that are obtained by fitting with ab initio all-electron atomic data. 
The projectors ${\cal{P}}_{lj}$ in equation (\ref{PPs}) enforce the condition that each core-shell characterised by a different combination of the angular-momentum quantum numbers $l$ and $j=l\pm 1/2$ is included in the PP with different parameters.   
Following equation (\ref{PPs}), PP can be written as linear combination of spherical symmetric GFs acting on functional spaces of different symmetry $g=(l,j)$ (e.g. $s_{1/2},p_{1/2},p_{3/2},d_{3/2},d_{5/2}\cdots$), that is: 
\begin{equation}\label{PPsymm}
\tilde{V}_{g}\left(r\right)=\sum_{i}\tilde{c}_{i,g}e^{-\alpha_{i}r^{2}}
\end{equation}
We remind that in a fully relativistic picture, shells characterised by the same orbital angular momentum $l$ differ in energy owing to the spin-orbit interaction.\\
\indent Since $l\cdot s= \frac{j(j+1)-l(l+1)-s(s+1)}{2}$, from a semi-relativistic quantum formulation in spherical symmetry one has:
\begin{equation}\label{poten}
\tilde{V}_{g}\left(r\right)=V_{l}\left(r\right)+\frac{1}{2m_{e}^{2}c^{2}}\frac{1}{r}\frac{dV_{so,l\pm1}}{dr}\frac{\left(l\pm\frac{1}{2}\right)\left(l\pm\frac{1}{2}+1\right)-l(l+1)-\frac{1}{2}\left(\frac{1}{2}+1\right)}{2}
\end{equation}
which can be specified for the different shells ($l>0$) as follows:
\begin{eqnarray}\label{potenspe}
\tilde{V}_{l,l-1/2}\left(r\right) & =V_{l}\left(r\right)-\frac{\left(l+1\right)}{4m_{e}^{2}c^{2}}\frac{1}{r}\frac{dV_{so,l-1}}{dr}\nonumber \\
\tilde{V}_{l,l+1/2}\left(r\right) & =V_{l}\left(r\right)+\frac{l}{4m_{e}^{2}c^{2}}\frac{1}{r}\frac{dV_{so,l+1}}{dr}\nonumber \\
\tilde{V}_{l+1,l+1/2}\left(r\right) & =V_{l+1}\left(r\right)-\frac{\left(l+2\right)}{4m_{e}^{2}c^{2}}\frac{1}{r}\frac{dV_{so,l}}{dr}\nonumber \\
\tilde{V}_{l-1,l-1/2}\left(r\right) & =V_{l-1}\left(r\right)+\frac{\left(l-1\right)}{4m_{e}^{2}c^{2}}\frac{1}{r}\frac{dV_{so,l}}{dr}
\end{eqnarray}
By subtracting the second and first equation above, and by summing the third and the fourth, one obtains the following relations:
\begin{eqnarray}
\frac{1}{4m_{e}^{2}c^{2}}\frac{1}{r}\frac{d\left[lV_{so,l+1}+\left(l+1\right)V_{so,l-1}\right]}{dr}=\tilde{V}_{l,l+1/2}\left(r\right)-\tilde{V}_{l,l-1/2}\left(r\right)\nonumber \\
\left(l-1\right)V_{l+1}\left(r\right)+\left(l+2\right)V_{l-1}\left(r\right)=\left(l-1\right)\tilde{V}_{l+1,l+1/2}\left(r\right)+\left(l+2\right)\tilde{V}_{l-1,l-1/2}\left(r\right)
\end{eqnarray}
Since for a given $l$, the scalar $\mathcal{V}_{s}$ and vector  $\mathcal{V}_{v}$ relativistic potentials are related to $V_{l}$ e $V_{so,l}$ by (see the upper and lower diagonal elements of the Dirac equation (\ref{DHFexplicit}) in matrix form): 
\begin{eqnarray}\label{vecsapot}
\mathcal{V}_{v,l}+\mathcal{V}_{s,l} & =V_{l}\nonumber \\
\mathcal{V}_{v,l}-\mathcal{V}_{s,l} & =V_{so,l}
\end{eqnarray}
one has:
\begin{equation}\label{vepoto}
3V_{0}=3\tilde{V}_{s_{1/2}},\quad V_{3}+4V_{1}=\tilde{V}_{f_{5/2}}+4\tilde{V}_{p_{3/2}},\quad 2V_{4}+5V_{2}=2\tilde{V}_{g_{7/2}}+5\tilde{V}_{d_{5/2}},\quad 6V_{3}=6\tilde{V}_{f_{7/2}},\quad 7V_{4}=7\tilde{V}_{g_{9/2}}
\end{equation}
Furthermore, by defining  $\tilde{I}_{l,l\pm1/2}=4m_{e}^{2}c^{2}\int r\tilde{V}_{l,l\pm1/2}dr$
we obtain:
\begin{eqnarray}
& V_{so,2}+2V_{so,0}=\tilde{I}_{p_{3/2}}-\tilde{I}_{p_{1/2}},\quad2V_{so,3}+3V_{so,1}=\tilde{I}_{d_{5/2}}-\tilde{I}_{d_{3/2}},\quad 3V_{so,4}+4V_{so,2}=\tilde{I}_{f_{7/2}}-\tilde{I}_{f_{5/2}}\nonumber \\
& 5V_{so,3}=\tilde{I}_{g_{9/2}}-\tilde{I}_{g_{7/2}},\quad6V_{so,4}=0
\end{eqnarray}
and, finally,
\begin{equation}\label{sopot}
V_{so,0}=0,\quad 3V_{so,1}=\tilde{I}_{d_{5/2}}-\tilde{I}_{d_{3/2}}-\frac{2}{5}\left(\tilde{I}_{g_{9/2}}-\tilde{I}_{g_{7/2}}\right),\quad 4V_{so,2}=\tilde{I}_{f_{7/2}}-\tilde{I}_{f_{5/2}}
\end{equation}

\begin{table}[htp!]
\begin{center}
{\scriptsize{}}%
\begin{tabular}{|c|c|c|c|c|c|c|c|}
\hline 
{\scriptsize{}Level} & \multicolumn{6}{c|}{{\scriptsize{}Gaussian}} & {\scriptsize{}Radial grid}\tabularnewline
\hline 
  & {\scriptsize{}$HF$} & {\tiny{}$DHF\,\,(ANO-R)$} & {\tiny{}$DHF\,\,(ANO-R)\,\,$mod.}{\scriptsize{} } & {\scriptsize{}Tempered} & %
\begin{minipage}[t]{1.8cm}%
{\scriptsize{}$aug-cc-pVTZ-PP$}{\scriptsize\par}  
{\scriptsize{}}{\scriptsize\par}%
\end{minipage} &
\begin{minipage}[t]{1.8cm}
{\scriptsize{}$cc-pVTZ-PP$}{\scriptsize\par} {\scriptsize{}}{\scriptsize\par}%
\end{minipage} & {\scriptsize{}$Loc.\,\,exch.$}
\tabularnewline
\hline 
{\scriptsize{}$1s_{1/2}$} & {\scriptsize{}$-2707.111$} & {\scriptsize{}$-2960.001$} & {\scriptsize{}$-2957.496$} & {\scriptsize{}$-2957.543$} &  & &  {\scriptsize{}$-2998.986$}\tabularnewline
\hline 
{\scriptsize{}$2s_{1/2}$} & {\scriptsize{}$-457.459$} & {\scriptsize{}$-532.437$} & {\scriptsize{}$-529.119$} & {\scriptsize{}$-529.566$} &  & & {\scriptsize{}$-534.299$}\tabularnewline
\hline 
{\scriptsize{}$2p_{1/2}$} & \multirow{2}{1.2cm}{\scriptsize{}$-439.183$} & {\scriptsize{}$-504.453$} & {\scriptsize{}$-501.095$} & {\scriptsize{}$-501.523$} &  & & {\scriptsize{}$-514.560$}\tabularnewline
\cline{1-1} \cline{3-7} \cline{4-7} \cline{5-7} \cline{6-7} \cline{7-7} 
{\scriptsize{}$2p_{3/2}$} &  & {\scriptsize{}$-443.371$} & {\scriptsize{}$-440.020$} & {\scriptsize{}$-440.443$} &  & & {\scriptsize{}$-445.810$}\tabularnewline
\hline 
{\scriptsize{}$3s_{1/2}$} & {\scriptsize{}$-109.538$} & {\scriptsize{}$-131.239$} & {\scriptsize{}$-127.760$} & {\scriptsize{}$-128.038$} &  &  & {\scriptsize{}$-128.102$}\tabularnewline
\hline 
{\scriptsize{}$3p_{1/2}$} & \multirow{2}{1.2cm}{\scriptsize{}$-100.900$} & {\scriptsize{}$-119.602$} & {\scriptsize{}$-116.176$} & {\scriptsize{}$-116.417$} &  & &  {\scriptsize{}$-118.838$}\tabularnewline
\cline{1-1} \cline{3-7} \cline{4-7} \cline{5-7} \cline{6-7} \cline{7-7} 
{\scriptsize{}$3p_{3/2}$} &  & {\scriptsize{}$-106.277$} & {\scriptsize{}$-102.803$} & {\scriptsize{}$-103.002$} &  &  & {\scriptsize{}$-103.588$}\tabularnewline
\hline 
{\scriptsize{}$3d_{3/2}$} & \multirow{2}{1.2cm}{\scriptsize{}$~-84.996$} & {\scriptsize{}$-88.911$} & {\scriptsize{}$-85.573$} & {\scriptsize{}$-85.898$} &  & &  {\scriptsize{}$-87.770$}\tabularnewline
\cline{1-1} \cline{3-7} \cline{4-7} \cline{5-7} \cline{6-7} \cline{7-7} 
{\scriptsize{}$3d_{5/2}$} &  & {\scriptsize{}$-86.498$} & {\scriptsize{}$-82.958$} & {\scriptsize{}$-83.272$} &  & &  {\scriptsize{}$-84.430$}\tabularnewline
\hline 
{\scriptsize{}$4s_{1/2}$} & {\scriptsize{}$-24.630$} & {\scriptsize{}$-31.421$} & {\scriptsize{}$-29.315$} & {\scriptsize{}$-29.289$} &  & &  {\scriptsize{}$-29.070$}\tabularnewline
\hline 
{\scriptsize{}$4p_{1/2}$} & \multirow{2}{1.2cm}{\scriptsize{}$~-20.848$} & {\scriptsize{}$-26.476$} & {\scriptsize{}$-24.487$} & {\scriptsize{}$-24.418$} &  & &  {\scriptsize{}$-24.990$}\tabularnewline
\cline{1-1} \cline{3-7} \cline{4-7} \cline{5-7} \cline{6-7} \cline{7-7} 
{\scriptsize{}$4p_{3/2}$} &  & {\scriptsize{}$-23.285$} & {\scriptsize{}$-21.383$} & {\scriptsize{}$-21.312$} &  & &  {\scriptsize{}$-21.292$}\tabularnewline
\hline 
{\scriptsize{}$4d_{3/2}$} & \multirow{2}{1.2cm}{\scriptsize{}$~-13.933$} & {\scriptsize{}$-15.721$} & {\scriptsize{}$-14.032$} & {\scriptsize{}$-14.039$} &  & &  {\scriptsize{}$-14.459$}\tabularnewline
\cline{1-1} \cline{3-7} \cline{4-7} \cline{5-7} \cline{6-7} \cline{7-7} 
{\scriptsize{}$4d_{5/2}$} &  & {\scriptsize{}$-15.331$} & {\scriptsize{}$-13.505$} & {\scriptsize{}$-13.514$} &  & &  {\scriptsize{}$-13.754$}\tabularnewline
\hline 
{\scriptsize{}$4f_{5/2}$} & \multirow{2}{1.2cm}{\scriptsize{}$~-4.606$} & {\scriptsize{}$-5.158$} & {\scriptsize{}$-4.157$} & {\scriptsize{}$-4.317$} &  &  & {\scriptsize{}$-4.837$}\tabularnewline
\cline{1-1} \cline{3-7} \cline{4-7} \cline{5-7} \cline{6-7} \cline{7-7} 
{\scriptsize{}$4f_{7/2}$} &  & {\scriptsize{}$-1.036$} & {\scriptsize{}$-4.069$} & {\scriptsize{}$-4.224$} &  & & {\scriptsize{}$-4.677$}\tabularnewline
\hline 
{\scriptsize{}$5s_{1/2}$} & {\scriptsize{}$-4.086$} & {\scriptsize{}$-5.310$} & {\scriptsize{}$-4.970$} & {\scriptsize{}$-4.954$} & {\scriptsize{}$-4.990$} & {\scriptsize{}$-4.991$} & {\scriptsize{}$-5.070$}\tabularnewline
\hline 
{\scriptsize{}$5p_{1/2}$} & \multirow{2}{1.2cm}{\scriptsize{}$~-2.815$} & {\scriptsize{}$-3.659$} & {\scriptsize{}$-3.360$} & {\scriptsize{}$-3.333$} & {\scriptsize{}$-3.182$} & {\scriptsize{$-3.192$}} & {\scriptsize{}$-3.697$}\tabularnewline
\cline{1-1} \cline{3-7} \cline{4-7} \cline{5-7} \cline{6-7} \cline{7-7} 
{\scriptsize{}$5p_{3/2}$} &  & {\scriptsize{}$-3.075$} & {\scriptsize{}$-2.897$} & {\scriptsize{}$-2.879$} & {\scriptsize{}$-2.898$} & {\scriptsize{}$-2.899$} & {\scriptsize{}$-3.040$}\tabularnewline
\hline 
{\scriptsize{}$5d_{3/2}$} & \multirow{2}{1.2cm}{\scriptsize{}$~-0.801$} & {\scriptsize{}$-0.897$} & {\scriptsize{}$-0.791$} & {\scriptsize{}$-0.791$} & {\scriptsize{}$-0.795$} & {\scriptsize{}$-0.792$} & {\scriptsize{}$-1.120$}\tabularnewline
\cline{1-1} \cline{3-7} \cline{4-7} \cline{5-7} \cline{6-7} \cline{7-7} 
{\scriptsize{}$5d_{5/2}$} &  & {\scriptsize{}$-0.775$} & {\scriptsize{}$-0.746$} & {\scriptsize{}$-0.748$} & {\scriptsize{}$-0.746$} & {\scriptsize{}$-0.745$} & {\scriptsize{}$-1.044$}\tabularnewline
\hline 
{\scriptsize{}Total energy} &  & {\scriptsize{}$-18941.867$} & {\scriptsize{}$-18977.683$} & {\scriptsize{}$-18961.916$} &  & & \tabularnewline
\hline 
\end{tabular} 
\caption{Mean-field energy levels (Koopmans) of the Au$^+$ ion obtained by a non-relativistic approach ($HF$), a fully quantum mechanical relativistic approach ($DHF$) using $ANO-R$ contracted ($DHF\,\,(ANO-R)$), $ANO-R$ uncontracted ($DHF\,\,(ANO-R)\,\,$ mod.), and tempered Gaussian basis sets. Third and second-last columns report the relativistic values of $\epsilon_i$ using the pseudopotential method with the aug-cc-PVTZ-PP uncontracted and cc-PVTZ-PP uncontracted Gaussian basis sets for the valence electrons \cite{figgen2005a}. Last column reports the results obtained by using a radial grid with the local-exchange interaction (see equation \ref{rho13}). Data in a.u.}\label{molel}
\end{center}
\end{table}
The coefficients $\left({c}_{i,g},\alpha_{i}\right)$ of equation (\ref{PPsymm}) (or $\left(B_{lj,k}, \beta_{lj,k}\right)$ of equation \ref{PPs}) are tabulated in Ref. \cite{figgen2005a}, and so the scalar and vector relativistic potentials (\ref{vecsapot}) can be assessed from equations (\ref{vepoto}-\ref{sopot}).
To show the impact of relativistic effects on the energy levels (and ionization potential) we also carried out the mean-field calculations using the non-relativistic approximation (standard HF from the Schr{\"o}dinger picture).

\begin{figure}[htp!]\label{denwf}
\centering
\includegraphics[width=0.49\linewidth,trim=2cm 2cm 2cm 2cm, clip]{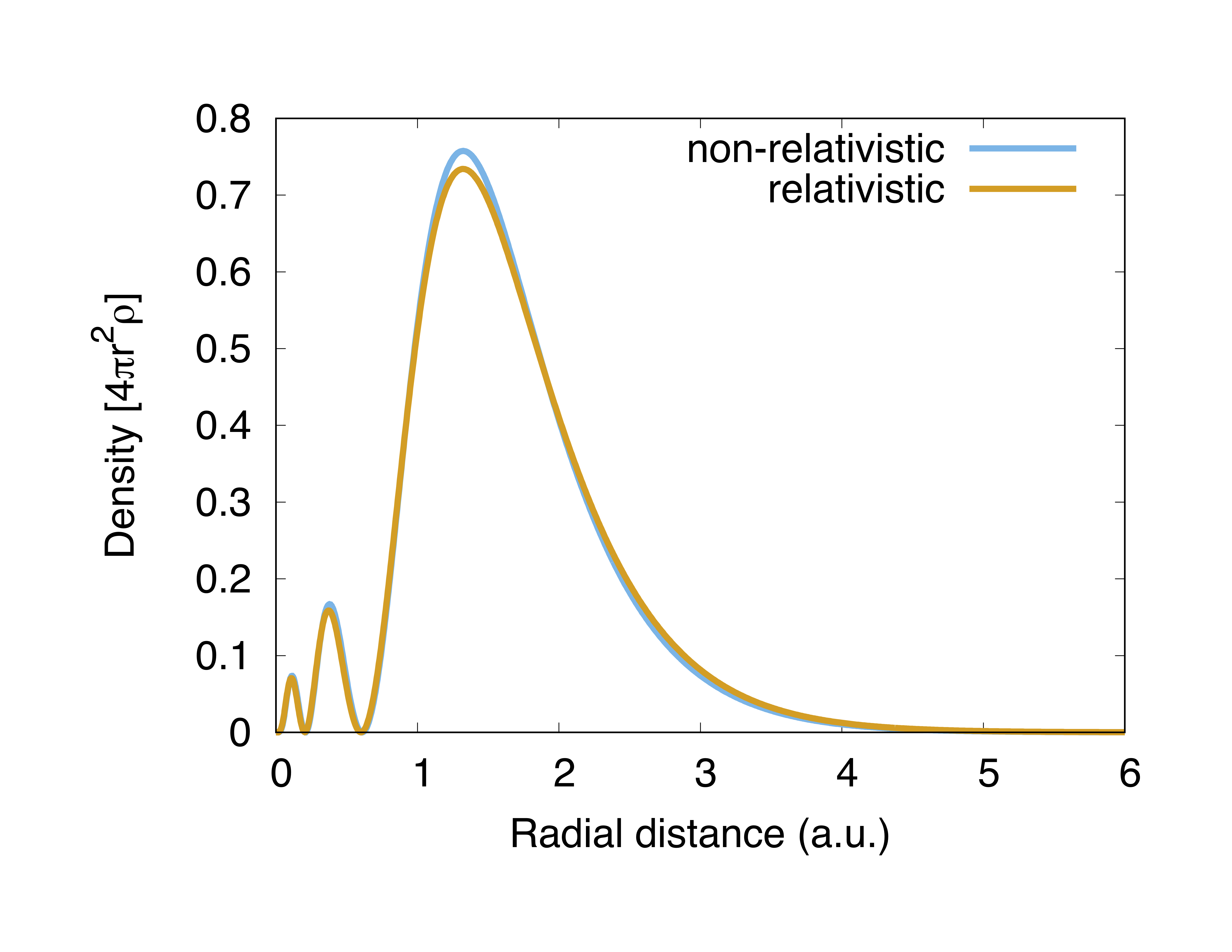}
\includegraphics[width=0.49\linewidth,trim=2cm 2cm 2cm 2cm, clip]{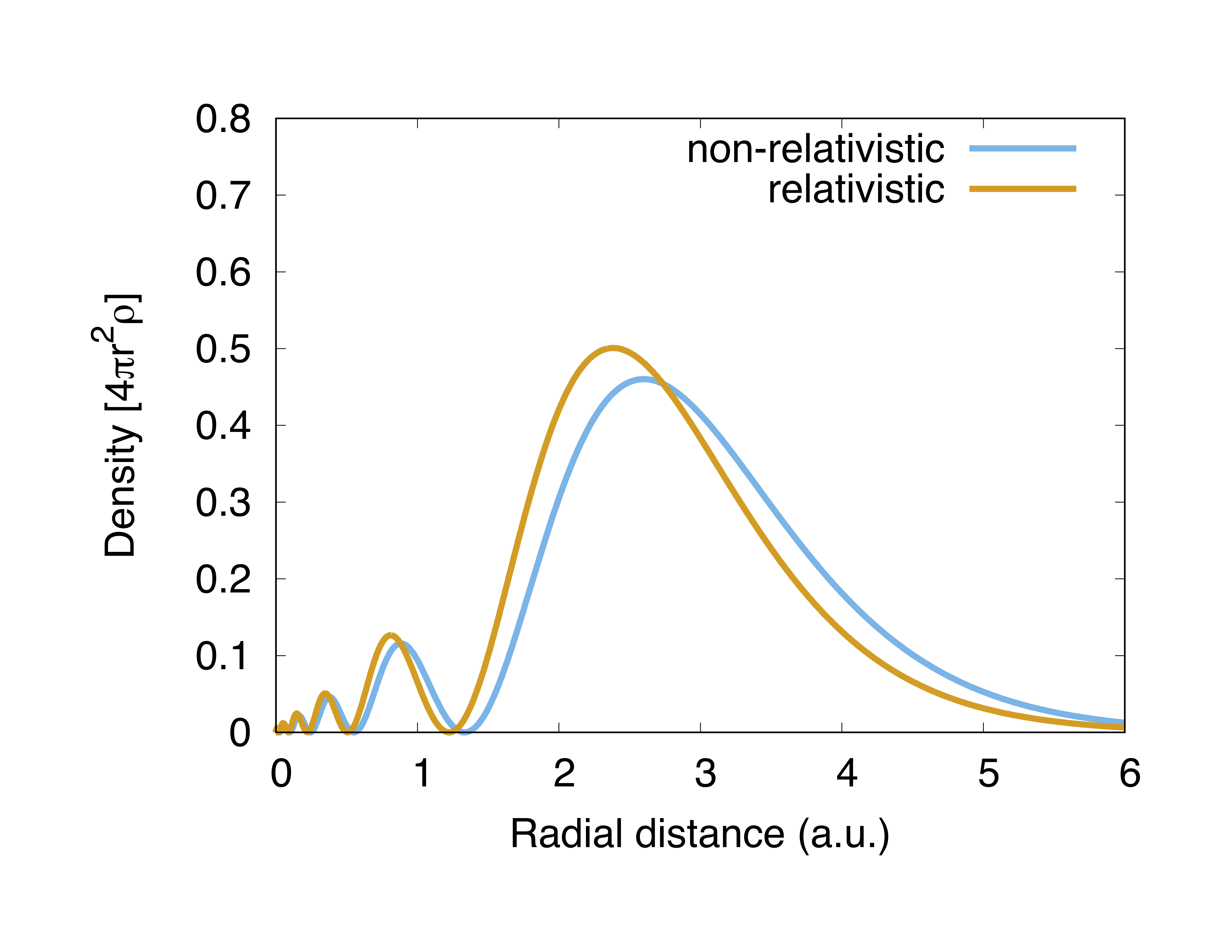}
\includegraphics[width=0.49\linewidth,trim=2cm 2cm 2cm 2cm, clip]{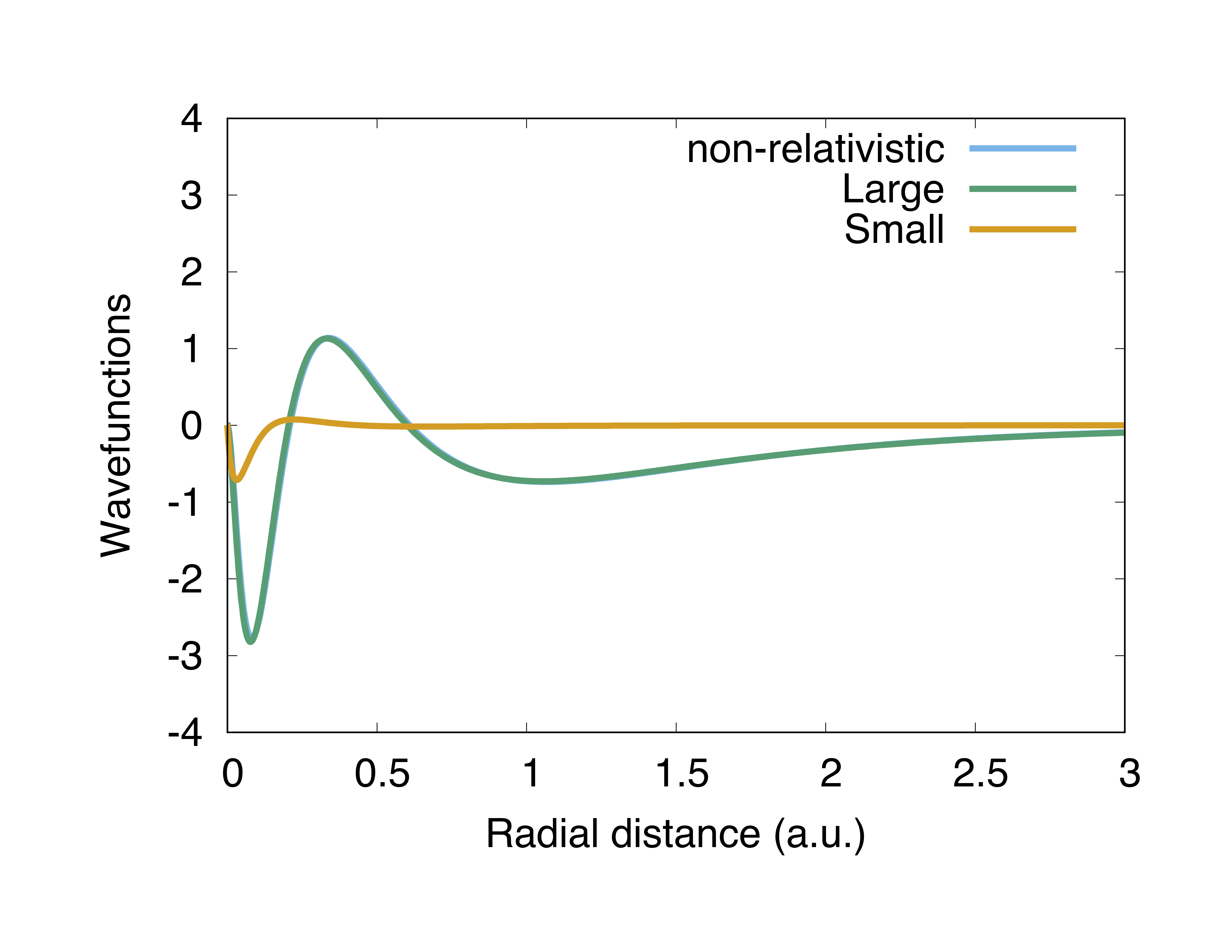}
\includegraphics[width=0.49\linewidth,trim=2cm 2cm 2cm 2cm, clip]{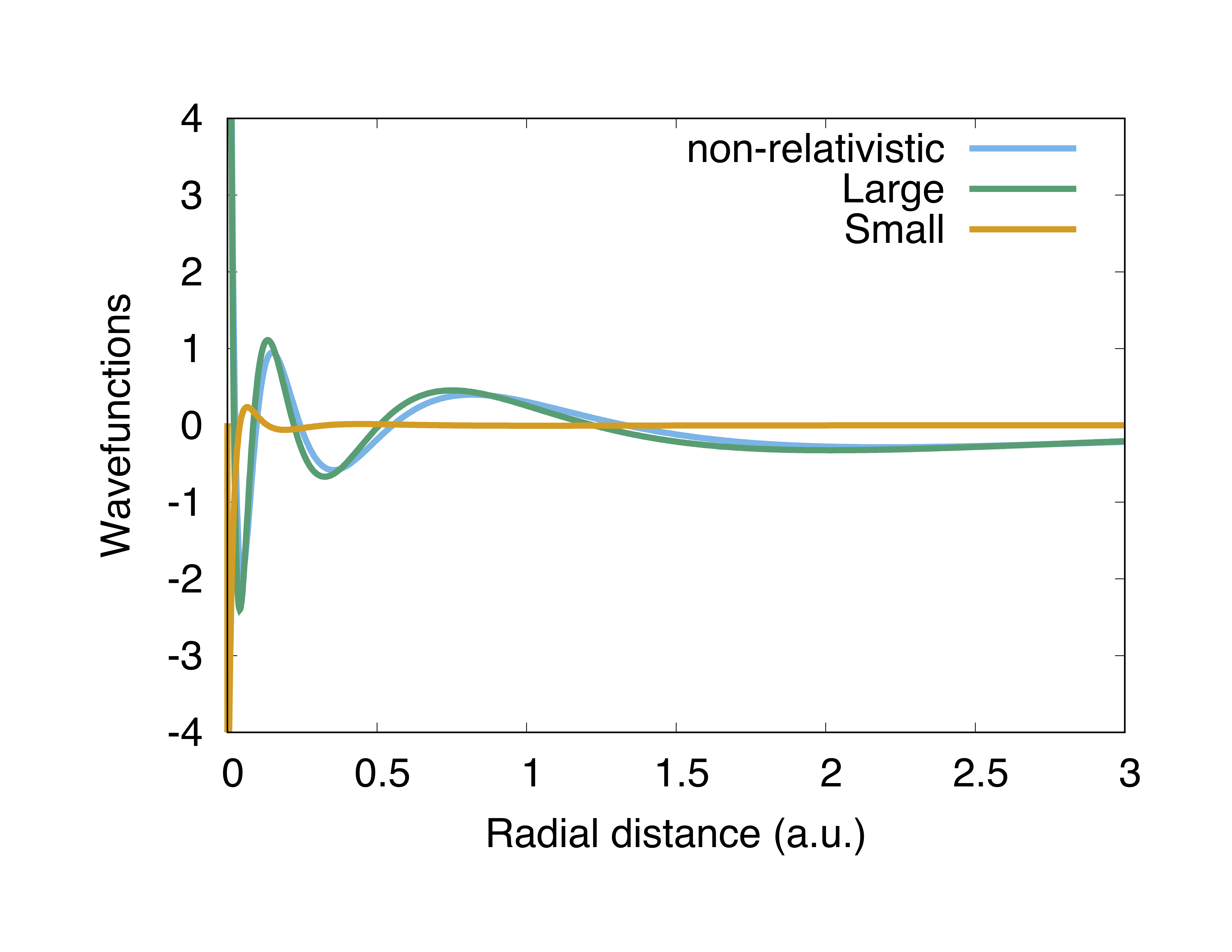}
\caption[width =0.7\textwidth]{Left: relativistic $5d_{3/2}$ vs. non relativistic $5d$ valence
wavefunctions (bottom panel, in green the large part of the Dirac spinor, in orange the small one) and densities (top panel, in orange color the DHF results while in cyan the HF calculations) of Au$^+$ represented in the radial grid. The non-relativistic $5d$ density is similar to the relativistic $5d_{3/2}$ density. Right: relativistic $6s_{1/2}$ vs. non relativistic $6s$ valence
wavefunctions (bottom panel, in green the large part of the Dirac spinor, in orange the small one) and densities (top panel, in orange color the DHF results while in cyan the HF calculations) of Au$^+$. The non relativistic $6s$ density differs from the relativistic $6s_{1/2}$ density for a small shift at larger distances, but shows similar oscillations.}
\end{figure}

\begin{figure}[htp!]
\centering
\includegraphics[width=0.7\linewidth]{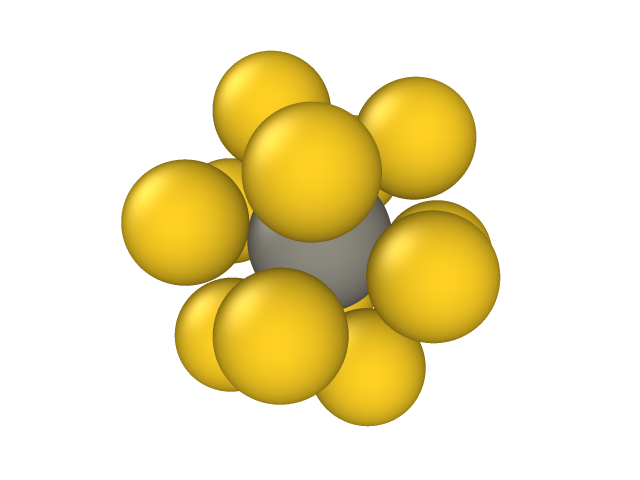}
\caption[width = 0.7\textwidth]{Optimized molecular structure of the W@Au$_{12}$ cage.}
\label{molecules}
\end{figure}

\begin{figure}[htp!]
\centering
\includegraphics[width=0.98\linewidth]{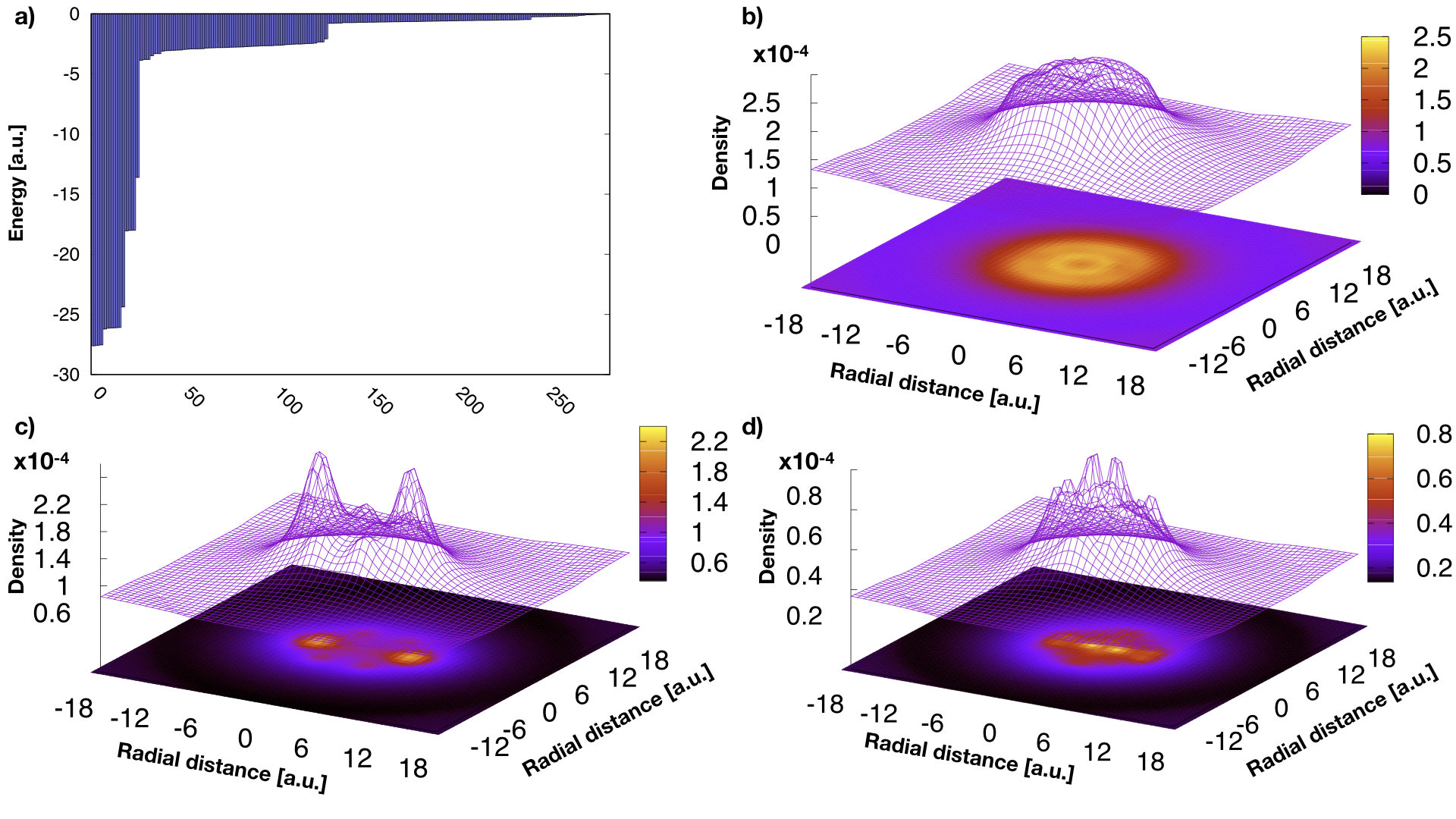}
\caption[width = 0.7\textwidth]{a) Pseudo-atomic orbitals energy-level diagram of the W@Au$_{12}$ nanocage from DHF calculations. b) Virtual state density at -12.7 eV. c) HOMO density. d) LUMO density. The coordinate (0,0) from which the radial distance is measured corresponds to the position of the W atom.}
\label{orbitalsplot}
\end{figure}
In Tab. \ref{molel} we report the values of the independent-particle eigenvalues $\epsilon_i$ of the mean-field operator at different levels of approximations, using either different Gaussian basis sets or the radial grid method. We stress that within the Gaussian basis set approach, we included in the relativistic Hamiltonian also the Gaunt term \cite{rehher2009} for unretarded electron interactions. The latter was at variance neglected in the radial grid calculations. We notice that the DHF total energy is the lowest for the $ANO-R$ modified basis set.
\begin{table}[htp!]
\begin{center}
{\scriptsize{}}%
\begin{tabular}{|c|c|c|c|c|c|c|}
\hline 
{\scriptsize{}Level} & \multicolumn{4}{c|}{{\scriptsize{}Gaussian}} & {\scriptsize{}Radial grid}\tabularnewline
\hline 
  & {\scriptsize{}$HF$} & {\tiny{}$DHF\,\,(ANO-R)$} & {\tiny{}$DHF\,\,(ANO-R)\,\,$mod.}{\scriptsize{} } &
\begin{minipage}[t]{1.8cm}%
{\scriptsize{}$aug-cc-pVTZ-PP$}{\scriptsize\par}  
{\scriptsize{}}{\scriptsize\par}%
\end{minipage} & {\scriptsize{}$Loc.\,\,exch.$}
\tabularnewline
\hline 
{\scriptsize{}$1s_{1/2}$} & {\scriptsize{}$-2362.726$} & {\scriptsize{}$-2568.121$} & {\scriptsize{}$-2563.117$} &  &  {\scriptsize{}$-2582.223$}\tabularnewline
\hline 
{\scriptsize{}$2s_{1/2}$} & {\scriptsize{}$-393.356$} & {\scriptsize{}$-453.526$} & {\scriptsize{}$-447.916$} &  &  {\scriptsize{}$-450.206$}\tabularnewline
\hline 
{\scriptsize{}$2p_{1/2}$} & \multirow{2}{1.2cm}{\scriptsize{}$-376.458$} & {\scriptsize{}$-432.118$} & {\scriptsize{}$-426.482$} &  &  {\scriptsize{}$-432.437$}\tabularnewline
\cline{1-1} \cline{3-7} \cline{4-7} \cline{5-7} \cline{6-7} \cline{7-7} 
{\scriptsize{}$2p_{3/2}$} &  & {\scriptsize{}$-382.822$} & {\scriptsize{}$-377.199$} &  &  {\scriptsize{}$-381.707$}\tabularnewline
\hline 
{\scriptsize{}$3s_{1/2}$} & {\scriptsize{}$-91.813$} & {\scriptsize{}$-111.215$} & {\scriptsize{}$-105.459$} &  &   {\scriptsize{}$-105.165$}\tabularnewline
\hline 
{\scriptsize{}$3p_{1/2}$} & \multirow{2}{1.2cm}{\scriptsize{}$-83.950$} & {\scriptsize{}$-102.003$} & {\scriptsize{}$-96.163$} &  &   {\scriptsize{}$-96.916$}\tabularnewline
\cline{1-1} \cline{3-7} \cline{4-7} \cline{5-7} \cline{6-7} \cline{7-7} 
{\scriptsize{}$3p_{3/2}$} &  & {\scriptsize{}$-91.025$} & {\scriptsize{}$-85.334$} &  &   {\scriptsize{}$-85.875$}\tabularnewline
\hline 
{\scriptsize{}$3d_{3/2}$} & \multirow{2}{1.2cm}{\scriptsize{}$~-69.482$} & {\scriptsize{}$-76.124$} & {\scriptsize{}$-70.260$} &  &   {\scriptsize{}$-71.626$}\tabularnewline
\cline{1-1} \cline{3-7} \cline{4-7} \cline{5-7} \cline{6-7} \cline{7-7} 
{\scriptsize{}$3d_{5/2}$} &  & {\scriptsize{}$-73.575$} & {\scriptsize{}$-67.943$} &  &   {\scriptsize{}$-69.174$}\tabularnewline
\hline 
{\scriptsize{}$4s_{1/2}$} & {\scriptsize{}$-19.489$} & {\scriptsize{}$-27.810$} & {\scriptsize{}$-22.934$} &   &  {\scriptsize{}$-29.070$}\tabularnewline
\hline 
{\scriptsize{}$4p_{1/2}$} & \multirow{2}{1.2cm}{\scriptsize{}$~-16.146$} & {\scriptsize{}$-24.001$} & {\scriptsize{}$-19.078$} &  &   {\scriptsize{}$-22.534$}\tabularnewline
\cline{1-1} \cline{3-7} \cline{4-7} \cline{5-7} \cline{6-7} \cline{7-7} 
{\scriptsize{}$4p_{3/2}$} &  & {\scriptsize{}$-21.207$} & {\scriptsize{}$-16.535$} &   &  {\scriptsize{}$-16.421$}\tabularnewline
\hline 
{\scriptsize{}$4d_{3/2}$} & \multirow{2}{1.2cm}{\scriptsize{}$~-10.078$} & {\scriptsize{}$-15.032$} & {\scriptsize{}$-10.284$} &   &  {\scriptsize{}$-10.479$}\tabularnewline
\cline{1-1} \cline{3-7} \cline{4-7} \cline{5-7} \cline{6-7} \cline{7-7} 
{\scriptsize{}$4d_{5/2}$} &  & {\scriptsize{}$-14.275$} & {\scriptsize{}$-9.794$} &  &  {\scriptsize{}$-9.992$}\tabularnewline
\hline 
{\scriptsize{}$4f_{5/2}$} & \multirow{2}{1.2cm}{\scriptsize{}$~-2.094$} & {\scriptsize{}$-6.566$} & {\scriptsize{}$-1.880$} &    & {\scriptsize{}$-2.379$}\tabularnewline
\cline{1-1} \cline{3-7} \cline{4-7} \cline{5-7} \cline{6-7} \cline{7-7} 
{\scriptsize{}$4f_{7/2}$} &  & {\scriptsize{}$-3.078$} & {\scriptsize{}$-1.784$} &  & {\scriptsize{}$-2.281$}\tabularnewline
\hline 
{\scriptsize{}$5s_{1/2}$} & {\scriptsize{}$-2.807$} & {\scriptsize{}$/$} & {\scriptsize{}$-3.433$} &  {\scriptsize{}$-3.259$}  &{\scriptsize{}$-3.515$} \tabularnewline
\hline 
{\scriptsize{}$5p_{1/2}$} & \multirow{2}{1.2cm}{\scriptsize{}$~-1.746$} & {\scriptsize{}$/$} & {\scriptsize{}$-2.272$} &  {\scriptsize{}$-2.391$} &  {\scriptsize{}$-2.413$}\tabularnewline
\cline{1-1} \cline{3-7} \cline{4-7} \cline{5-7} \cline{6-7} \cline{7-7} 
{\scriptsize{}$5p_{3/2}$} &  & {\scriptsize{}$/$} & {\scriptsize{}$-1.773$} & {\scriptsize{}$-1.932$} &  {\scriptsize{}$-2.008$}\tabularnewline
\hline 
{\scriptsize{}$5d_{3/2}$} & \multirow{2}{1.2cm}{\scriptsize{}$~-0.107$} & {\scriptsize{}$/$} & {\scriptsize{}$-0.350$} &  {\scriptsize{}$-0.378$} & {\scriptsize{}$-0.528$}\tabularnewline
\cline{1-1} \cline{3-7} \cline{4-7} \cline{5-7} \cline{6-7} \cline{7-7} 
{\scriptsize{}$5d_{5/2}$} &  & {\scriptsize{}$/$} & {\scriptsize{}$/$} &  {\scriptsize{}$/$} & {\scriptsize{}$/$}\tabularnewline
\hline 
{\scriptsize{}Total energy} &   {\scriptsize{}$-15286.935$} & & {\scriptsize{}$-16139.352$} &  & \tabularnewline
\hline 
\end{tabular} 
\caption{Mean-field energy levels (Koopmans) of the W atoms obtained by a non-relativistic approach ($HF$), a fully quantum mechanical relativistic approach ($DHF$) using $ANO-R$ contracted ($DHF\,\,(ANO-R)$), $ANO-R$ uncontracted ($DHF\,\,(ANO-R)\,\,$ mod.) Gaussian basis sets. Third and second-last columns report the relativistic values of $\epsilon_i$ using the pseudopotential method with the aug-cc-PVTZ-PP uncontracted and cc-PVTZ-PP uncontracted Gaussian basis sets for the valence electrons \cite{figgen2005a}. Last column reports the results obtained by using a radial grid with the local-exchange interaction (see equation \ref{rho13}). Data in a.u.}\label{molec}
\end{center}
\end{table}

Furthermore, in Fig. (\ref{denwf}) we plot the wavefuntions (bottom panels) and the densities (top panels) for the $5d_{3/2}$ (left) and $6s_{1/2}$ (right) states, respectively, obtained by using a radial mesh of 2000 points. By comparing the non-relativistic HF calculations (cyan lines) with the DHF results (orange lines), we notice that the non-relativistic $5d$ density is similar to the relativistic $5d_{3/2}$ density.\\
\indent A similar calculation was performed in the case of tungsten, which is the foreign atom hosted in the gold cage. In Tab. \ref{molec} we report the single-level energies in a.u. using either different Gaussian basis sets or the radial mesh.

\subsection{The W@Au$_{12}$ superatomic system}

The atomic positions of the W@Au$_{12}$ nanocluster have been optimized by using the VASP program suite \cite{PhysRevB.54.11169,KRESSE199615}. To find the equilibrium structure of W@Au$_{12}$ we used the PBE-PAW pseudo-potential \cite{PhysRevB.50.17953,Kresse_1994,PhysRevB.59.1758} with a plane-wave cut-off equal to 250 eV to treat the ion-valence Coulomb interaction, and a Gaussian smearing of the partial occupancies with a width $\sigma=0.05$ eV. The simulation supercell is a cubic box with 15~\AA~ side, which is large enough to avoid spurious interactions between periodic images (long-range van der Waals forces are not present in this calculation, which makes it equivalent to an infinite box with an affordable computational cost). The Brillouin zone was sampled only at the $\Gamma$ point. The equilibrium geometry of the W@Au$_{12}$ nanocage is represented in Figure (\ref{molecules}). W@Au$_{12}$ presents a icosahedral molecular geometry, having W at the center of the cage with W to Au equilibrium bond distance equal to 2.8~\AA~ and Au-Au 2.95~\AA.\\
\indent The stability of the W@Au$_{12}$ nanocage, and in general of gold clusters, is improved by encapsulating hetero-atoms owing to increased intermolecular metallophilic interactions. Several theoretical and experimental studies have been carried out on gold nano-cages hosting foreign atoms \cite{https://doi.org/10.1002/anie.200290048,C5NR07246H,C8CP04651D,PhysRevLett.95.253401,Bulusu8326,doi:10.1021/jp811103u}. The existence of the icosahedral W@Au$_{12}$ cluster was first theoretically predicted \cite{doi:10.1021/cr00085a006} and soon after successfully synthesized under gas-phase conditions \cite{https://doi.org/10.1002/anie.200290048}. Indeed the Au$_{12}$-cage itself is unstable, thus its stability was attributed to the aurophilic attraction, in turns generated by the strong relativistic effects in gold complexes, and the 18-valence electron rule (1 electron from each of the twelve 6$s^1$ orbitals of Au, and 6 unpaired electrons from the $6s^25d^4$ shells of W). Indeed, the electronic structure of this compact nanocage is characterised by a closed-shell configuration. In our DHF calculation 19 valence electrons for each Au atom and 14 electrons for the W atom, populating the $n$=5,6 orbitals, have been explicitly included in the model Hamiltonian, while the remaining ones (all orbitals up to $n$=4) are treated via the frozen-core pseudo-potentials \cite{figgen2005a}, as outlined in the previous section.
The choice of a small-core PP, where all 5$spd$ orbitals of Au hosting 19 electrons (14 for W) with a similar $\langle r \rangle$ are considered explicitly as valence space, increases the accuracy of the calculations due to a better treatment of the electron correlation using a reduced ``non-chemical'' core \cite{figgen2005a}.
To perform the DHF calculations we used $aug-cc-PVTZ-PP$ Gaussian basis sets, including triple-zeta plus polarization functions ($PTZ$), for the valence orbitals of all the atoms.
In Fig. \ref{orbitalsplot}a) we report the energy-levels diagram and, furthermore, a plot of the density (square of the wavefunctions) for several levels, including a virtual state (Fig. \ref{orbitalsplot}b), the HOMO (Fig. \ref{orbitalsplot}c), and the LUMO (Fig. \ref{orbitalsplot}d)). We notice that the HOMO has a character mainly derived by the Au$6s$ and $5d$ (two major lobes) orbitals with a small contribution from W (central lobe), while the LUMO is composed primarily of W$5d$ orbitals (central lobe). The virtual state is a mixture of Au$ns$ and $nd$ orbitals delocalized all over the cage. The HOMO-LUMO gap from our DHF calculations is around 6 eV, which is an overestimation of the value of 3 eV obtained from previous relativistic density functional calculations using the generalized-gradient Perdew-Wang exchange-correlation functional \cite{https://doi.org/10.1002/anie.200290048} and B3LYP/LANL1DZ \cite{Pyykko} with a similar value for the electron affinity. It is well known that the values of these observables are typically overstimated using the HF approximation. 

\section{Application to the $\beta$-decay of heavy atoms}

\subsection{General theory of $\beta$-decay}

$\beta$-decays occur in unstable nuclei, whereby an excess of neutrons (protons) leads to the conversion into a proton (neutron) with the emission of an electron (positron) and an electron antineutrino (neutrino).
A typical $\beta^{-}$-decay process is the following:

\begin{equation}\label{nbmd}
{}^{A}_{Z}X_{N} \to {}^{A}_{Z+1}X^{'}_{N-1}  + e^{-}+\bar \nu_{e}
\end{equation}

\noindent where the energy released upon decay ($Q$-value), defined as the difference between the initial and final nuclear mass energies, determines whether the transition can (or cannot) occur. In particular, a positive $Q$-value means that the transition is energetically allowed. Other constraints are related to the conservation of the electric charge, lepton number and baryon number of the initial parent nucleus (${}^{A}_{Z}X_{N}$) and of the reaction final products (${}^{A}_{Z+1}X^{'}_{N-1}  + e^{-}+\bar \nu_{e}$). 
In the traditional approach, beta-decay spectra of allowed and forbidden unique
transitions are interpreted by using the following expression \cite{Behrens}:

\begin{equation}\label{rateMg}
    \frac{dN}{dW} \propto pWq^2 F(Z,W) C(W)
\end{equation}
where 
\begin{itemize}
    \item $pWq^2$ is a phase-space factor that accounts for both the relativistic total kinetic energy ($W$, including thus the electron rest mass) and the momentum sharing between the $\beta$-electron ($p$) and  (anti-)neutrino ($q$);
    \item a Fermi function $F(Z,W)$ that accounts for the static corrections due to the Coulomb field of the nucleus;
    \item a shape factor $C(W)$ that accounts for the coupling between the nuclear and lepton dynamics. Typically, this factor reads:
    \begin{equation}\label{cw}
    C(W)=(2L'-1)! \sum_{k=1}^{L'} \lambda_k \frac{p^{2(k-1)}q^{2(L'-k)}}{(2k-1)![2(L'-k)+1]!}
     \end{equation} 
     where $\lambda_k=\frac{(\alpha_{-k}^2+\alpha_k^2)}{\alpha_{-1}^2+\alpha_1^2}$, $\alpha_k$s are the Coulomb amplitudes of the electron wave functions, and $L'$ is related to the order of forbiddance (e.g. $L'=1$ if $\Delta J=0$ for an allowed transition or
$L'=\Delta J$ for any ($L'-1$)th forbidden unique transition). $C(W)$ is assumed independent of the nuclear structure details of the decaying atom \cite{behrens2013}. However, nuclear structure effects cannot be neglected when dealing with forbidden non-unique transitions \cite{Mougeot2015}, and there is no such a simple relation for $C(W)$ in that case as given by Equation (\ref{cw}).
\end{itemize}
To simplify Equation (\ref{cw}) one generally assume all $\lambda_k=1$. Despite convincing 
theoretical arguments \cite{Schopper1966}, a rigorous approach to model forbidden non-unique transitions is still missing \cite{RevModPhys.90.015008}.
Our relativistic approach can help to unravel the physics of beta-decay of these transitions.\\
\indent In our model, the $\beta$-decay rate from an initial state $|i>$ to a final state $|f>$ can be described by using the Fermi golden rule as:
\begin{equation}\label{rate}
N_{i\rightarrow f}=2\pi\frac{\left|\left\langle i|V|f\right\rangle \right|^{2}}{\left\langle i|i\right\rangle \left\langle f|f\right\rangle }\delta(E_{i}-E_{f})
\end{equation}
where
\begin{equation}\label{interact}
\hat{V}=\int d{\bf{r}}\frac{G_{F}}{\sqrt{2}}\left[\hat{\bar{\psi}}_{p}({\bf{r}})\gamma^{\mu}\left(1-x\gamma^{5}\right)\hat{\psi_{n}}(\bf{r})\right]\left[\hat{\bar{\psi}}_{e}(\bf{r})\gamma_{\mu}\left(1-\gamma^{5}\right)\hat{\psi_{\nu}}(\bf{r})\right]=\int d{\bf{r}}V(\bf{r})
\end{equation}
is the phenomenological weak interaction of the Standard Model of particles \cite{Hayen2018b} with
$x=1.26\pm0.02$ and $G_{F}=1.16637\times10^{-5}GeV^{-2}$,
$\hat{\psi_{p}}(\bf{r})$, $\hat{\psi_{n}}(\bf{r})$, $\hat{\psi_{e}}(\bf{r})$,
$\hat{\psi_{\nu}}(\bf{r})$ are the field operators 
that annhilate a proton, a neutron, and electron, and a neutrino, respectively, and  $\gamma_{\mu},\gamma^{\mu}$ are the Dirac matrices ($\hat{\bar{\psi}}=\hat{\psi}^{+}\gamma_{0}$) \cite{rehher2009}.
In particular, we are interested in reckoning the probability per unit time that the atomic system decays 
from a statistical mixture of initial states $\hat{\rho}_i=p_i|i><i|$ to a mixture of final states $\hat{\rho}_f=p_f|f><f|$, that is: 
\begin{equation}\label{rate2}
N_{i\rightarrow f}=2\pi\mathrm{Tr}(\hat{\rho}_i \hat{V}P_f\hat{V})\delta(E_i-E_f)+h.c.
\end{equation}
where $P_f=\sum_{f}|f><f|$ is the projector onto the final states.
In the typical approximation to the general theory of $\beta$-decay the initial and final states can be written as:
\begin{eqnarray}
|i>=|h_i>\otimes |e_i>\\
|f>=|h_f>\otimes |e_f>\otimes |\bar{\nu}_f>
\end{eqnarray}
where $|h_{i,f}>$ are initial and final multi-nucleon states,  $|e_{i,f}>$ are initial and final multi-electron states, and $ |\bar{\nu}_f>$ is the final anti-neutrino state. We point out
that the initial and final multi-nucleon states and the initial multi-electron states are
characterised by a discrete spectrum, while the final multi-electron state, which describes the $\beta$-emission, is of course a continuum state that can be written as linear combination
of external product $ |e_f>=\sum_{j} I_{j,f} \wedge |\eta_{j,f}> $ where $|\eta_{j,f}>$ describes the $\beta$-electron continuum wavefunction.\\
\indent 
Using this, the evaluation of the transition operator
\begin{equation}
V_{fi}=\int d^{3}r\left\langle f|\frac{G_{F}}{\sqrt{2}}\left[\bar{\psi}_{p}(\vec{r})\gamma^{\mu}\left(1-x\gamma^{5}\right)\psi_{n}(\vec{r})\right]\left[\bar{\psi}_{e}(\vec{r})\gamma_{\mu}\left(1-\gamma^{5}\right)\psi_{\nu}(\vec{r})\right]|i\right\rangle 
\end{equation}
can be simplified to:
\begin{eqnarray}
V_{fi} & \simeq & \int d^{3}r\frac{G_{F}}{\sqrt{2}}\left\langle f_{h}|\left[\bar{\psi}_{p}(\vec{r})\gamma^{\mu}\left(1-x\gamma^{5}\right)\psi_{n}(\vec{r})\right]|i_{h}\right\rangle \left\langle f_{l}|\left[\bar{\psi}_{e}(\vec{r})\gamma_{\mu}\left(1-\gamma^{5}\right)\psi_{\nu}(\vec{r})\right]|i_{l}\right\rangle \nonumber \\
 & =& \int d^{3}r\frac{G_{F}}{\sqrt{2}}J_{i\rightarrow f}^{H,\mu}(\vec{r})J_{i\rightarrow f,\mu}^{L}(\vec{r})\label{eq:mat_el_current}
\end{eqnarray}
that is, it can be factorized in the product of leptonic 
\begin{equation} \label{lept}
J_{i\rightarrow f,\mu}^{L}(\vec{r})=\psi_{f,e}^{+}(\vec{r})\gamma_{0}\gamma_{\mu}\left(1-\gamma^{5}\right)\psi_{i,\nu}(\vec{r})
\end{equation} 
and hadronic 
\begin{equation}\label{hadr}
J_{i\rightarrow f,\mu}^{H}(\vec{r})=\psi_{f,p}^{+}(\vec{r})\gamma_{0}\gamma^{\mu}\left(1-x\gamma^{5}\right)\psi_{i,n}(\vec{r}) 
\end{equation}
currents. This is ultimately due to the large rest mass of the $W$ boson, which is the vector that mediates the weak interaction.
We can safely assume that electrons and neutrinos are not coupled and thus the leptonic current can be further factorised into the independent product of the neutrino and electron field operators (or wavefunctions). Furthermore, also the hadronic current can be factorised into the product of neutron and proton wavefunctions in systems where the mean-field approximation is expected to work rather well \cite{morresi2018nuclear} (such as in the case of the odd-even nuclei $^{63}$Ni, $^{129}$I, $^{241}$Pu). Odd-odd nuclei, where many-body effects are expected to play an important role (such as in the case of $^{36}$Cl, $^{210}$Bi, and $^{138}$La), are more complex to model and need a more correlated approach to the nucleon-nucleon binding energy \cite{morresi2018nuclear}. Indeed, the separability of the hadronic current requires basically that only one nucleon is involved in the decay (one neutron to one proton) and it is independent of the remaining nucleons (which build the ``core'', typically approximated by a linear combination of angular momentum wavefunctions). Therefore, in our approach the hadronic current is only partially correlated via the presence of a core of nucleons that has the only goal of recovering the total angular momentum of the parent reactant and of the final daughter nucleus.
However, this does not represent an intrinsic limit of our method, as the hadronic current can be assessed via more accurate first-principles approaches \cite{HjorthJensen:391525}.   

\subsection{Ab-initio calculation of the hadronic and leptonic currents}

In this work we consider a $\beta$-decaying system composed by an isolated nucleus
surrounded by atomic electrons (which can be partially stripped, depending on temperature). Thus, we can safely assume to deal with a spherical symmetric problem. Here we show how to reckon the fermion wavefunctions appearing in the expression of the leptonic and hadronic currents, equations (\ref{lept}) and (\ref{hadr}), respectively.

\subsubsection{Leptonic current}

\paragraph{Neutrino wavefunction}

The neutrino mass is negligible, therefore
we can assume to deal with a free-particle, whose energy-normalized
wave functions 
\begin{equation}
\frac{u_{\kappa}(r)}{r}=k_{\nu}\sqrt{\frac{1}{\pi c}}j_{l_{\kappa}}(k_{\nu}r),\qquad\frac{v_{\kappa}(r)}{r}=k_{\nu}S_{\kappa}\sqrt{\frac{1}{\pi c}}j_{l_{-\kappa}}(k_{\nu}r)
\end{equation}
can be found by solving the free-particle Dirac equation (equation (\ref{DHF}) with $V=0$).
For an emitted antineutrino ($\beta^-$ process), the energy-normalized
solution reads:

\begin{equation}
\psi_{i,\nu}({\bf r})\equiv\psi_{\kappa_{\nu},m_{\nu}}^{-}({\bf r})=\frac{1}{r}\left(\begin{array}{c}
k_{\nu}\sqrt{\frac{1}{\pi c}}j_{l_{\kappa_{\nu}}}(k_{\nu}r)\chi_{\kappa_{\nu},m_{\nu}}(\Omega)\\
-iS_{\kappa_{\nu}}k_{\nu}\sqrt{\frac{1}{\pi c}}j_{l_{-\kappa_{\nu}}}(k_{\nu}r)\chi_{-\kappa_{\nu},m_{\nu}}(\Omega)
\end{array}\right)
\end{equation}

\paragraph{Electron (corrected) wavefunction}\label{subsec:Electron-(corrected)-wavefunctio}

In a spherical symmetric atomic system the spinor wavefunctions $\psi$
can be labelled by means of two azimutal quantum numbers $\kappa$
and $m$ as follows:
\begin{equation}
\psi_{\kappa,m}({\bf r})=\frac{1}{r}\left(\begin{array}{c}
u_{\kappa}(r)\chi_{\kappa,m}(\Omega)\\
iv_{\kappa}(r)\chi_{-\kappa,m}(\Omega)
\end{array}\right)\label{eq: sphsol}
\end{equation}
 where $u_{\kappa}(r)$ and $v_{\kappa}(r)$ are radial functions and
$\chi_{\kappa,m}(\Omega)$ are the spin spherical harmonics.
The solution of the Dirac equation for the atomic electrons is thus simplified by the spherical symmetry but made it cumbersome by
the interaction with the other electrons (in both the initial and in the final states).
Therefore, several level of approximations are in place.
\begin{enumerate}
\item The electron wavefunction is found by solving self-consistently the DHF equation in spherical symmetry (see equation \ref{DHF}) for electrons interacting via a mean-field potential (see equation \ref{potdhf}).

\item 
Asymptotically, when the potential is negligible, they will behave as free particles:
\begin{equation}
\frac{u_{\kappa}(r)}{r}\simeq c_{u}\frac{\sin\left(kr-l_{\kappa}\frac{\pi}{2}+\varphi\right)}{kr},\qquad\frac{v_{\kappa}(r)}{r}\simeq c_{v}\frac{\sin\left(x-l_{-\kappa}\frac{\pi}{2}+\varphi\right)}{kr}
\end{equation}
with normalization condition given by 
\begin{equation}
\frac{\pi}{2k^{2}}\left(\left|c_{u}\right|^{2}+\left|c_{v}\right|^{2}\right)\delta\left(k-k^{\prime}\right)
\end{equation}
which requires that 
\begin{equation}
\left(\left|c_{u}\right|^{2}+\left|c_{v}\right|^{2}\right)=\frac{2k^{2}}{\pi}
\end{equation}
and 
\begin{equation}
c_{u}=k\sqrt{\frac{E+mc^{2}}{\pi E}}A\qquad,\qquad c_{v}=kS_{\kappa}\sqrt{\frac{E-mc^{2}}{\pi E}}A
\end{equation}

\item Inclusion of the non-orthogonality 
between the bound initial $\varphi_{f_i}^{b}$ and final $\varphi_{f_f}^{b}$ orbitals, which are obtained by solving the DHF equation (\ref{DHF}) with two different atomic numbers. 
The continuum electron wavefunction $\psi_{f,e}(\bf{r})$ is then calculated in the field produced
by the nucleus and the surrounding electrons and thus modified by their presence. 
Finally, taking into account the Pauli's principle the bare Coulomb continuum wavefunction $\psi_{f}^{c}(\bf{r})$ corrected for the non-orthogonality between initial and final states can be written as a Slater determinant:  
\begin{equation}
\psi_{f,e}({\bf r})=\left|\begin{array}{ccccc}
\langle\varphi_{f,1}^{b}\big|\varphi_{i,1}^{b}\rangle & \langle\varphi_{f,1}^{b}\big|\varphi_{i,2}^{b}\rangle & \ldots & \langle\varphi_{f,1}^{b}\big|\varphi_{i,N}^{b}\rangle & \varphi_{f,1}^{b}(\bf{r})\\
\langle\varphi_{f,2}^{b}\big|\varphi_{i,1}^{b}\rangle & \langle\varphi_{f,2}^{b}\big|\varphi_{i,2}^{b}\rangle & \ldots & \langle\varphi_{f,2}^{b}\big|\varphi_{i,N}^{b}\rangle & \varphi_{f,2}^{b}(\bf{r})\\
\vdots & \vdots & \ddots & \vdots & \vdots\\
\langle\varphi_{f,N}^{b}\big|\varphi_{i,1}^{b}\rangle & \langle\varphi_{f,N}^{b}\big|\varphi_{i,2}^{b}\rangle & \ldots & \langle\varphi_{f,N}^{b}\big|\varphi_{i,N}^{b}\rangle & \varphi_{f,N}^{b}(\bf{r})\\
\langle\psi_{f}^{c}\big|\varphi_{i,1}^{b}\rangle & \langle\psi_{f}^{c}\big|\varphi_{i,2}^{b}\rangle & \ldots & \langle\psi_{f}^{c}\big|\varphi_{i,N}^{b}\rangle & \psi_{f}^{c}(\bf{r})
\end{array}\right|\label{eq: correction}
\end{equation}

\item Inclusion of the shake-up and shake-off states of the final ion. Indeed,
while we expect that at room conditions the (both nuclear and electronic) ground-to-ground $\beta$-decay gives a major contribution to the rate, to obtain the total rate one has to include 
all possible electronic (and, in principle, also nuclear) excitations that the final ion can undergo. Typically, the inclusion of shake-up and shake-off has a sizable effect on the $\beta$-spectrum only at low energy.
\end{enumerate}

\noindent For a more detailed description of the calculation of the leptonic matrix elements on  a real-space grid we refer the reader to Ref. \cite{morresi2018nuclear}.

\subsubsection{Hadronic current}

\paragraph{Neutron and proton wavefunction calculations}

While we expect that the effective shape of the $\beta$ spectra is almost independent of the behaviour of the hadronic current, however explicit numerical solutions of the Dirac equations for both protons and neutrons must be found in order to asses the transition matrix elements (\ref{eq:mat_el_current}). Assuming that the recoil energy of the final nucleus is small with respect to their rest mass, we can solve the Dirac equation (\ref{DHF}), with a semi-empirical scalar and vector relativistic Wood-Saxon potential to describe the nucleon-nucleon interaction \cite{Woodsaxon}. This potential represents essentially a screened Coulomb interaction as follows:
\begin{equation}\label{WS}
V_C(r)=-V_C\left[ 1 + \exp{\left ( \frac{r-R}{a}\right)} \right]^{-1}
\end{equation}

\noindent where $V_C$ represents the isospin-dependent total strength:
\begin{equation}\label{VWS}
V_C=V_0\left ( 1 \pm \chi\frac{N-Z}{A} \right)
\end{equation}

\noindent and the upper sign is for protons and the lower one for neutrons, while $\chi$ is a constant to be optimized case-by-case. In equation (\ref{WS}) $R=R_0 A^{1/3}$ ($R_0$ is a constant, $A$ is the mass number) and $a$=constant are the size of nuclear potential and diffuseness of the surface, respectively. However, the nuclear force is also spin-dependent. Thus, 
one has to include the nuclear spin-orbit contribution, which can be modelled by:
\begin{equation}\label{WSSO}
\tilde{V}_{SO}(r)=\tilde{V}_{SO}\left[ 1 + \exp{\left ( \frac{r-R_{SO}}{a_{SO}}\right)} \right]^{-1}
\end{equation}
where $\tilde{V}_{SO}$ is the strength, while $R_{SO}=R_{0,SO} A^{1/3}$ and $a_{SO}$=constant are the radius and the diffuseness of the spin-orbit term (typically $a=a_{SO}$).
The spin-orbit interaction strength is determined by $\tilde{V}_{SO}=\lambda V_C$, where $\lambda$ is a proportionality constant.
We point out that the parameters $V_0$, $\chi$, $\lambda$, $a=a_{SO}$, $R_0,R_{0,SO}$ must be optimized by minimizing the difference between experimental and calculated orbital energies. 
In particular, the values that have been used in our simulations are the following $V_0=52.06~$ MeV, $\chi=0.639$, $R_0=1.260~$ fm, $R_{0,SO}=1.160~$ fm, $\lambda=24.1$, and $a=a_{SO}=0.662~$ fm. \\
\indent For a more detailed description of the calculation of the hadronic current term in a real-space radial grid we refer the reader to Ref. \cite{morresi2018nuclear}.

\subsection{Test cases: the $\beta$-decay spectra of $^{31}$Si and $^{89}$Sr}

\begin{figure}[htp!]
\includegraphics[width=.48\linewidth]{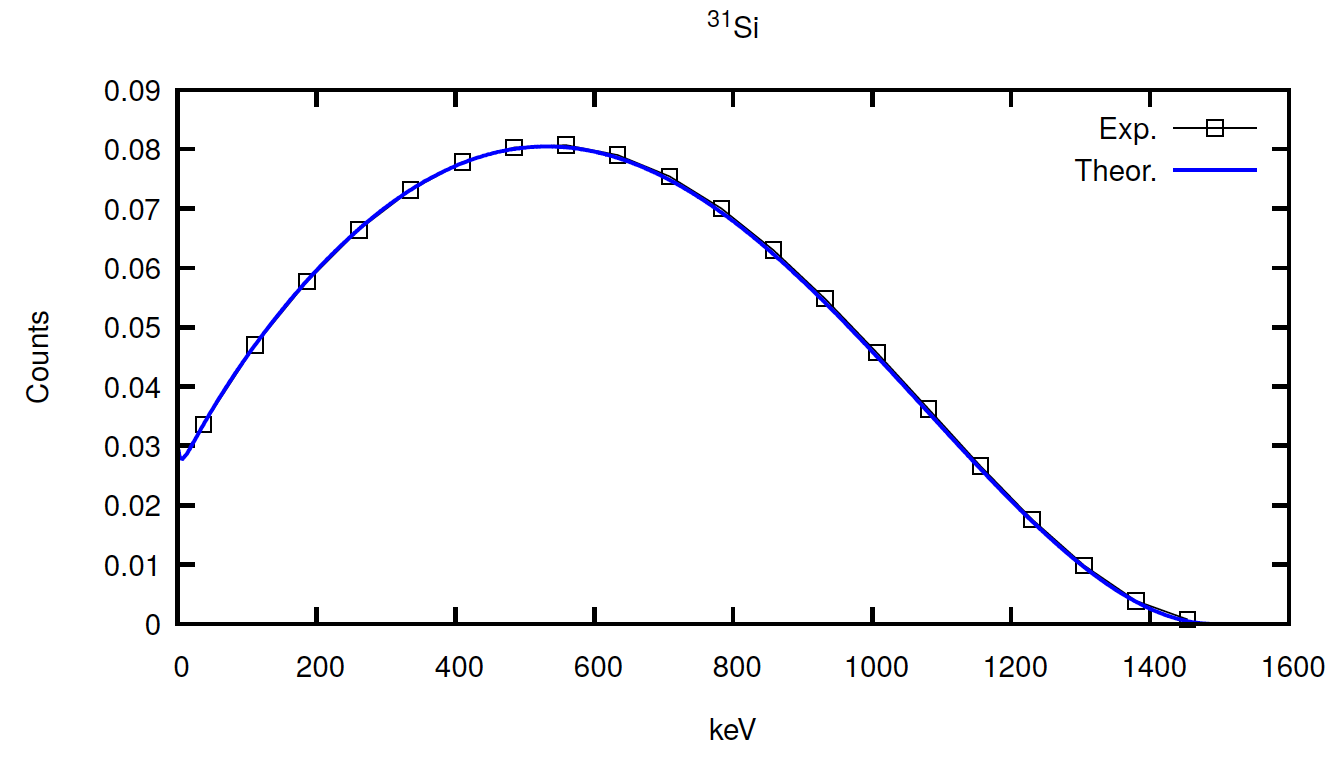}
\includegraphics[width=.48\linewidth]{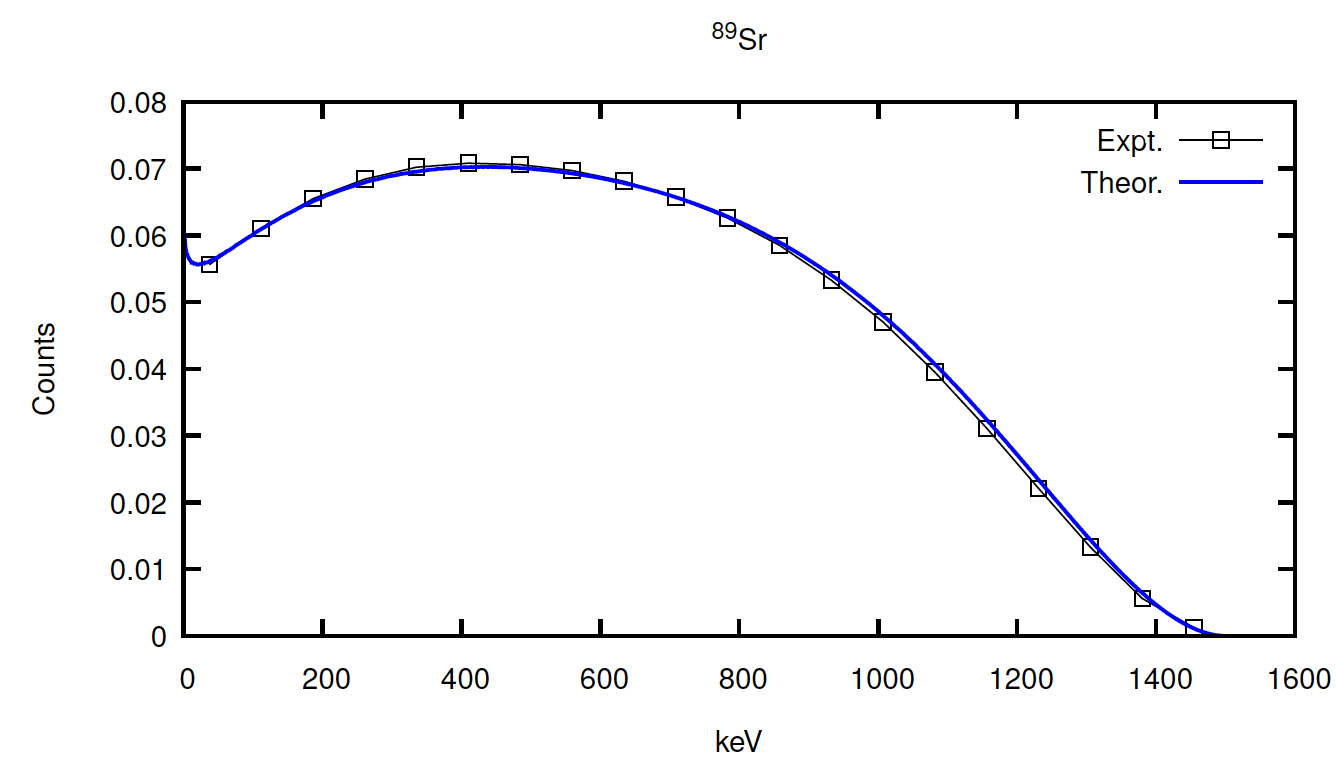}
\caption[]{Left panel: allowed $\beta$ spectrum of $^{31}$Si. Experimental data (empty squares) from Ref. \cite{Eckerman}. Right panel: First forbidden unique $\beta$ spectrum of $^{89}$Sr. Experimental data (empty squares) from Ref. \cite{Eckerman}.}
\label{fig:betadic}
\end{figure}

As a test case of our relativistic approach to $\beta$-decay of heavy-mass nuclei, we show in Fig. \ref{fig:betadic} a comparison between our model and the experimental measurements recorded for the allowed $\beta$ transition of ${}^{31}_{14}$Si$_{17}$ (left panel) and for the first forbidden unique decay of ${}^{89}_{38}$Sr$_{51}$ (right panel). We stress that the agreement between experimental data \cite{Eckerman} and calculations is rather good.

The ${}^{31}_{14}$Si$_{17}$ (odd neutron, even proton) nucleus $\beta^-$-decays 100\% as follows \cite{MARTIN20131497}:

\begin{equation}\label{nbmd1}
^{31}_{14}\mathrm{Si}_{17} \to {}^{31}_{15}\mathrm{P}_{16} + e^{-}+\bar \nu_{e}
\end{equation}

\noindent with a ground state to ground (GS-GS) state $Q$-value (it first decays to the first excited state of P) equal to 1491.50 keV and half-life of 157.36 minutes at room conditions. The parent-nucleus ground state is characterized by the quantum number $3/2^+$ ($J=3/2$ is the total angular momentum and $\pi=+$ the parity of the wavefunction). Protons are in a closed-shell configuration, while one neutron is uncoupled: a mean-field single-particle approximation can describe rather accurately its wavefunction. In the final state, before undergoing a $\gamma$ transition to the ground state, the ${}^{31}_{15}\mathrm{P}_{16}$ nucleus is still in the $3/2^+$ configuration. Thus, parity and total angular momentum are both conserved: this is an allowed transition \cite{morresi2018nuclear}.

The ${}^{89}_{38}$Sr$_{51}$ (odd neutron, even proton) nucleus $\beta^-$-decays 100\% as follows \cite{MARTIN20131497}:

\begin{equation}\label{nbmd2}
^{89}_{38}\mathrm{Sr}_{51} \to {}^{89}_{39}\mathrm{Y}_{50} + e^{-}+\bar \nu_{e}
\end{equation}

\noindent with a ground state to ground state $Q$-value (it first decays to the first excited state of Y) equal to 1500.9 keV and half-life of 50.563 days at room conditions. The parent nucleus ground state is characterized by the quantum numbers $J=5/2^+$. Protons are in a closed-shell configuration, while one neutron is uncoupled: again mean-field single-particle approximation can describe rather accurately its wavefunction. In the final state, before undergoing a $\gamma$ transition to the ground state, the ${}^{89}_{39}\mathrm{Y}_{50}$ nucleus is in the $9/2^+$ configuration. Thus, parity is conserved while the total angular momentum changes by two units: this is first forbidden unique transition \cite{morresi2018nuclear}.

\subsection{$\beta$-decay in astrophysical scenarios: the case of $^{134}$Cs}

\begin{figure}[htp!]
\includegraphics[width=.4\linewidth]{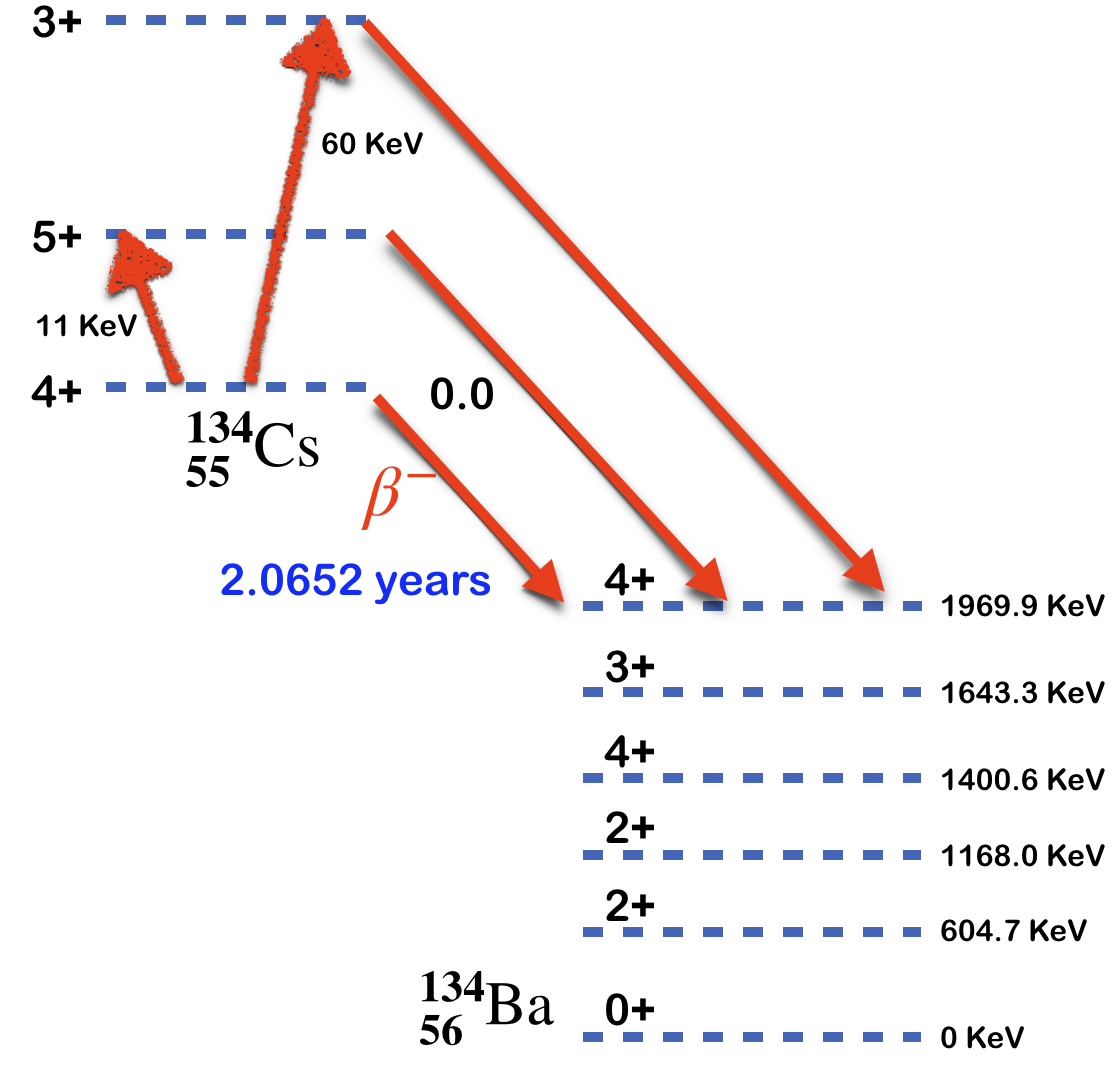}
\includegraphics[width=.55\linewidth]{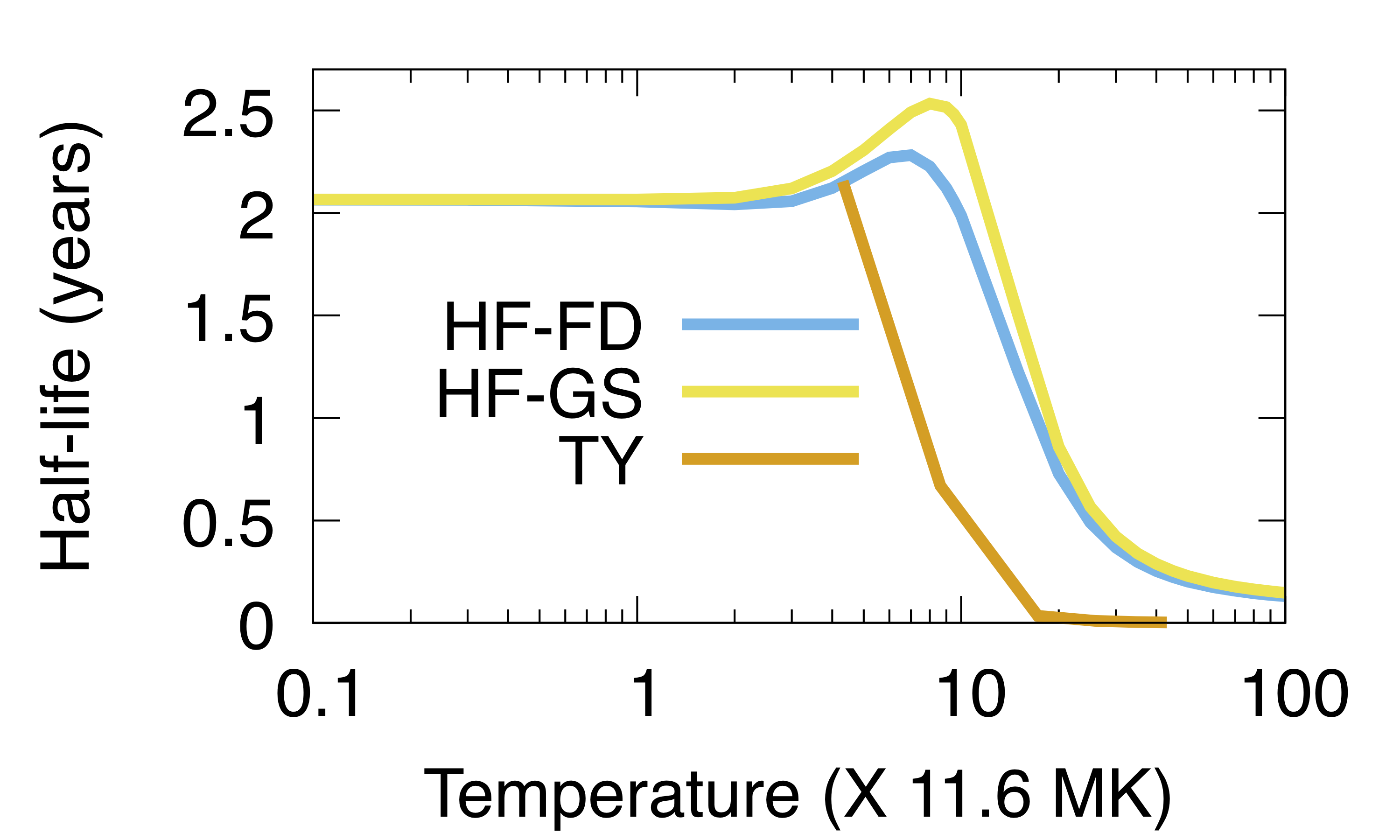}
\caption[]{Left panel: decay scheme of $^{134}_{55}$Cs. Data from \cite{nudat}.
Right panel: half-life (years) in $log~ x-y$ scale of $^{134}_{55}$Cs obtained by using DHF with FD probability distribution for electrons (cyan line), without including the decay to bound states (yellow line, no Fermi-Dirac temperature, orbitals are occupied in their HF ground state). Brown line: Takahashi-Yokoi (TY) calculations \cite{TAKAHASHI1987375,TAKAHASHI1983578} based on systematics  \cite{osti_672449}. Nuclear excited state decays are included in all simulations. The electron density is 10$^{26}$ electrons/cm$^3$.}
\label{trascheme}
\end{figure}

Our understanding of the cosmic chemical abundances is tightly bound to the advancements in measuring or modelling the rates at which nuclear reactions, and in particular $\beta$-decays, occur in stellar environment. However, experimental set-up and procedures to reproduce the (ionized) plasma density and temperature in astrophysical scenarios aiming to measure $\beta$-decay rates 
are still in its infancy \cite{PANDORA,refId0}. 
Therefore, to model the isotopic abundances of chemical elements in stars we must rely on theoretical modelling \cite{Simonucci_2013,vescovi2019effects}.\\
\indent 
It is known that the abundance of Ba in evolved (AGB) stars is determined by a chain of processes that include the slow neutron-capture process occurring in the stable isotope $^{133}$Cs, whose presence is in turn affected by the concentration of Xe in the solar system \cite{lodders2019solar}. Here, we discuss 
specifically the following $\beta$-decay:
\begin{equation}
_{55}^{134}\mathrm{Cs}~ \rightarrow  _{56}^{134}\mathrm{Ba} + e^- + \bar{\nu}
    \nonumber
\end{equation}
$^{134}_{55}$Cs decays to several nuclear excited states of $^{134}_{56}$Ba characterised by different total angular momentum. A transition scheme is reported in the left panel of Fig. \ref{trascheme} \cite{MARTIN20131497}.

The GS-GS $Q$-value is 2058.7 keV and in Earth's conditions the $\beta$-decay occurs in 2.0652 years to the stable nucleus of $^{134}_{56}$Ba.
Additionally to the $4^+$ GS, several nuclear excited states of $^{134}_{55}$Cs can be identified, the lowest energy being $5^+$ (11 keV above the GS) and possibly $3^+$ (60 keV above the GS, unsafe attribution). Given the high temperature in stellar environment and assuming a degeneracy-weighted Boltzmann distribution $p_i=e^{(-E_i/K_{\mathrm{B}} T)}$, where $E_i$ is the energy of the nuclear level, and $T$ the temperature of the stellar plasma ($1$ keV $\approx$ 11.6 MK), Cs nuclear excited states may be populated and contribute to the decay. The excited state level population can modify dramatically the isotopic half-life (typically reducing it by several orders of magnitude).\\
\indent 
The calculation of the $\beta$ decay has been carried out then by solving the DHF equations for both the electron liquid (see section 5.2.1) and the nucleus (see section 5.2.2), using the following approximations:
\begin{itemize}
\item Our rate has been renormalized at all temperatures by a constant factor, which has been obtained so as to recover the experimental $log(ft)$ in our room temperature simulations \cite{TAKAHASHI1987375,TAKAHASHI1983578,osti_672449}. 
\item We assumed that the neutron in the parent Cs nucleus weak decays from the $2d_{3/2}$ shell to the $1g_{7/2}$ shell of daughter Ba (according to the nuclear shell model).
\item The electronic levels of the Cs atom have not been re-optimized at each temperature and are populated according the Fermi-Dirac (FD) distribution for fermions $F(T,\mu)=\frac{1}{1+e^{(\epsilon_i-\mu_{e^{-}})/(K_{\mathrm{B}}T)}}$, where the energies $\epsilon_i$ of the levels $i$ are obtained by the  self-consistent solution of the DHF equation for electrons. 
\item The chemical potentials of electrons and positrons are calculated as ideal Fermi gases in a box using the relativistic energy-momentum dispersion $E^2=c^2p^2+m_e c^4$ by the following relation: 
\begin{equation} n_p=\int_0^{\infty}dp~ p^2/\pi^2\times(F((c\times \sqrt(p^2+c^2)-\mu_e)/K_{\mathrm{B}}T)-F((c\times\sqrt(p^2+c^2)+\mu_e)/K_{\mathrm{B}}T))
\end{equation}
At variance, protons are non-relativistic particles. The density of electrons in the charge-neutral plasma is taken 10$^{26}$ electrons/cm$^3$.
\end{itemize}

Following the decay scheme reported in  the left panel of Fig. \ref{trascheme}, we included several possible decay paths from the $4^+, 5^+,3^+$ nuclear states of Cs, that is Cs$(4^+)\rightarrow \mathrm{Ba}(4^+,3^+,4^+)$; $\mathrm{Cs}(5^+)\rightarrow \mathrm{Ba}(4^+,3^+,4^+)$, and $\mathrm{Cs}(3^+)\rightarrow \mathrm{Ba}(4^+,3^+,4^+,2^+,2^+)$.
In the right panel of Fig. \ref{trascheme} we show the cumulative half-life, taking into account both electron and nuclear excitations, using a FD and a degeneracy-weighted Boltzmann distribution for electrons and nuclei, respectively. \\
\indent We notice that the half-life of the $^{134}_{55}$Cs isotope is dramatically affected by the presence of the three nuclear excited states of Cs (see the fast decrease of the cyan line reported in the right panel of Fig. \ref{trascheme} above 10 keV, 1 keV=11.6 MK) as well as by the electronic temperature (compare the yellow and cyan line shown in the right panel of Fig. \ref{trascheme} without and with electronic temperature, respectively), which may lead to partial or total ionization.
We observe that the nuclear excited state dynamics, which takes into account the population of fast-decaying nuclear excited states, is the most relevant of the two concurrent effects, as it can decrease the rate by a factor of 15 at 100 keV (1 GK) to 23 at 1000 keV with reference to room temperature conditions. This effect is of course in place at higher temperatures ($>$10 keV) according to the high nuclear excitation energies of Cs (11 and 60 keV, respectively).

\section{Electron elastic cross section in liquid water}

Cancer treatment based on fast ion beams -- the so-called hadrontherapy -- is much more efficient than conventional radiotherapy that uses photons as energy carriers \cite{Kraft2000}. Indeed, in the former approach  a substantial fraction of the ion initial kinetic energy is deposited in a region near the end of their trajectories -- the so-called Bragg peak -- while in the latter a majority of the radiation dose is delivered by photons near the entrance and then smoothly distributed along the path inside the body, irradiating also the healthy tissues and causing undesired biodamage especially in the therapy of deep-seated tumors \cite{Nikjoo2016RepProgPhys}.\\
\indent In hadrontherapy the energy lost by the ions in their way within the medium typically results in the emission of secondary electrons ejected via ionizations of the constituents. Most of these secondary electrons are produced with very low energies (below 100 eV) and specific angular pattern \cite{PhysRevB.96.064113,doi:10.1021/cr030453x}. The assessment of the scattering cross sections for the different collisional processes, such as excitation, ionization or elastic scattering \cite{taioli2020relative}, initiated by the ion beam inside the living tissue is of paramount importance to determine the generated secondary electron cascade, the concentration of energy deposition and the damaging events in biotargets.
Typically human tissue is well mimicked by liquid water, which is its main constituent. Here we show how our relativistic approach based on HGBF can help our understanding of the elastic scattering in this medium.\\
\indent 
Elastic scattering of electrons by atomic or molecular centers \cite{gorfinkiel2005electron} can be dealt either with the relativistic Mott theory \cite{Mott1929} or with the direct self-consistent solution of the Dirac equation. \\
\indent The former approach solves the Dirac equation typically using either a semi-empirical electron-electron interaction central potential or a best fit of data from Hartree-Fock (HF) simulations \cite{Dapor2020book}. Then, the differential elastic scattering cross section -- defined as the ratio of the number of particles elastically scattered into the solid angle $d\Omega$ per unit time divided by the incident flux -- is obtained for an unpolarized electron beam by the following expression \cite{Dapor2020book}:
\begin{equation}\label{Mott}
\frac{d\sigma_{\mathrm{el}}}{d\Omega}\,=\,|f(\theta)|^2+|g(\theta)|^2
\end{equation}
where $f(\theta), g(\theta)$ are the direct and spin-flip scattering amplitudes, and $\theta$ represents the scattering angle between the incident and the emitted electron.
Finally, by integrating over all solid angles one obtains the total elastic scattering cross section:
\begin{equation}\label{eqtotel}
 \sigma_{\mathrm{el}}(T) = \int_{\Omega} \frac{d \sigma_{\mathrm{el}}(T,\theta)}{d \Omega} {d \Omega}.
\end{equation}
and from the latter observable the elastic mean free path reads
$\lambda_{el}=\frac{1}{{\cal{N}}\sigma_{el}}$, where $\cal{N}$ is the target atomic or molecular number density.
Equation (\ref{Mott}) is valid to model the elastic scattering of an unpolarized beam of electrons impinging on a single water molecule, which is the typical of the gas phase. To take into account the presence of randomly distributed water molecules in the liquid phase extensions have been proposed \cite{Dapor2020book}.\\
\indent The second approach computes the differential and total elastic cross sections by directly solving the DHF equation (\ref{DHFexplicit}) or (equation (\ref{hamnosimm}) if HGBF are used) for a single molecule or a cluster of water molecules.
Water is a tri-atomic system, thus the numerical solution will be searched by using our fully relativistic quantum mechanical method with HGBF basis sets centered on the atomic nuclei of the molecule(s). We project out the potential interaction only, so as to recover the elastic continuum (see equations \ref{kinterm} and \ref{projectpot}). In particular, the electronic wavefunctions and the mean-field electron-electron interaction potential were expanded using the {\it aug-cc-pVTZ} Gaussian functions basis set optimized for both hydrogen and oxygen atoms \cite{pritchard2019a}. Mono-- and bi-- electronic molecular integrals of the bare Coulomb and exchange interaction are computed at each SCF cycle as outlined in sections 3.2-3.5. \\
\indent 
Once the scattering stationary states $|\psi_{\bf{k}}^+(E)\rangle$ of the projected Dirac Hamiltonian (\ref{hamnosimm}) are obtained at energy $E$,
the differential elastic cross section for solid angle unit is finally reckoned via the Fermi Golden rule as follows:
\begin{equation}
\frac{\mbox{d}\sigma_{el}}{\mbox{d}\Omega}=\frac{m^2}{4\pi^2}|\langle \phi_{k\hat{n}}|T^+(E)|\phi_{\bf{k}}\rangle|^2=\frac{m^2}{4\pi^2}|\langle \phi_{k\hat{n}}|V|\psi_{\bf{k}}^+\rangle|^2 
\label{ESDCS}
\end{equation}
where $\phi_{k\hat{n}}$ is the impinging electron wavefunction represented by a plane-wave with momentum $k$ ($E=\frac{k^2}{2m}$) along the direction of motion $\hat{n}$, $\phi_{\bf{k}}$ is the outgoing free plane wave elastically scattered in the direction $\bf{k}$, and $T^+(E)$ is the on-shell $T$-matrix defined by $T|\phi_{\bf{k}}=V|\psi_{\bf{k}}^+\rangle$. The scattering wavefunction $\psi_\textbf{k}^+(E)$ is characterised by the so-called outgoing $(+)$ wave boundary conditions, that is the eigensolutions of the Dirac equation are matched asymptotically by Coulomb functions \cite{taioli2010electron,taioli2015computational,morresi2018nuclear} defining the normalization condition. 
When the electron kinetic energy is large, the equation (\ref{ESDCS})
can be simplified by using the first-order Born approximation, whereby the $T$-operator expansion in powers of the molecular relativistic potential $T=V+VG_0V+VG_0VG_0V+....$ ($V$ is obtained via the self-consistent solution of the Dirac equation, $G_0=\frac{1}{E-H_0-i\epsilon}$ is the free Green's function, see equation \ref{kinterm}) consists only of the first term, that is $T=V$. Adopting the first-order Born approximation, the formula (\ref{ESDCS}) is equivalent to the Mott cross section (\ref{Mott}).
To assess the differential and total elastic cross section for electrons moving within liquid water one should also consider the presence of neighboring molecules. Thus, we trimmed a cluster of 6 water molecules  after carrying out molecular dynamics (MD) simulations with several thousands molecules reproducing the experimental water density at room conditions ($\rho=1$ g/cm$^3$), using the empirical TIP3P force-field \cite{doi:10.1021/jp973084f} implemented in the LAMMPS package \cite{LAMMPS}. The presence of a number of water molecules takes into account the possibility of multiple scattering effects resulting in possible destructive interference between the escaping waves.
\begin{figure}[htp!]
\includegraphics[width=.49\linewidth,trim=2cm 2cm 2cm 2cm, clip]{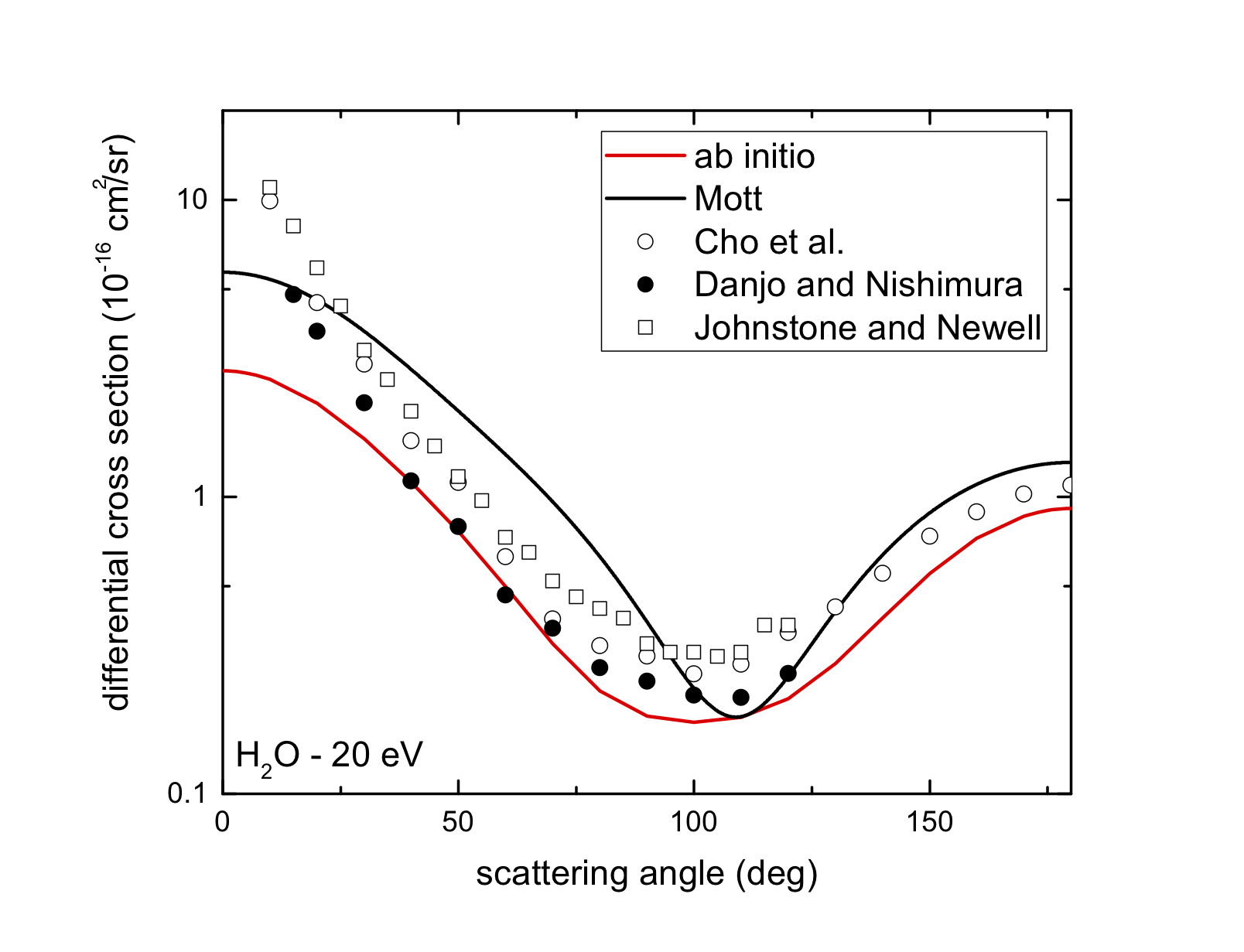}
\includegraphics[width=.49\linewidth,trim=1cm 2cm 2cm 2cm, clip]{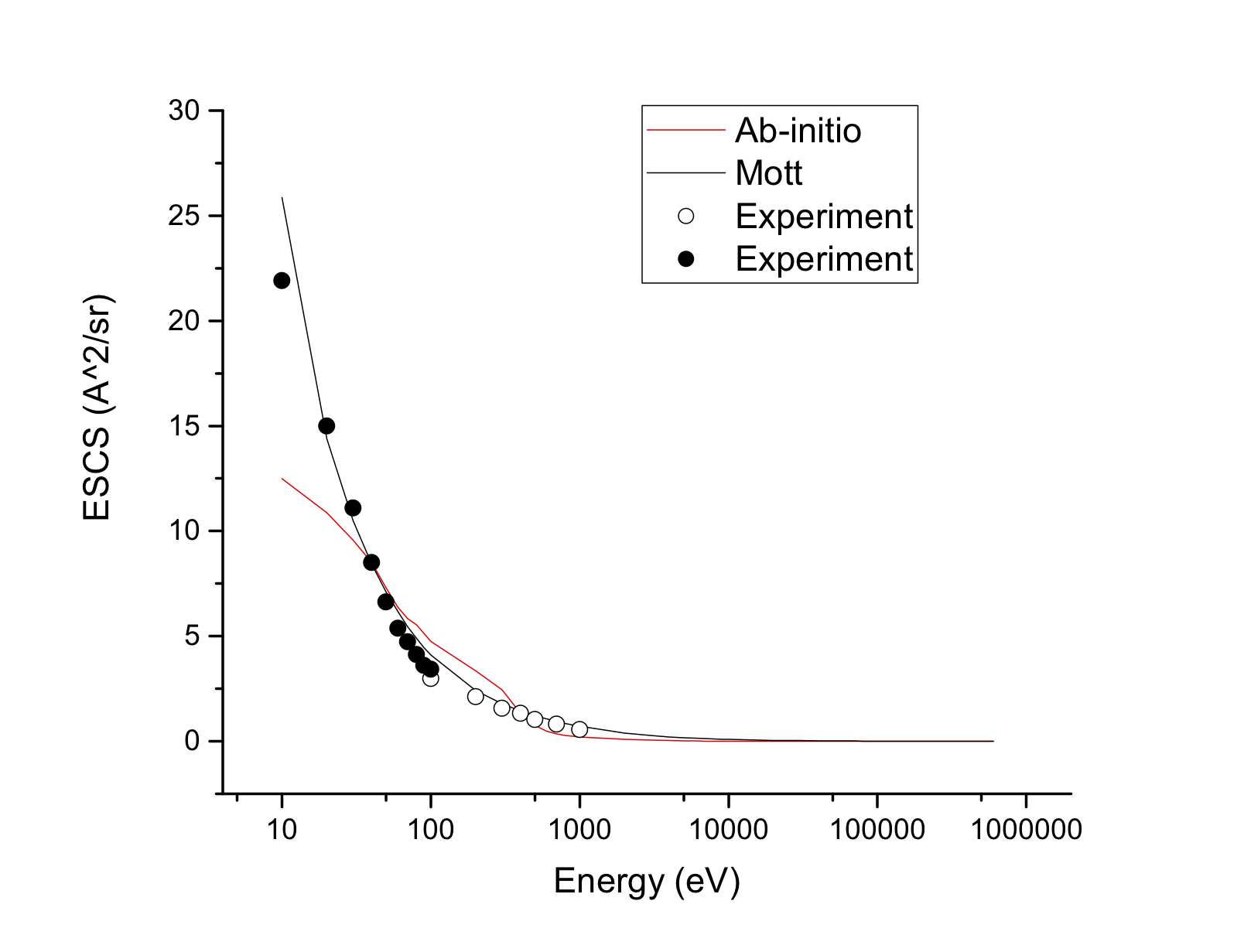}
\caption[]{Left panel: differential elastic scattering cross section (DESCS) of an electron beam impinging with kinetic energy of 20 eV on liquid water as a function of the scattering angle. Calculations from the DHF approach for a cluster of 6 water molecules (see equation \ref{ESDCS}) is reported in solid red line \cite{taioli2020relative}. Calculations from the Mott approach for a single water molecule (see equation \ref{Mott}) is reported in solid black line \cite{taioli2020relative}. Symbols represent experimental data recorded on water vapour from: Cho \cite{Cho2004} (squares), Danjo \cite{Danjo1985} (circles), Johnstone \cite{Johnstone1991} (triangles) and Katase \cite{Katase1986} (diamonds). Right panel: total elastic scattering cross section (ESCS) as a function of the incoming electron beam kinetic energy. Red line reports our ab-initio data for a cluster of 6 water molecules, while black line is obtained by the Mott theory with one water molecule. Experimental data from Refs. \cite{Katase1986,Itikawa2005}.}
\label{fig:ESDCS}
\end{figure} 
In the left panel of Fig. \ref{fig:ESDCS} we report the differential elastic cross sections (DESCS) of an electron beam incident on liquid water with kinetic energy equal to 20 eV. Red and black solid lines correspond to calculations based on 
equation (\ref{ESDCS}) with a water molecule cluster and on equation (\ref{Mott}) with a single water molecule, respectively \cite{taioli2020relative}, while symbols represent experimental data \cite{Cho2004,Danjo1985,Johnstone1991,Katase1986}. In equation (\ref{Mott}) the atomic potential consists of a superposition of Yukawa functions, whose parameters were set according to a best fit of data from Hartree-Fock simulations \cite{doi:10.1063/1.1595653}, while exchange effects were taken into account via the Furness and McCarthy formula \cite{Furness_1973}. We notice that particularly at small scattering angles the ab-initio calculations performed by solving numerically the Dirac equation for a cluster of molecule is in better agreement with experimental data (even though the latter measurements are performed on water vapour). In the right panel we report a comparison between the total elastic cross section obtained by the equation (\ref{eqtotel}) using the fully relativistic approach (red line) for the 6 water molecule cluster, the Mott approach on single water molecule (black line), and the experimental data. 
We notice that the condensed phase nature of liquid water emerges as a significant deviation from the single water molecule particularly at low energy, where the elastic cross section assessed on the cluster is appreciably reduced with respect to the single molecule, stressing the importance of including multiple scattering in the modelling.

\section{Conclusions}

This chapter is aimed at describing the relativistic theory of many-fermion systems, which stems from the merging of the two most fundamental theory of modern physics, that is quantum mechanics and the special theory  of relativity. 
In particular, we focused on the recent theoretical and computational advances of a method for solving the DHF equation for interacting relativistic fermions. This approach can make use of both radial and Gaussian basis sets, which allows one to study any nuclear, atomic, and molecular system with controllable and systematically improvable accuracy.\\
\indent The attractiveness of the HGBF method is due to the  the possibility of writing analytical expressions of the integrals needed for computing the matrix elements of the Coulomb potential for both bound and scattering states. Furthermore, this approach can be also easily applied to multicentric systems, such as molecules and clusters. We showed that HGFB sets, typically created for bound states of quantum systems within the Schr{\"o}dinger picture (thus, in the limit of an infinitely large speed of light), should be modified by uncontraction procedures to account properly for the four-dimensional nature of electrons and nucleons, which ultimately leads to the spin-orbit relativistic effects.  \\
\indent
On the other hand, the spherical symmetry, typically found in atomic and nuclear systems, allows one to formulate the many-fermion problem in the radial coordinate only (in absence of symmetry-destroying fields), resulting in the solution of a mono-dimensional Dirac equation. The choice between these two computational schemes basically depends on the symmetry of the problem under investigation and on the accuracy needed, keeping in mind that the radial basis approach scales almost linearly with the number of mesh points while the HGBF approach grows in between a cubic (due to matrix diagonalization) and quartic (due to the bi-electronic integrals calculation) manner as a function of the number of basis functions.
In this regard, we discussed a few applications of our method to both atomic and molecular systems based on gold, where relativistic affects dictate the electronic structure properties. \\
\indent
The computational effort increasing with system size is faced with the implementation of a relativistic pseudopential approach, where the interaction between ion cores and outer electrons is treated by replacing the core electrons by a weaker pseudopotential that acts on a set of pseudo wavefunctions with the aim to suppress the strong oscillations in the Coulomb potential close to the nucleus that require several Gaussian functions to be accurately described at an increased computational cost.  \\
\indent Furthermore, we showed the extension of our relativistic approach based on these two complementary numerical solutions to deal with i) the elastic continuum for describing the electron-molecule scattering; ii) the $\beta$-decay emission from heavy elements in both Earth and astrophysical scenarios by including the electron temperature via the FD distribution and the nuclear temperature via the Boltzmann population. \\
\indent We point out that in all the case studies discussed in this chapter the description and the  accuracy achievable with our numerical methods rely essentially on the interplay between the so-called relativistic effects, connected to the Dirac four-dimensional representation with respect to the one-dimensional nature of the Schr{\"o}dinger quantum mechanics, and the electron-electron correlation.\\
\indent Finally, desirable and prospective developments of our theoretical and computational schemes will be sought along two different lines. On the theoretical side, we plan to move from the solution of electronic structure problems to the study of electron spectroscopy, most notably to deliver a relativistic theory of the Auger effect in heavy-elements compounds within the framework of the formal theory of scattering. Owing to the paramount importance of the many-body interaction in Auger and resonant decays, this goal requires to devise and develop tailor-made methods for treating dynamic correlation effects beyond the HF approximation. Furthermore, to predict quantitatively the spectral lineshapes one needs to account on the same level of theory not only for the intrinsic dynamical properties of the perturbed system, but also for the feature of the incident beam, be it x-ray photon or electron beams. A further notable application of our approach, which will require serious methodological development, will be in the field of nuclear astrophysics, where nucleosynthesis models still show severe discrepancies with the observations, possibly due to inaccurate (if not wrong) nuclear input data (e.g. the cosmological abundance of $^7$Li). These experiments along with our theoretical and computational tools will also be useful to support the conclusions drawn so far by using Standard Model of particles or to search for a new physics beyond the Standard Model (BSM).

\section{Acknowledgments}

S.S. and S.T. acknowledge funding from the Pandora project provided by the National Institute of Nuclear Physics. 
S.T. acknowledges fruitful discussions with F. Stella. 


\bibliographystyle{elsarticle-num.bst}

\end{document}